\documentclass{article}

\usepackage{amsfonts}
\usepackage{makeidx}
\usepackage{latexsym,amsmath,amssymb,amscd}
\usepackage{makecell}
\usepackage{xcolor}
\usepackage{endnotes}

\let\footnote=\endnote

\newtheorem{theorem}{Theorem}

\newtheorem{proposition}[theorem]{Proposition}

\allowdisplaybreaks

\begin{document}

\title{Quadratic first integrals of autonomous conservative dynamical systems}

\author{Michael Tsamparlis$^{1,a)}$ and Antonios Mitsopoulos$^{1,b)}$ \\
%EndAName
{\ \ }\\
$^{1}${\textit{Faculty of Physics, Department of
Astronomy-Astrophysics-Mechanics,}}\\
{\ \textit{University of Athens, Panepistemiopolis, Athens 157 83, Greece}}
\vspace{12pt}
%EndAName
\\
$^{a)}$Email: mtsampa@phys.uoa.gr
%EndAName
\\
$^{b)}$Email: antmits@phys.uoa.gr
}

\date{}

\maketitle

\begin{abstract}
An autonomous holonomic dynamical system is described by a system of
second order differential equations whose solution gives the trajectories of the system.
The solution is facilitated by the use of first integrals which are used to
reduce the order of the system of differential equations and, if there are
enough of them, to determine the solution. Therefore in the study of
dynamical systems it is important that there exists a systematic method to
determine first integrals of second order differential equations. On the
other hand a system of second order differential equations defines (as a
rule) a kinetic energy (or Lagrangian)\ which provides a symmetric second
order tensor which we call the kinetic metric. This metric via its
symmetries (or collineations) brings into the scene the Differential
Geometry which provides numerous results and methods concerning the
determination of these symmetries. It is apparent that if one manages to
provide a systematic way which will relate the determination of the first
integrals of a given dynamical system with the symmetries of the kinetic
metric defined by this very system, then one will have at his/her disposal
the powerful methods of Differential Geometry in the determination of the
first integrals and consequently the solution of the dynamical equations.
This was also a partial aspect of Lie's work on the symmetries of
differential equations. The subject of the present work is to provide a
theorem which realizes this scenario. The method we follow has been
considered previously in the literature and consists of the following steps.
Consider the generic quadratic first integral of the form $%
I=K_{ab}(t,q^{c})\dot{q}^{a}\dot{q}^{b}+K_{a}(t,q^{c})\dot{q}^{a} +K(t,q^{c})$
where $K_{ab}(t,q^{c}), K_{a}(t,q^{c}), K(t,q^{c})$ are unknown tensor
quantities and require $dI/dt=0.$ This condition leads to a system of
differential equations involving the coefficients $%
K_{ab}(t,q^{c}),K_{a}(t,q^{c}),K(t,q^{c})$ whose solution provides all
possible quadratic first integrals of this form. We demonstrate the
application of the theorem in the classical cases of the geodesic equations
and the generalized Kepler potential in which we obtain all the known
results in a systematic way. We also obtain and discuss the time dependent FIs which are
as important as the autonomous FIs determined by other methods.
\end{abstract}

\section{Introduction}

The dynamical equations of a general holonomic dynamical system have the
functional form
\begin{equation}
\ddot{q}^{a} = \omega^{a}(t,q,\dot{q})  \label{FL.0}
\end{equation}%
where $\omega^{a}= Q^{a}\left( t,q,\dot{q}\right) - \Gamma^{a}_{bc} \dot{q}%
^{b} \dot{q}^{c} - V^{,a}$, $Q^{a}$ are the generalized (non-conservative)
forces, $\Gamma_{bc}^{a}$ are the Riemannian connection coefficients determined from
the kinetic metric $\gamma_{ab}$ (kinetic energy), $-V^{,a}$
are the conservative forces and Einstein summation convention is used.
 Equation (\ref{FL.0}) defines in the jet space $%
J^{1}\left\{ t,q^{a},\dot{q}^{a}\right\} $ the Hamiltonian vector field
\begin{equation}
\mathbf{\Gamma} =\frac{d}{dt}=\frac{\partial }{\partial t}+\dot{q}^{a}\frac{\partial
}{\partial q^{a}}+\omega ^{a}\frac{\partial }{\partial \dot{q}^{a}}.
\label{FI.3}
\end{equation}%
A Lie symmetry with generator $\mathbf{X} =\xi(t,q,\dot{q}) \partial_{t} +\eta^{a}(t,q,\dot{q})\partial_{q^{a}}$ is a point transformation in the jet
space $J^{1}\{t,q^{a},\dot{q}^{a}\}$ which preserves the set of solutions of
(\ref{FL.0}). The mathematical condition for $\mathbf{X}$ to be a Lie symmetry of (%
\ref{FL.0}) is that there exists a function $\lambda \left( t,q,\dot{q}%
\right) $ such that
\begin{equation}
\left[ \mathbf{X}^{\left[ 1\right] },\mathbf{\Gamma} \right] =\lambda (t,q,\dot{q}) \mathbf{\Gamma}
\label{FL.0.1}
\end{equation}%
where $\mathbf{X}^{[1]}=\xi (t,q,\dot{q})\partial _{t}+\eta ^{a}(t,q,\dot{q})\partial
_{q^{a}}+\left( \dot{\eta}^{a}-\dot{q}^{a}\dot{\xi}\right) \partial _{\dot{q}%
^{a}}$ is the first prolongation\footnote{This is the complete lift of $X$ in $TM$.} of $\mathbf{X}$ in $J^{1}\left\{ t,q^{a},%
\dot{q}^{a}\right\}$. Equivalently we have
\begin{equation}
\mathbf{X}^{[2]}H^{a}=0\implies \mathbf{X}^{[1]}H^{a}+ \eta^{a[2]}=0   \label{FL.0.2}
\end{equation}%
where $H^{a}\equiv \ddot{q}^{a}-\omega ^{a}$, $\eta ^{a[2]}=\ddot{\eta}^{a}-2%
\ddot{q}^{i}\dot{\xi}-\dot{q}^{i}\ddot{\xi}$ and $\mathbf{X}^{[2]}= \mathbf{X}^{[1]}+\eta
^{a[2]}\partial _{\ddot{q}^{a}}$ is the second prolongation of $\mathbf{X}$ in $%
J^{2}\left\{ t,q^{a},\dot{q}^{a},\ddot{q}^{a}\right\} .$

The standard method to determine the first integrals (FIs) of a Lagrangian
dynamical system is to use special Lie symmetries, the Noether symmetries. A
Noether symmetry is a Lie symmetry which in addition satisfies the Noether condition\footnote{M. Lutzky, J. Phys. A 11, 249 (1978).\label{Lutzky}}$^{,}$\footnote{W. Sarlet and F. Cantrijin, J. Phys. A: Math. Gen. 14, 479 (1981).\label{Sarlet}}$^{,}$\footnote{W. Sarlet and F. Cantrijin, SIAM Review 23, 467 (1981).\label{Sarlet Cantrijn 81}}$^{,}$\footnote{T.M. Kalotas and B.G. Wybourne, J. Phys. A: Math. Gen. 15, 2077 (1982).\label{kalotas}}
\begin{equation}
\mathbf{X}^{[1]}L+\frac{d\xi }{dt}L=\frac{df}{dt}  \label{FI.1}
\end{equation}%
where $f(t,q,\dot{q})$ is the gauge or the Noether function. According to
Noether's theorem\footnote{E. Noether, Nachr. d. K\"{o}nig. Gesellsch. d. Wiss. zu G\"{o}ttingen, Math-phys. Klasse, 235 (1918) (translated in English by
M.A. Tavel [physics/0503066]). \label{noe1}}$^{,}$\footnote{G.P. Flessas, P.G.L. Leach and S. Cotsakis, Can. J. Phys. 73, 543 (1995).\label{leachnoe1}} to every Noether symmetry there corresponds the FI
\begin{equation}
I=\xi \left( \dot{q}^{a}\frac{\partial L}{\partial \dot{q}^{a}}-L\right)
-\eta ^{a}\frac{\partial L}{\partial \dot{q}^{a}}+f  \label{FI.4}
\end{equation}%
which can easily be determined if one knows the generator $\mathbf{X}$ of the Lie
symmetry. The FIs are used to reduce the order of the dynamical
equations. It has been shown in Ref. \ref{Sarlet Cantrijn 81} that the FIs
 of Noether symmetries of holonomic autonomous Lagrangian
systems are invariants of the prolonged vector field $\mathbf{X}^{[1]},$ therefore
they satisfy the two conditions $\frac{dI}{dt}= \mathbf{X}^{[1]}(I)=0$, which provide
the reduction of the dynamical equations by two.

Noether symmetries being special Lie symmetries may be considered in two classes:

a. Noether point symmetries resulting from Lie point symmetries of the form $\xi (t,q)$, $\eta ^{a}(t,q)$ and

b. Dynamical Noether symmetries resulting from dynamical Lie symmetries for
which $\xi (t,q,\dot{q},\ddot{q},...)$, \newline $\eta ^{a}(t,q,\dot{q},\ddot{q},...). $

In the following we restrict our discussion to dynamical Noether symmetries
in $J^{1}\{t,q^{a},\dot{q}^{a}\}$ therefore $\xi (t,q,\dot{q})$, $\eta^{a}(t,q,\dot{q})$.

Noether point symmetries form a finite dimensional Lie algebra and dynamic
Noether symmetries an infinite dimensional Lie algebra.

In dynamical Lie symmetries one has an extra degree of freedom which is
removed if one demands an extra condition in which case one works with the
so-called gauged dynamical Lie symmetries. In this respect one usually
requires the gauge condition $\xi =0$ so that the generator is simplified to
$\mathbf{X}=\eta ^{a}(t,q,\dot{q})\partial _{q^{a}}.$ This gauge condition will be
tacitly assumed in the following.

Concerning the geometric nature of Noether symmetries it has been shown\footnote{M. Tsamparlis and A. Paliathanasis, J. Phys. A: Math. Theor. 45, 275202 (2012).\label{lew2}}$^{,}$\footnote{A. Paliathanasis and M. Tsamparlis, J. Geom. Phys. 62, 2443
(2012).\label{p3}} that the generators of Noether point symmetries of
autonomous holonomic dynamical systems with a regular Lagrangian (i.e. $\det\frac{\partial ^{2}L}{\partial \dot{q}^{a}\dot{q}^{b}}\neq 0$) of the form $%
L=\frac{1}{2}\gamma _{ab}\dot{q}^{a}\dot{q}^{b}-V(q)$, where $\gamma _{ab}=%
\frac{\partial ^{2}L}{\partial \dot{q}^{a} \partial\dot{q}^{b}}$ is the kinetic
metric defined by the Lagrangian, are elements of the homothetic algebra of $%
\gamma _{ab}.$ A similar firm result does not exist for dynamical Noether
symmetries beyond the fact that their generators form an infinite
dimensional Lie algebra.

In this paper we consider the questions: \emph{To what extent the first
integrals of (\ref{FL.0}) are covered by Noether symmetries, that is, are
there non-Noetherian first integrals? Furthermore}, \emph{how and to what
extent one can `geometrize' the dynamical Noether symmetries?}

This question is not new. It was raised for the first time by Darboux\footnote{G. Darboux, Archives Neerlandaises (ii) 6, 371 (1901).\label{Darboux}} and Whittaker\footnote{E.T. Whittaker, ``A Treatise on the Analytical Dynamics of Particles and Rigid Bodies'', Cambridge University Press, Ch. 12,

(1937).\label{Whittaker}} who considered the Newtonian
autonomous holonomic systems with two degrees of freedom and determined most
potentials $V(q)$ for which the system has a quadratic first integral other
than the Hamiltonian (energy). The complete answer to this problem was given much later by G. Thompson\footnote{G. Thompson, J. Math. Phys. 25, 3474 (1984). \label{Thompson 1984 II}}$^{,}$\footnote{G. Thompson, J. Phys. A: Math Gen 17, 985 (1984). \label{Thompson 1984}}.

The same problem for the general autonomous dynamical system in a Riemannian space has also been considered more recently$^{\ref{kalotas}}$$^{,}$\footnote{H. Stephani, ``Differential Equations: Their Solutions using Symmetry'', Cambridge University Press, New York, (1989).\label{StephaniB}}. In this latter approach one assumes the generic quadratic
first integral (QFI) to be of the form%
\begin{equation}
I=K_{ab}\dot{q}^{a}\dot{q}^{b}+K_{a}\dot{q}^{a}+K  \label{FL.5}
\end{equation}%
where the coefficients $K_{ab},K_{a},K$ are tensors depending on the
coordinates $t, q^{a}$ and imposes the condition $\frac{dI}{dt}=0.$ This
condition leads to a system of differential equations involving the unknown
quantities $K_{ab},K_{a},K$ whose solution provides the QFIs (\ref{FL.5}).

In all occasions considered so far the system of these conditions has been solved for specific cases only.

The aim of the present work\footnote{A recent preliminary work along this line is the following: L. Karpathopoulos, M. Tsamparlis and A. Paliathanasis, J. Geom.
Phys. 133, 279 (2018). \label{Karp}} is the following:

a. To give in the case of autonomous conservative equations the general solution of the system resulting from the condition $%
\frac{dI}{dt}=0$.

b. To geometrize the answer to the maximum possible degree.

c. To determine the generalized/point Noether symmetries which admit the
resulting first integrals as Noether integrals.

In order to do that we work in a similar way with the previous authors. That
is we consider a not (necessarily Lagrangian) dynamical system and determine
the system of equations involving the unknown quantities $K_{ab},K_{a},K.$
We solve this system and determine the quadratic first integrals of the form
(\ref{FL.5}). Nowhere we use the concept of symmetry. Subsequently for each
first integral we compute the generator of the transformation (not
necessarily a Noether transformation) in the jet bundle $J^{1}\{t,q^{a},\dot{%
q}^{a}\}$ which generates this integral.

In the case the dynamical system is Lagrangian - which is almost always the
case because we can always take the Lagrangian to be  the Kinetic energy -
we can associate a gauged generalized Noether symmetry whose Noether
integral is the considered first integral. It will be shown in section \ref%
{sec.generators} that the generators of these Noether symmetries are read
directly from the expression of the FI with no further calculations. It
follows that all quadratic first integrals of the form (\ref{FL.5}) are
Noetherian, provided the Lagrangian is regular.

\section{The conditions for a quadratic first integral}

\subsection{The case of a general dynamical system}

In this section we consider a dynamical system defined by the equations of
motion
\begin{equation}
\ddot{q}^{a}=Q^{a}\left( t,q,\dot{q}\right) - \Gamma^{a}_{bc} \dot{q}^{b}%
\dot{q}^{c}-V^{,a}  \label{FL.0.3}
\end{equation}%
where $Q^{a}$ are the non-conservative forces, $\Gamma^{a}_{bc}$ are the Riemannian
connection coefficients determined form the kinetic metric $\gamma_{ab}$
defined by the kinetic energy and $-V^{,a}$ are the conservative forces.

We consider next a function $I(t,q^{a},\dot{q}^{a})$ which is linear and
quadratic in the velocities with coefficients which depend only on the
coordinates $t,q^{a}$ ,that is, $I$ it is of the form
\begin{equation}
I=K_{ab}(t,q)\dot{q}^{a}\dot{q}^{b}+K_{a}(t,q)\dot{q}^{a}+K(t,q)
\label{FI.5}
\end{equation}%
where $K_{ab}$ is a symmetric tensor, $K_{a}$ is a vector and $K$ is an
invariant.

We demand that $I$ is a FI of (\ref{FL.0.3}). This
requirement leads to the condition
\begin{equation}
\frac{dI}{dt}=0  \label{DS1.10a}
\end{equation}%
which gives a system of equations for the coefficients $K_{ab},$ $K_{a}$ and
$K$. Using the dynamical equations (\ref{FL.0.3}) to replace $\ddot{q}^{a}$
whenever it appears we find\footnote{%
Round brackets indicate symmetrization of the enclosed indices. A comma
indicates partial derivative and a semicolon Riemannian covariant derivative.%
}
\begin{align}
\frac{dI}{dt}& =K_{(ab;c)}\dot{q}^{a}\dot{q}^{b}\dot{q}^{c}+\left(
K_{ab,t}+K_{a;b}\right) \dot{q}^{a}\dot{q}^{b}+2K_{ab}\dot{q}%
^{(b}(Q^{a)}-V^{,a)})+\left( K_{a,t}+K_{,a}\right) \dot{q}^{a}+  \notag \\
& \quad +K_{a}(Q^{a}-V^{,a})+K_{,t}.  \label{eq.veldep3}
\end{align}%
In order to get a working environment we restrict our considerations to
linear generalized forces, that is we consider the case $Q^{a}=A_{b}^{a}(q)%
\dot{q}^{b}$. Then the general result (\ref{eq.veldep3}) becomes
\begin{align*}
0& =K_{(ab;c)}\dot{q}^{a}\dot{q}^{b}\dot{q}^{c}+\left(
K_{ab,t}+K_{a;b}+2K_{c(b}A_{a)}^{c}\right) \dot{q}^{a}\dot{q}^{b}+\left(
K_{a,t}+K_{,a}-2K_{ab}V^{,b}+\right. \\
& \quad \left. +K_{b}A_{a}^{b}\right) \dot{q}^{a}+K_{,t}-K_{a}V^{,a}
\end{align*}%
from which follows the system of equations
\begin{eqnarray}
K_{(ab;c)} &=&0  \label{eq.veldep4.1} \\
K_{ab,t}+K_{(a;b)}+2K_{c(b}A_{a)}^{c} &=&0  \label{eq.veldep4.2} \\
-2K_{ab}V^{,b}+K_{a,t}+K_{,a}+K_{b}A_{a}^{b} &=&0  \label{eq.veldep4.3} \\
K_{,t}-K_{a}V^{,a} &=&0.  \label{eq.veldep4.4}
\end{eqnarray}%
Condition $K_{(ab;c)}=0$ implies that $K_{ab}$ is a Killing tensor (KT) of order
2 (possibly zero) of the kinetic metric $\gamma _{ab}$. Because $\gamma_{ab} $ is autonomous the condition $K_{(ab;c)}=0$ is satisfied if $K_{ab}$ is of the form
\begin{equation*}
K_{ab}(t,q)=g(t)C_{ab}(q)
\end{equation*}%
where $g(t)$ is an arbitrary analytic function and $C_{ab}(q)$ ($C_{ab}=C_{ba}$) is a Killing tensor of order 2 of the metric $\gamma _{ab}.$
This choice of $K_{ab}$ and equation (\ref{eq.veldep4.2}) indicate that we
set
\begin{equation*}
K_{a}(t,q)=f(t)L_{a}(q)+B_{a}(q)
\end{equation*}%
where $f(t)$ is an arbitrary analytic function and $L_{a}(q),B_{a}(q)$ are
arbitrary vectors. With these choices the system of equations (\ref%
{eq.veldep4.1}) -(\ref{eq.veldep4.4}) becomes
\begin{eqnarray}
g(t)C_{(ab;c)} &=&0  \label{eq.veldep5} \\
g_{,t}C_{ab}+f(t)L_{(a;b)}+B_{(a;b)}+2g(t)C_{c(b}A_{a)}^{c} &=&0
\label{eq.veldep6} \\
-2g(t)C_{ab}V^{,b}+f_{,t}L_{a}+K_{,a}+(fL_{b}+B_{b})A_{a}^{b} &=&0
\label{eq.veldep7} \\
K_{,t}-(fL_{a}+B_{a})V^{,a} &=&0.  \label{eq.veldep8}
\end{eqnarray}

Conditions (\ref{eq.veldep5}) - (\ref{eq.veldep8}) must be supplemented with
the integrability conditions $K_{,at}=K_{,ta}$ and $K_{,[ab]}=0$ for the
scalar function $K$. The integrability condition $K_{,at}=K_{,ta}$ gives -
if we make use of (\ref{eq.veldep7}) and (\ref{eq.veldep8}) - the equation
\begin{equation}
f_{,tt}L_{a}+f_{,t}L_{b}A_{a}^{b}+f\left( L_{b}V^{;b}\right) _{;a}+\left(
B_{b}V^{;b}\right) _{;a}-2g_{,t}C_{ab}V^{,b}=0.  \label{eq.veldep9}
\end{equation}
Condition $K_{,[ab]}=0$ gives the equation
\begin{equation}
2g\left( C_{[a\left\vert c\right\vert }V^{,c}\right) _{;b]}-f_{,t}L_{\left[
a;b\right] }-(fL_{c;[b}+B_{c;[b})A_{a]}^{c}-(fL_{c}+B_{c})A_{[a;b]}^{c}=0
\label{eq.veldep10}
\end{equation}
which is known as the second order Bertrand-Darboux equation.

Finally the system of equations which we have to solve consists of equations
(\ref{eq.veldep5}) - (\ref{eq.veldep10}).

\subsection{The case of autonomous conservative dynamical systems}

\label{sec.tables.theorem}

We restrict further our considerations to the case of autonomous
conservative dynamical systems so that $V=V(q)$ and $Q^{a}=0$. In this case
the system of equations (\ref{eq.veldep5}) - (\ref{eq.veldep10}) reduces as
follows
\begin{eqnarray}
gC_{(ab;c)} &=&0  \label{FL.1.a1} \\
g_{,t}C_{ab}+fL_{(a;b)}+B_{(a;b)} &=&0  \label{FL.1.a} \\
-2gC_{ab}V^{,b}+f_{,t}L_{a}+K_{,a} &=&0  \label{FL.1.b} \\
K_{,t}-fL_{a}V^{,a}-B_{a}V^{,a} &=&0  \label{FL.1.c} \\
f_{,tt}L_{a}+f(L_{b}V^{,b})_{;a}+(B_{b}V^{,b})_{;a}-2g_{,t}C_{ab} V^{,b} &=&0
\label{FL.1.d} \\
2g\left( C_{[a\left\vert c\right\vert }V^{,c}\right) _{;b]}-f_{,t}L_{[a;b]}
&=&0.  \label{FL.1.e}
\end{eqnarray}

These equations have been found before by e.g. Kalotas (see in Ref. \ref{kalotas} equations (12a) - (12d) ) who considered their solution in certain special cases.

Obviously the solution of this system of equations is quite involved and
requires the consideration of many cases and subcases. The general solution
of the system is stated in the following Theorem (the proof is given in
Appendix A).

\begin{theorem}
\label{The first integrals of an autonomous holonomic dynamical system}

We assume that the functions $g(t),f(t)$ are analytic so that they may be
represented by polynomial functions as follows
\begin{equation}  \label{eq.thm1}
g(t) = \sum^n_{k=0} c_k t^k = c_0 + c_1 t + ... + c_n t^n
\end{equation}
\begin{equation}  \label{eq.thm2}
f(t) = \sum^m_{k=0} d_k t^k = d_0 + d_1 t + ... + d_m t^m
\end{equation}
where $n, m \in \mathbb{N}$, or may be infinite, and $c_k, d_k \in \mathbb{R}
$. Then the independent QFIs  of an autonomous conservative dynamical system
are the following: \bigskip

\textbf{Integral 1.}
\begin{equation*}
I_{1} = -\frac{t^{2}}{2} L_{(a;b)}\dot{q}^{a}\dot{q}^{b} + C_{ab}\dot{q}^{a} \dot{q}^{b} + t L_{a} \dot{q}^{a} + \frac{t^{2}}{2} L_{a}V^{,a} + G(q)
\end{equation*}
where $C_{ab}$, $L_{(a;b)}$ are KTs, $\left(L_{b}V^{,b}\right)_{,a} =
-2L_{(a;b)} V^{,b}$ and $G_{,a}= 2C_{ab}V^{,b} - L_{a}$.

\textbf{Integral 2.}
\begin{equation*}
I_{2} = -\frac{t^{3}}{3} L_{(a;b)}\dot{q}^{a}\dot{q}^{b} + t^{2} L_{a} \dot{q%
}^{a} + \frac{t^{3}}{3} L_{a}V^{,a} - t B_{(a;b)} \dot{q}^{a}\dot{q}^{b} +
B_{a}\dot{q}^{a} + tB_{a}V^{,a}
\end{equation*}
where $L_{a}$, $B_{a}$ are such that $L_{(a;b)}$, $B_{(a;b)}$ are KTs, $%
\left(L_{b}V^{,b}\right)_{,a} = -2L_{(a;b)} V^{,b}$ and $\left(B_{b}V^{,b}%
\right)_{,a} = -2B_{(a;b)} V^{,b} - 2L_{a}$.

\textbf{Integral 3.}
\begin{equation*}
I_{3} = -e^{\lambda t} L_{(a;b)}\dot{q}^{a}\dot{q}^{b} + \lambda e^{\lambda
t} L_{a} \dot{q}^{a} + e^{\lambda t} L_{a} V^{,a}
\end{equation*}
where $\lambda \neq 0$, $L_{a}$ is such that $L_{(a;b)}$ is a KT and $\left(L_{b}V^{,b}
\right)_{,a} = -2L_{(a;b)} V^{,b} - \lambda^{2} L_{a}$.
\end{theorem}

\bigskip

For easier reference in the following Tables we collect the LFIs (Linear First Integrals) and the
QFIs  of Theorem \ref{The first integrals of an autonomous holonomic dynamical system}
where KV stands for Killing vector.
\bigskip

\begin{tabular}{|l|l|}
\multicolumn{2}{l}{Table 1: The QFIs of Theorem \ref{The first integrals of an autonomous holonomic dynamical system}.} \\
\hline
QFI & Conditions \\ \hline
\makecell[l]{$I_{1} = -\frac{t^{2}}{2} L_{(a;b)}\dot{q}^{a}\dot{q}^{b} + C_{ab}\dot{q}^{a} \dot{q}^{b} + t L_{a} \dot{q}^{a} +$ \\ \qquad \enskip $+ \frac{t^{2}}{2} L_{a}V^{,a} + G(q)$} & \makecell[l]{$C_{ab}, L_{(a;b)}$ are KTs, $\left(L_{b}V^{,b}\right)_{,a} = -2L_{(a;b)} V^{,b}$, \\ $G_{,a}=2C_{ab}V^{,b} - L_{a}$} \\
\makecell[l]{$I_{2} = -\frac{t^{3}}{3} L_{(a;b)}\dot{q}^{a}\dot{q}^{b} +
t^{2} L_{a} \dot{q}^{a} + \frac{t^{3}}{3} L_{a}V^{,a} -$ \\ \qquad \enskip
$- t B_{(a;b)} \dot{q}^{a}\dot{q}^{b} + B_{a}\dot{q}^{a} + tB_{a}V^{,a}$} & %
\makecell[l]{$L_{(a;b)}, B_{(a;b)}$ are KTs, $\left(L_{b}V^{,b}\right)_{,a}
= -2L_{(a;b)} V^{,b}$, \\ $\left(B_{b}V^{,b}\right)_{,a} = -2B_{(a;b)}
V^{,b} - 2L_{a}$} \\
$I_{3} = e^{\lambda t} \left( -L_{(a;b)}\dot{q}^{a}\dot{q}^{b} + \lambda
L_{a} \dot{q}^{a} + L_{a}V^{,a} \right)$ & $L_{(a;b)} = KT$, $%
\left(L_{b}V^{,b}\right)_{,a} = -2L_{(a;b)} V^{,b} - \lambda^{2} L_{a}$ \\
\hline
\end{tabular}

\bigskip

\begin{tabular}{|l|l|}
\multicolumn{2}{l}{Table 2: The LFIs of Theorem \ref{The first integrals of an autonomous holonomic dynamical system}.} \\
\hline
LFI & Conditions \\ \hline
$I_{1}=-tG_{,a}\dot{q}^{a}-\frac{s}{2}t^{2}+G(q)$ & $G_{,a}=KV$, $G_{,a}V^{,a}=s$ \\
$I_{2}=(t^{2}L_{a}+B_{a})\dot{q}^{a}+\frac{s}{3}t^{3}+tB_{a}V^{,a}$ & $%
L_{a},B_{a}$ are KVs, $L_{a}V^{,a}=s$, $\left( B_{b}V^{,b}\right)
_{,a}=-2L_{a}$ \\
$I_{3}=e^{\lambda t}\left( \lambda L_{a}\dot{q}^{a}+L_{a}V^{,a}\right) $ & $%
L_{a}=KV$, $\left( L_{b}V^{,b}\right) _{,a}=-\lambda ^{2}L_{a}$ \\ \hline
\end{tabular}

\bigskip

We note that all the QFIs reduce to LFIs when the Killing tensor $K_{ab}$
vanishes.

It can be checked that the LFIs of the second Table produce all the
potentials\footnote{
M. Tsamparlis and A. Paliathanasis, J. Phys. A: Math. Theor. 44(17), 175 (2011). \label{TsaPal1}
} which admit a LFI given in Ref. \ref{TsaPal1} which are due to Noether  point
symmetries.

\section{The gauged generalized Noether symmetry associated with an
independent QFI}

\label{sec.generators}

We compute the generators of the gauged (i.e. $\xi =0$) Noether symmetries
which admit the first integrals of Theorem \ref{The first integrals of an
autonomous holonomic dynamical system} listed in the first Table of section \ref{sec.tables.theorem} as Noether
integrals. The Noether integral (\ref{FI.4}) for a gauged Noether symmetry
(in the gauge $\xi =0!)$ becomes%
\begin{equation}
I=f-\frac{\partial L}{\partial \dot{q}^{a}}\eta ^{a}.  \label{eq.Noe2}
\end{equation}%
Replacing $L=T-V(q)$ we find%
\begin{equation}
I=f-\eta ^{a}\gamma _{ab}\dot{q}^{b}=f-\eta _{a}\dot{q}^{a}  \label{eq.Noe3}
\end{equation}%
and using (\ref{FI.5}) it follows%
\begin{equation}
\eta _{a}=-K_{ab}\dot{q}^{b}-K_{a},\enskip f=K  \label{eq.Noe4}
\end{equation}%
that is \emph{we obtain directly the Noether generator and the Noether
function from the FI }$I$\emph{\ by reading the coefficients }$K_{ab}(t,q)$%
\emph{, }$K_{a}(t,q)$\emph{\ and }$K(t,q)$\emph{\ respectively}\footnote{%
It is easy to show that the set $\{-K_{ab}\dot{q}^{b}-K_{a};K\}$ for a
general QFI $I$ given by (\ref{FI.5}) does satisfy the gauged Noether
condition $\mathbf{X}^{[1]}L=\frac{df}{dt}$.}.  Therefore the (gauged
generalized) Noether symmetry associated with a given QFI $I$ follows
trivially from the  FI\ $I.$ Equivalently, all QFIs are (gauged generalized)
Noether integrals.

\section{A detour on Killing tensors}

The first coefficient of the quadratic first integral is $%
K_{ab}=g(t)C_{ab}(q)$ where $C_{ab}(q)$ is a second order KT. Therefore it
will be useful to recall briefly some results on the second order KTs and their
relation to symmetries which are necessary for the application of the
Theorem \ref{The first integrals of an autonomous holonomic dynamical system}
in practical examples.

\subsection{Projective collineations and their reductions}

A projective collineation (PC) is a point transformation generated by a
vector field $\eta^{a}$ satisfying the condition $L_{\eta}
\Gamma_{bc}^{a}=2\delta _{(b}^{a}\phi ,_{c)}$ where $\phi $ is the
projection function of $\eta^{a}$. The PC is called special iff $\phi
_{(;ab)}=0$ that is $\phi _{,a}$ is a gradient KV. Using the identity
\begin{equation}
L_{\eta }\Gamma^{a}_{bc} = \eta^{a}{}_{;bc} - R^{a}{}_{bcd} \eta^{d}
\label{FL9.a.1}
\end{equation}%
we have that the condition for a PC\ is
\begin{equation}
\eta^{a}{}_{;bc} - 2\delta _{(b}^{a}\phi ,_{c)} = R^{a}{}_{bcd} \eta^{d}.
\label{FL9.a.2}
\end{equation}%
When $\phi =0$ the PC is called an Affine Collineation (AC). The condition
which defines an AC is%
\begin{equation}
\eta^{a}{}_{;bc} - R^{a}{}_{bcd} \eta^{d} = 0.  \label{FL9.a.3}
\end{equation}

Gradient KVs and the homothetic vector (HV) are obviously ACs. An AC which is not generated from
neither KVs nor the HV is called a proper AC.

PCs can be defined by the gradient KVs and the HV. We have the result:

\emph{If in a space there exist }$m$\emph{\ gradient KVs }$S_{I,a}$\emph{\ }$%
I=1,2,...,m$\emph{\ and the gradient HV }$H^{,a}$\emph{\ with homothetic
factor }$\psi $, \emph{\ then the vectors }$S_{I}H^{,a}$\emph{\ are
non-gradient special PCs with projection function } $\psi S_I$.

ACs can be defined by the KVs. We have the following easily established
result:

\emph{If a space admits }$m$\emph{\ gradient KVs }$S_{I,a}$\emph{\ }$%
I=1,2,...,m$\emph{\ one defines }$m^{2}$\emph{\ non-gradient ACs by the
formula }$S_{I}S_{J,a}$.

\subsection{Killing tensors}

A KT of order $m$ is a totally symmetric tensor of type $(0,m)$ defined by
the requirement
\begin{equation*}
C_{(a_{1}a_{2}...a_{m};k)}=0.
\end{equation*}

A Killing tensor $C_{ab}$ of order 2 is defined by the condition $C_{(ab;c)}=0$. We have the following result:$^{\ref{Thompson 1984}}$$^{,}$\footnote{G. Thompson, J. Math. Phys. 27, 2693 (1986). \label{Thompson 1986}}$^{,}$\footnote{E.G. Kalnins and W.M. Miller, SIAM J. Math. Anal. 11, 1001 (1980).\label{KIalinis 1980}}

\begin{proposition}
\label{pro.KT} On a general (pseudo-Riemannian) manifold of dimension $n$,
the (vector) space of Killing tensors of order $m$ has dimension less than
or equal to%
\begin{equation*}
\frac{(n+m-1)!(n+m)!}{(n-1)!n!m!(m+1)!}
\end{equation*}%
and the equality is attained if and only if M is of constant curvature. If
the latter space is flat then since on flat spaces there are not proper ACs,
PCs the Killing tensors are generated by just the Killing vectors.
\end{proposition}

We deduce that in a space of dimension $n$ the number of KTs of order 2 are
less or equal to
\begin{equation*}
\frac{(n+1)!(n+2)!}{12(n-1)!n!}=\frac{n(n+1)^{2}(n+2)}{12}.
\end{equation*}

From the AC condition (\ref{FL9.a.3}) and the property $R_{abcd}=-R_{bacd}$
it follows that
\begin{equation*}
\eta _{(a;bc)}=0\implies \eta _{((a;b);c)}=0.
\end{equation*}%
Therefore an AC $\eta _{a}$ defines a Killing tensor of order 2 of the form $%
\eta _{(a;b)}$ which implies that in a space which admits $m$ gradient KVs $%
S_{I,a}$ and a HV one can define $m^{2}+1$ KTs of order 2. The $m^{2}$ KTs $%
S_{I,(a}S_{|J|,b)}$ are defined by the $m^{2}$ proper ACs resulting from the
$m$ gradient KVs and the remaining KT (which is the metric $g_{ab}$) is
defined by the HV.

Besides these KTs, it is possible to define new KTs of order 2 by the
following recipe:

\emph{Consider} $m$ \emph{gradient KVs} $S_{I,a}$ \emph{where} $I=1,...,m$
\emph{and} $r$ \emph{non-gradient KVs} $M_{Aa}$ \emph{where} $A=1,...,r$.
\emph{Then i) the vectors} $\xi_{IAa} = S_{I}M_{Aa}$ \emph{define the} $mr$
\emph{KTs} $\xi_{IA(a;b)} = S_{I,(a} M_{|A|b)}$\emph{; and ii) the} $r^{2}$
\emph{quantities} $M_{A(a} M_{|B|b)}$ \emph{are KTs.}

From the above results we infer the following:

If a space admits $m$ gradient KVs $S_{J,a}$ and $r$ non-gradient KVs $%
M_{Aa} $, then we can construct the following $m(m+r) +r^{2}$ KTs of order 2
\begin{equation}
C_{ab} = \alpha^{IJ} S_{I,(a} S_{|J|,b)} + \beta^{IA} S_{I,(a} M_{|A|b)} +
\gamma^{AB} M_{A(a} M_{|B|b)}.  \label{FL.9a.2}
\end{equation}

By introducing the vector\footnote{%
In the vector $L_{a}$ given by (\ref{FL.9a}) AC is only the first part $%
S_{I}S_{J,a}$ whereas the second part $S_{I}M_{Aa}$ is not an AC because it
does not satisfy the AC condition (\ref{FL9.a.3}). The proof is as follows.
\begin{equation*}
(SM_{a})_{;bc}=-2S_{,(b}M_{c);a}+SM_{a;b;c}  \label{N.1}
\end{equation*}%
where we have used that $M_{a}$ is a non-gradient KV that is $%
M_{a;b}=M_{[a;b]}.$ It is easy to show the identity $M_{(a;bc)}=0$. This is
written%
\begin{equation*}
M_{a;bc}+M_{a;cb}+\left( M_{b;ac}-M_{b;ca}\right) +\left(
M_{c;ab}-M_{c;ba}\right) =0.
\end{equation*}%
Using Ricci identity
\begin{equation*}
M_{a;bc}-M_{a;cb}=R_{dabc}M^{d}=-R_{bcad}M^{d}
\end{equation*}%
and replacing in $(SM_{a})_{;bc}=-2S_{,(b}M_{c);a}+SM_{a;b;c}$ we find
\begin{equation*}
(SM^{a})_{;bc}-R^{a}{}_{bcd}(SM^{d})=2M_{;(b}^{a}S_{,c)} \label{n1}
\end{equation*}
which shows that the vector $S_{I}M_{Ja}$ does not satisfy the AC condition (%
\ref{FL9.a.3}).}
\begin{equation}
L_{a}= \alpha^{IJ} S_I S_{J,a} + \beta^{IA} S_I M_{Aa}  \label{FL.9a}
\end{equation}
the KTs (\ref{FL.9a.2}) are written
\begin{equation}
C_{ab} = L_{(a;b)} + \gamma^{AB} M_{A(a} M_{|B|b)}.  \label{FL.9a.2c}
\end{equation}
Hence the constraint $C_{ab}=L_{(a;b)}$ which appears in Theorem \ref{The first integrals of an autonomous holonomic dynamical system} implies that $\gamma^{AB}=0$ and the KTs (\ref{FL.9a.2}) reduce to
\begin{equation}
C_{ab} = \alpha^{IJ} S_{I,(a} S_{|J|,b)} + \beta^{IA} S_{I,(a} M_{|A|b)}.
\label{FL.9a.1}
\end{equation}

For recent works\footnote{D. Garfinkle and E.N. Glass, Class. Quantum Grav. 27, 095004 (2010).\label{Class KTs}}$^{,}$\footnote{R. Rani, S. Brian Edgar and A. Barnes, Class. Quantum Grav 20, 301 (2003).\label{Barnes 2003}} on KTs see Refs. \ref{Class KTs}, \ref{Barnes 2003} and references cited therein.

\subsection{Projective Collineations and KTs of order 2 in maximally
symmetric spaces}

The special projective Lie algebra of a maximally symmetric space consists
of the vector fields of Table 3 $(I,J=1,2,...,n)$.

\begin{center}
\begin{tabular}{|l|l|l|}
\multicolumn{3}{l}{Table 3: Collineations of Euclidean space $E^{n}$} \\
\hline
\multicolumn{1}{|l|}{Collineation} & Gradient & Non-gradient \\ \hline
\multicolumn{1}{|l|}{Killing vectors (KV)} & $\mathbf{K}_{I}= \delta_{I}^{i} \partial_{i}$ & $\mathbf{X}_{IJ}=\delta _{\lbrack I}^{j}\delta
_{j]}^{i}x_{j}\partial _{i}$ \\ \hline
\multicolumn{1}{|l|}{Homothetic vector (HV)} & $\mathbf{H}=x^{i}\partial
_{i}$ &  \\ \hline
\multicolumn{1}{|l|}{Affine Collineations (AC)} & $\mathbf{A}%
_{II}=x_{I}\delta _{I}^{i}\partial _{i}$ & $\mathbf{A}_{IJ}=x_{J}\delta
_{I}^{i}\partial _{i}$, $I\neq J$ \\ \hline
\multicolumn{1}{|l|}{Special Projective collineations (SPC)} &  & $\mathbf{P}_{I}=x_{I}\mathbf{H}$ \\ \hline
\end{tabular}
\end{center}

That is a maximally symmetric space, or space of constant curvature, of
dimension $n$ admits

- $n$ gradient KVs and $\frac{n(n-1)}{2}$ non-gradient KVs

- 1 gradient HV

- $n^{2}$ non-proper ACs

- $n$ PCs which are special (that is the partial derivative of the projective function is a gradient KV). \vspace{12pt}

It is well-known (see Ref. \ref{Thompson 1986}) that in manifolds of constant
curvature all KTs of order $2$ are the ones of the form (\ref{FL.9a.2}); and KTs of the form $%
L_{(a;b)}$ are given by  (\ref{FL.9a.1}) where the vectors $L_{a}$ are of the form (%
\ref{FL.9a}). These results are summarized in the following proposition.

\begin{proposition}
\label{prop1} In a space $V^{n}$ the vector fields of the form
\begin{equation}
L_{a}=c_{1I}S_{I,a}+c_{2A}M_{Aa}+c_{3}HV_{a}+ c_{4}AC_{a}+c_{5IJ}S_{I}S_{J,a}+2c_{6IA}S_{I}M_{Aa}+ c_{7K}(PC_{Ka} + CKV_{Ka})
\label{FL.20}
\end{equation}%
where $S_{I,a}$ are the gradient KVs, $M_{Aa}$ are the non-gradient KVs, $HV_{a}$ is the homothetic vector, $AC_{a}$ are the proper ACs, $S_{I}S_{J,a}$ are non-proper ACs, $PC_{Ka}$ are proper PCs with a projective factor $\phi_{K}$ and $CKV_{Ka}$ are conformal KVs with conformal factor $-2\phi_{K}$, produce the KTs of order 2 of the form $C_{ab}=L_{(a;b)}$. In the case of maximally symmetric spaces
there do not exist proper PCs and proper ACs therefore only the vectors
generated by the KVs are necessary. In this case (\ref{FL.20}) takes the
form
\begin{equation}
L_{a}=c_{1I}S_{I,a}+c_{2K}M_{Ka}+c_{3}HV_{a}+c_{5IJ}S_{I}S_{J,a}+ 2c_{6IK}S_{I}M_{Ka}.
\label{FL.21}
\end{equation}%
The general KT of order 2 is given by the formula (\ref{FL.9a.2}) and the
KTs of the form $C_{ab}=L_{(a;b)}$ are given by the formula (\ref{FL.9a.1})
where the vector $L_{a}$ is given by (\ref{FL.9a}).\footnote{The vectors $L_a$ of the form (\ref{FL.9a}) can be called master symmetries. They can be defined covariantly
via the Schouten bracket as $[g,[g,L]]$.}
\end{proposition}

The constraint $C_{ab}=L_{(a;b)}$ in the important cases of $E^{2}$ and $E^{3}$ which concern Newtonian systems
has the following quantities.

\subsection{The case of $E^{2}$}

\label{sec.KTE2}

The general KT of order 2 is
\begin{equation}
C_{ab}=\left(
\begin{array}{cc}
\gamma y^{2}+2ay+A & -\gamma xy-ax-\beta y+C \\
-\gamma xy-ax-\beta y+C & \gamma x^{2}+2\beta x+B%
\end{array}%
\right).  \label{FL.14b}
\end{equation}%
The vector $L^{a}$ generating the KT $C_{ab}=L_{(a;b)}$ is
\begin{equation}
L^{a}=\left(
\begin{array}{c}
-2\beta y^{2}+2axy+Ax+a_{8}y+a_{11} \\
-2ax^{2}+2\beta xy+a_{10}x+By+a_{9}%
\end{array}%
\right)  \label{FL.14}
\end{equation}%
and the generated KT is
\begin{equation}
C_{ab}=\left(
\begin{array}{cc}
2ay+A & -ax-\beta y+C \\
-ax-\beta y+C & 2\beta x+B%
\end{array}%
\right)   \label{FL.14.1}
\end{equation}%
where $2C=a_{8}+a_{10}$. This is a subcase of (\ref{FL.14b}) for $\gamma =0.$

\subsection{The case of $E^{3}$}

\label{sec.KTE3}

In $E^{3}$ the general KT of order 2 has independent components%
\begin{eqnarray}
C_{11} &=&\frac{a_{6}}{2}y^{2}+\frac{a_{1}}{2}%
z^{2}+a_{4}yz+a_{5}y+a_{2}z+a_{3}  \notag \\
C_{12} &=&\frac{a_{10}}{2}z^{2}-\frac{a_{6}}{2}xy-\frac{a_{4}}{2}xz-\frac{%
a_{14}}{2}yz-\frac{a_{5}}{2}x-\frac{a_{15}}{2}y+a_{16}z+a_{17}  \notag \\
C_{13} &=&\frac{a_{14}}{2}y^{2}-\frac{a_{4}}{2}xy-\frac{a_{1}}{2}xz-\frac{%
a_{10}}{2}yz-\frac{a_{2}}{2}x+a_{18}y-\frac{a_{11}}{2}z+a_{19}  \label{FL.E3}
\\
C_{22} &=&\frac{a_{6}}{2}x^{2}+\frac{a_{7}}{2}%
z^{2}+a_{14}xz+a_{15}x+a_{12}z+a_{13}  \notag \\
C_{23} &=&\frac{a_{4}}{2}x^{2}-\frac{a_{14}}{2}xy-\frac{a_{10}}{2}xz-\frac{%
a_{7}}{2}yz-(a_{16}+a_{18})x-\frac{a_{12}}{2}y-\frac{a_{8}}{2}z+a_{20}
\notag \\
C_{33} &=&\frac{a_{1}}{2}x^{2}+\frac{a_{7}}{2}%
y^{2}+a_{10}xy+a_{11}x+a_{8}y+a_{9}  \notag
\end{eqnarray}%
where $a_{I}$ with $I=1,2,...,20$ are arbitrary real constants.

The vector $L^{a}$ generating the KT $C_{ab}=L_{(a;b)}$ is
\begin{equation}
L_{a}=\left(
\begin{array}{c}
-a_{15}y^{2}-a_{11}z^{2}+a_{5}xy+a_{2}xz+2(a_{16}+a_{18})yz+a_{3}x +2a_{4}y+2a_{1}z+a_{6}
\\
-a_{5}x^{2}-a_{8}z^{2}+a_{15}xy-2a_{18}xz+a_{12}yz+ 2(a_{17}-a_{4})x+a_{13}y+2a_{7}z+a_{14}
\\
-a_{2}x^{2}-a_{12}y^{2}-2a_{16}xy+a_{11}xz+a_{8}yz+2(a_{19}- a_{1})x+2(a_{20}-a_{7})y+a_{9}z+a_{10}%
\end{array}%
\right) \label{eq.Kep.5}
\end{equation}
and the generated KT is
\begin{equation}
C_{ab}=\left(
\begin{array}{ccc}
a_{5}y+a_{2}z+a_{3} & -\frac{a_{5}}{2}x-\frac{a_{15}}{2}y+a_{16}z+a_{17} & -%
\frac{a_{2}}{2}x+a_{18}y-\frac{a_{11}}{2}z+a_{19} \\
-\frac{a_{5}}{2}x-\frac{a_{15}}{2}y+a_{16}z+a_{17} & a_{15}x+a_{12}z+a_{13}
& -(a_{16}+a_{18})x-\frac{a_{12}}{2}y-\frac{a_{8}}{2}z+a_{20} \\
-\frac{a_{2}}{2}x+a_{18}y-\frac{a_{11}}{2}z+a_{19} & -(a_{16}+a_{18})x-\frac{%
a_{12}}{2}y-\frac{a_{8}}{2}z+a_{20} & a_{11}x+a_{8}y+a_{9}%
\end{array}%
\right)   \label{eq.Kep.8}
\end{equation}%
which is a subcase of the general KT (\ref{FL.E3}) for $%
a_{1}=a_{4}=a_{6}=a_{7}=a_{10}=a_{14}=0$.

Working in the same way we may compute the KTs in a space of constant
curvature of a larger dimension.

We note that the covariant expression of the most general KT $\Lambda _{ij}$ of order 2 of $E^{3}$ is\footnote{
M. Crampin, Rep. Math. Phys. 20, 31 (1984).
\label{Crampin}}$^{,}$\footnote{C. Chanu, L. Degiovanni and R.G. McLenaghan, J. Math Phys 47, 073506 (2006). \label{Chanu}}
\begin{equation}
\Lambda _{ij}=(\varepsilon _{ikm}\varepsilon _{jln}+\varepsilon
_{jkm}\varepsilon _{iln})A^{mn}q^{k}q^{l}+(B_{(i}^{l}\varepsilon
_{j)kl}+\lambda _{(i}\delta _{j)k}-\delta _{ij}\lambda _{k})q^{k}+D_{ij}
\label{CRA.46}
\end{equation}%
where $A^{mn},B_{i}^{l},D_{ij}$ are constant tensors all being symmetric and
$B_{i}^{l}$ also being traceless; $\lambda ^{k}$ is a constant vector. This
result is obtained from the solution of the Killing tensor equation in
Euclidean space.

Observe that $A^{mn}$, $D_{ij}$ have each 6 independent components; $%
B_{i}^{l}$ has 5 independent components; and $\lambda ^{k}$ has 3
independent components. Therefore $\Lambda _{ij}$ depends on $6+6+5+3=20$
arbitrary real constants, a result which is in accordance with the one found
earlier in the case of $E^{3}$.

Having given the above general results on KTs of order 2 in flat spaces $%
E^{2},E^{3}$ we continue with applications of Theorem \ref{The first
integrals of an autonomous holonomic dynamical system}.

\section{The quadratic first integrals of geodesic equations of an $n$
dimensional Riemannian space}

Concerning the first integrals of geodesic equations we have the following well-known result\footnote{L. P. Eisenhart, ``Riemannian Geometry'', Princeton University Press, (1949). \label{Eisenhart}}.

\begin{proposition}
\label{prop.Eisen} The geodesic equations admit $m$th order first integrals
of the form
\begin{equation}
A_{r_{1}...r_{m}}\lambda ^{r_{1}}...\lambda ^{r_{m}}=const  \label{FOIG.1}
\end{equation}%
where $\lambda ^{a}\equiv \dot{q}^{a}$ and $A_{r_{1}...r_{m}}$ is a KT, that
is
\begin{equation}
A_{(r_{1}...r_{m};k)}=0.  \label{FOIG.2a}
\end{equation}%
\end{proposition}

In order to determine the quadratic first integrals of the geodesic
equations in an $n$-dimensional Riemannian space with metric $\gamma_{ab}$ we apply
Theorem \ref{The first integrals of an autonomous holonomic dynamical system}
with $V=0.$ For each case of Theorem \ref{The first integrals of an autonomous holonomic dynamical system} we have:
\bigskip

\textbf{Integral 1.}
In that case $L_{a}=-G_{,a}$ and the FI is written
\begin{equation*}
I_{1} = \frac{t^{2}}{2} G_{;ab} \dot{q}^{a}\dot{q}^{b} + C_{ab}\dot{q}^{a} \dot{q}^{b} - t G_{,a} \dot{q}^{a}+ G(q)
\end{equation*}
where $C_{ab}$, $G_{;ab}$ are KTs.

The FI $I_{1}$ consists of the two independent FIs
\[
I_{1a} = C_{ab}\dot{q}^{a} \dot{q}^{b}, \enskip I_{1b} = \frac{t^{2}}{2} G_{;ab} \dot{q}^{a}\dot{q}^{b} - t G_{,a} \dot{q}^{a}+ G(q).
\]

\textbf{Integral 2.}
Since $V=0$ the condition $\left(B_{b}V^{,b}\right)_{,a} = -2B_{(a;b)} V^{,b} - 2L_{a}$ implies $L_{a}=0$. Therefore the FI is written
\begin{equation*}
I_{2} = - t B_{(a;b)} \dot{q}^{a}\dot{q}^{b} +
B_{a}\dot{q}^{a}
\end{equation*}
where $B_{a}$ is such that $B_{(a;b)}$ is a KT.

\textbf{Integral 3.}
Since $V=0$ and $\lambda\neq0$ the condition $\left(L_{b}V^{,b}\right)_{,a} = -2L_{(a;b)} V^{,b} - \lambda^{2} L_{a}$ implies $L_{a}=0$. Therefore the FI $I_{3}=0$.
\bigskip

We collect the above results in the following Table.
\bigskip

\begin{tabular}{|l|l|}
\multicolumn{2}{l}{Table 4: The QFIs of geodesic equations.} \\
\hline
QFI & Condition \\ \hline
$I_{1a} = C_{ab}\dot{q}^{a}\dot{q}^{b}$ & $C_{ab}=$ KT \\
$I_{1b} = \frac{t^{2}}{2} G_{;ab} \dot{q}^{a}\dot{q}^{b} - tG_{,a} \dot{q}^{a} + G(q)$ & $G_{;ab}=$ KT \\
$I_{2} = -tB_{(a;b)}\dot{q}^{a}\dot{q}^{b} + B_{a}\dot{q}^{a}$ & $B_{(a;b)}= $ KT \\ \hline
\end{tabular}

\section{The general Kepler problem $V=-\frac{k}{r^{\ell}}$}

\label{sec.GKepler}

This is a three dimensional Euclidean dynamical system with kinetic metric $%
\delta _{ij}=diag(1,1,1)$ and potential $V=-\frac{k}{r^{\ell }}$ where $k,\ell$ are non-zero real constants and $r=(x^{2}+y^{2}+z^{2})^{\frac{1}{2}}$. This dynamical
system reduces to the 3d harmonic oscillator for $k<0$, $\ell =-2$ (which is the probe dynamical system for checking the validity of arguments and
calculations) and to the classical Kepler problem considered earlier by
Kalotas (see Ref. \ref{kalotas}) for $\ell =1$. The Lagrangian of the system is
\begin{equation}
L=\frac{1}{2}(\dot{x}^{2}+\dot{y}^{2}+\dot{z}^{2})+\frac{k }{r^{\ell }}
\label{eq.GKep.1}
\end{equation}%
with equations of motion
\begin{equation}
\ddot{x}=-\frac{\ell k}{r^{\ell +2}}x,\enskip\ddot{y}=-\frac{\ell k }{r^{\ell +2}}y,\enskip\ddot{z}=-\frac{\ell k}{r^{\ell +2}}z.
\label{eq.GKep.1a}
\end{equation}

To determine the QFIs of the above dynamical system we apply Theorem \ref{The first integrals of an autonomous holonomic dynamical system}. \\
\bigskip
\textbf{Integral 1.}
\begin{equation*}
I_{1} = -\frac{t^{2}}{2} L_{(a;b)}\dot{q}^{a}\dot{q}^{b} + C_{ab}\dot{q}^{a} \dot{q}^{b} + t L_{a} \dot{q}^{a} + \frac{t^{2}}{2} L_{a}V^{,a} + G(q)
\end{equation*}
where $C_{ab}$, $L_{(a;b)}$ are KTs, $\left(L_{b}V^{,b}\right)_{,a} =
-2L_{(a;b)} V^{,b}$ and $G_{,a}= 2C_{ab}V^{,b} - L_{a}$.
\bigskip

Since $C_{ab}$, $L_{(a;b)}$ are KTs the results of the section \ref{sec.KTE3} imply
\begin{eqnarray*}
C_{11} &=&\frac{a_{6}}{2}y^{2}+\frac{a_{1}}{2}%
z^{2}+a_{4}yz+a_{5}y+a_{2}z+a_{3}  \\
C_{12} &=&\frac{a_{10}}{2}z^{2}-\frac{a_{6}}{2}xy-\frac{a_{4}}{2}xz-\frac{%
a_{14}}{2}yz-\frac{a_{5}}{2}x-\frac{a_{15}}{2}y+a_{16}z+a_{17} \\
C_{13} &=&\frac{a_{14}}{2}y^{2}-\frac{a_{4}}{2}xy-\frac{a_{1}}{2}xz-\frac{%
a_{10}}{2}yz-\frac{a_{2}}{2}x+a_{18}y-\frac{a_{11}}{2}z+a_{19}
\\
C_{22} &=&\frac{a_{6}}{2}x^{2}+\frac{a_{7}}{2}%
z^{2}+a_{14}xz+a_{15}x+a_{12}z+a_{13}  \\
C_{23} &=&\frac{a_{4}}{2}x^{2}-\frac{a_{14}}{2}xy-\frac{a_{10}}{2}xz -\frac{a_{7}}{2}yz-(a_{16}+a_{18})x-\frac{a_{12}}{2}y -\frac{a_{8}}{2}z +a_{20}
 \\
C_{33} &=&\frac{a_{1}}{2}x^{2}+\frac{a_{7}}{2}%
y^{2}+a_{10}xy+a_{11}x+a_{8}y+a_{9}
\end{eqnarray*}
\begin{equation*}
L_{a}=\left(
\begin{array}{c}
-b_{15}y^{2}-b_{11}z^{2}+b_{5}xy+b_{2}xz +2(b_{16}+b_{18})yz+b_{3}x+2b_{4}y+2b_{1}z+b_{6}
\\
-b_{5}x^{2}-b_{8}z^{2}+b_{15}xy-2b_{18}xz+b_{12}yz+ 2(b_{17}-b_{4})x+b_{13}y+2b_{7}z+b_{14}
\\
-b_{2}x^{2}-b_{12}y^{2}-2b_{16}xy+b_{11}xz+b_{8}yz+2(b_{19}- b_{1})x+2(b_{20}-b_{7})y+b_{9}z+b_{10}%
\end{array}%
\right)
\end{equation*}
and
\begin{equation*}
L_{(a;b)}=\left(
\begin{array}{ccc}
b_{5}y+b_{2}z+b_{3} & -\frac{b_{5}}{2}x-\frac{b_{15}}{2}y+b_{16}z+b_{17} & -%
\frac{b_{2}}{2}x+b_{18}y-\frac{b_{11}}{2}z+b_{19} \\
-\frac{b_{5}}{2}x-\frac{b_{15}}{2}y+b_{16}z+b_{17} & b_{15}x+b_{12}z+b_{13}
& -(b_{16}+b_{18})x-\frac{b_{12}}{2}y-\frac{b_{8}}{2}z+b_{20} \\
-\frac{b_{2}}{2}x+b_{18}y-\frac{b_{11}}{2}z+b_{19} & -(b_{16}+b_{18})x-\frac{%
b_{12}}{2}y-\frac{b_{8}}{2}z+b_{20} & b_{11}x+b_{8}y+b_{9}%
\end{array}%
\right).
\end{equation*}

Substituting in the condition $\left(L_{b}V^{,b}\right)_{,a} =
-2L_{(a;b)} V^{,b}$ and taking the integrability conditions $G_{,[ab]}=0$ of the constraint $G_{,a}= 2C_{ab}V^{,b} - L_{a}$ we find the following conditions:
\begin{equation*}
a_{16}=a_{18}=0,\enskip(\ell+2)a_{17}=0, \enskip(\ell+2)a_{19}=0,\enskip %
(\ell+2)a_{20}=0,\enskip(\ell-1)a_{2}=0,\enskip(\ell-1)a_{5}=0, \enskip (\ell-1)a_{11}=0,
\end{equation*}%
\begin{equation*}
a_{2}=a_{12},\enskip a_{5}=a_{8},\enskip a_{11}=a_{15},\enskip%
(\ell+2)(a_{3}-a_{13})=0,\enskip(\ell+2)(a_{3}-a_{9})=0, \enskip b_{3}=b_{9}=b_{13}, \enskip (\ell-2)b_{3}=0
\end{equation*}
and $b_{1}=b_{2}=b_{4}=b_{5}=b_{6}=b_{7}=b_{8}=b_{10}= b_{11}=b_{12}=b_{14}=b_{15}=b_{16}=b_{17}=b_{18}=b_{19}=b_{20} =0$.

The above conditions lead to the following four cases: a) $\ell=-2$ (3d harmonic oscillator); b) $\ell=1$ (the Kepler problem); c) $\ell=2$; and d) $\ell \neq -2,1,2$.
\bigskip

a) Case $\ell=-2$.

We have $L_{a}=0$ and
\begin{equation*}
a_{2}=a_{5}=a_{8}=a_{11}=a_{12}=a_{15}=a_{16}=a_{18}=0.
\end{equation*}%

Then the independent components of the KT $C_{ab}$ are
\begin{eqnarray*}
C_{11} &=&\frac{a_{6}}{2}y^{2}+\frac{a_{1}}{2}z^{2}+a_{4}yz+a_{3}  \\
C_{12} &=&\frac{a_{10}}{2}z^{2}-\frac{a_{6}}{2}xy-\frac{a_{4}}{2}xz-\frac{%
a_{14}}{2}yz+a_{17} \\
C_{13}&=&\frac{a_{14}}{2}y^{2}-\frac{a_{4}}{2}xy -\frac{a_{1}}{2}xz-\frac{a_{10}}{2}yz+a_{19} \\
C_{22} &=&\frac{a_{6}}{2}x^{2}+\frac{a_{7}}{2}z^{2}+a_{14}xz+a_{13} \\
C_{23} &=&\frac{a_{4}}{2}x^{2}-\frac{a_{14}}{2}xy-\frac{a_{10}}{2}xz -\frac{a_{7}}{2}yz+a_{20} \\
C_{33} &=& \frac{a_{1}}{2}x^{2}+\frac{a_{7}}{2}y^{2}+a_{10}xy+a_{9}.
\end{eqnarray*}

Substituting in $G_{,a}=2C_{ab}V^{,b}$ and integrating with respect to each coordinate we find
\begin{equation*}
G(x,y,z)=-2k(a_{3}x^{2}+a_{13}y^{2}+a_{9}z^{2}+2a_{17}xy+
2a_{19}xz+2a_{20}yz).
\end{equation*}

The first integral is
\begin{eqnarray*}
I_{1} &=& C_{ab}\dot{q}^{a}\dot{q}^{b} + G(x,y,z) \\ &=&\frac{a_{1}}{2}(z\dot{x}-x\dot{z})^{2}+\frac{a_{6}}{2} (y%
\dot{x}-x\dot{y})^{2}+\frac{a_{7}}{2}(z\dot{y}-y\dot{z})^{2} +a_{3}\left( \dot{x}^{2}-2kx^{2}\right) +a_{9}(\dot{z}^{2}-2kz^{2}) + \\
&&+a_{13}(\dot{y}^{2}- 2ky^{2}) + a_{4}(y\dot{x} - x\dot{y}) (z\dot{x} - x\dot{z}) + a_{10}(z\dot{x} - x\dot{z}) (z\dot{y} - y\dot{z}) - a_{14} (z\dot{y} - y\dot{z}) (y\dot{x} - x\dot{y}) + \\
&&+2a_{17}(\dot{x}\dot{y}-2kxy)+ 2a_{19}(\dot{x}\dot{z}-2kxz) +2a_{20}(\dot{y}\dot{z}-2kyz).
\end{eqnarray*}

The FI $I_{1}$ consists of the FIs
\[
L_{1} = y\dot{z} - z\dot{y}, \enskip L_{2}= z\dot{x} - x\dot{z}, \enskip L_{3}= x\dot{y} - y\dot{x}, \enskip B_{ij} = \dot{q}_{i} \dot{q}_{j} - 2k q_{i}q_{j}
\]
where $q_{i}=(x,y,z)$, $L_{i}$ are the components of the angular momentum and $B_{ij}=B_{ji}$ are the components of a symmetric tensor. From these nine FIs the maximal number of functional independent FIs is $2n-1=5$ where $n=3$ is the dimension of the system.

The total energy of the system is written
\begin{equation*}
H \equiv E = \frac{1}{2} (B_{11} + B_{22} + B_{33}) = \frac{1}{2}(\dot{x}^{2}+\dot{y}^{2} +\dot{z}^{2}) -kr^{2}.
\end{equation*}

For $k=-\frac{1}{2}$ the tensor $B_{ij}= \dot{q}_{i}\dot{q}_{j}+q_{i}q_{j}$ is the
Jauch-Hill-Fradkin tensor\footnote{
P.G.L. Leach and V.M. Gorringe, Phys. Lett. A 133, 289 (1988). \label{Leach 1988}}.

The Poisson brackets for the components of the angular momentum give the well-known relation
\begin{equation} \label{eq.angm.1}
\{L_{a},L_{b}\}= \varepsilon_{abc} L^{c}
\end{equation}
where $\varepsilon_{abc}$ is the totally-antisymmetric Levi-Civita symbol. This means that $L_{1},L_{2},L_{3}$ are not in involution and hence cannot be used to show Liouville integrability.

The system is integrable because the triplet $H,L_{a},B_{aa}$ is functionally independent (i.e. linearly independent gradients over the phase space) and in involution, i.e. $\{H,B_{aa}\}=$ $\{H,L_{a}\}=$ $\{B_{aa}, L_{a}\}=0$.

The set $\{B_{11},B_{22},B_{33}\}$ is functionally independent and in involution as well.

We compute
\[
\{L_{1},B_{22}\}=\{B_{33},L_{1}\}=2B_{23}, \enskip
\{L_{2},B_{33}\}=\{B_{11},L_{3}\}=2B_{13}, \enskip
\{L_{3},B_{11}\}=\{B_{22},L_{3}\}=2B_{12}.
\]

We infer that the system of the 3d harmonic oscillator is also superintegrable, because it is integrable and the set $H,L_{1},L_{2},L_{3},B_{aa}$ is functionally independent.

\bigskip

b) Case $\ell=1$.

We have $L_{a}=0$ and
\begin{equation*}
a_{16}=a_{17}=a_{18}=a_{19}=a_{20}=0,\enskip a_{2}=a_{12},\enskip a_{3}=a_{9}=a_{13},\enskip a_{5}=a_{8},\enskip a_{11}=a_{15}.
\end{equation*}
Then the independent components of the KT $C_{ab}$ are
\begin{eqnarray*}
C_{11} &=& \frac{a_{6}}{2}y^{2}+ \frac{a_{1}}{2}%
z^{2}+a_{4}yz+a_{5}y+a_{2}z+a_{3} \\
C_{12} &=&\frac{a_{10}}{2}z^{2}-\frac{a_{6}}{2}xy-\frac{a_{4}}{2}xz-\frac{%
a_{14}}{2}yz-\frac{a_{5}}{2}x-\frac{a_{11}}{2}y \\
C_{13} &=&\frac{a_{14}}{2}y^{2}-\frac{a_{4}}{2}xy-\frac{a_{1}}{2}xz-\frac{%
a_{10}}{2}yz-\frac{a_{2}}{2}x-\frac{a_{11}}{2}z \\
C_{22} &=&\frac{a_{6}}{2}x^{2}+\frac{a_{7}}{2}%
z^{2}+a_{14}xz+a_{11}x+a_{2}z+a_{3} \\
C_{23} &=&\frac{a_{4}}{2}x^{2}-\frac{a_{14}}{2}xy-\frac{a_{10}}{2}xz-\frac{%
a_{7}}{2}yz-\frac{a_{2}}{2}y-\frac{a_{5}}{2}z \\
C_{33} &=&\frac{a_{1}}{2}x^{2}+\frac{a_{7}}{2}%
y^{2}+a_{10}xy+a_{11}x+a_{5}y+a_{3}.
\end{eqnarray*}
Substituting in $G_{,a}=2c_{0}C_{ab}V^{,b}$ and integrating
with respect to each coordinate we find
\begin{equation*}
G(x,y,z)=-\frac{k}{r}(a_{11}x+a_{5}y+a_{2}z+2a_{3}).
\end{equation*}

The first integral is
\begin{eqnarray*}
I_{1} &=& C_{ab}\dot{q}^{a}\dot{q}^{b} + G(q) \\
&=&\frac{a_{1}}{2}(z\dot{x}-x\dot{z})^{2} +\frac{a_{6}}{2} (y\dot{x}-x\dot{y})^{2} + \frac{a_{7}}{2} (z\dot{y}-y\dot{z})^{2} + a_{4} (y\dot{x}-x\dot{y}) (z\dot{x}-x\dot{z}) +a_{10} (z\dot{x}-x\dot{z}) (z\dot{y}-y\dot{z}) - \\
&& - a_{14} (z\dot{y}-y\dot{z}) (y\dot{x}-x\dot{y}) +2a_{3}\left[
\frac{1}{2}(\dot{x}^{2}+\dot{y}^{2}+\dot{z}^{2}) -\frac{k}{r}\right] + a_{2}\left[ z(\dot{x}^{2}+\dot{y}^{2})-\dot{z}(x\dot{x}+y\dot{y}) -\frac{k}{r}z\right] + \\
&&+a_{5}\left[ y(\dot{x}^{2}+\dot{z}^{2})-\dot{y}(x\dot{x}+z\dot{z})-\frac{k}{%
r}y\right]+ a_{11}\left[ x(\dot{y}^{2}+\dot{z}%
^{2})-\dot{x}(y\dot{y}+z\dot{z})-\frac{k}{r}x\right].
\end{eqnarray*}

The FI $I_{1}$ contains the following FIs: \newline

i. The three components of the angular momentum $L_{1}=y\dot{z}-z\dot{y}$, $L_{2}=z\dot{x}-x\dot{z}$ and $L_{3}= x\dot{y}-y\dot{x}$. \newline

ii. The total energy (Hamiltonian) of the system $E \equiv \frac{1}{2}(\dot{x}^{2}+\dot{y}^{2}+\dot{z}^{2}) -\frac{k}{r}$.
\newline

iii. The three components of the Runge-Lenz vector
\[
R_{1} = x(\dot{y}^{2}+\dot{z}^{2})-\dot{x}(y\dot{y}+z\dot{z})- \frac{k}{r}x, \enskip R_{2} =y(\dot{x}^{2}+\dot{z}^{2})- \dot{y}(x\dot{x}+z\dot{z})-\frac{k}{r}y, \enskip R_{3} = z(\dot{x}^{2}+\dot{y}^{2})-\dot{z}(x\dot{x}+y\dot{y}) -\frac{k}{r}z
\]
which can be written in the compact form
\begin{equation}
R_{i}=(v^{j}v_{j})x_{i}-(x^{j}v_{j})v_{i}-\frac{k}{r}x_{i}. \label{eq.Kep.6}
\end{equation}
where $x_{i}=(x,y,z)$ and $v_{i}=\dot{x}_{i}= (\dot{x},\dot{y},\dot{z})$. The linear combination $\mu^{i}R_{i}$, where $\mu^{i}$ are arbitrary constants, is the Noether invariant found in Kalotas (see Ref. \ref{kalotas}).

Using the vector identity
\begin{equation*}
\mathbf{A}\times \left( \mathbf{B}\times \mathbf{C}\right) =\left( \mathbf{A}%
\cdot \mathbf{C}\right) \mathbf{B}-\left( \mathbf{A}\cdot \mathbf{B}\right)
\mathbf{C}
\end{equation*}%
equation (\ref{eq.Kep.6}) is written in the well-known vector form
\begin{equation}
\mathbf{R}=\mathbf{v}\times (\mathbf{x}\times \mathbf{v})-\frac{k}{r}\mathbf{%
x}.  \label{eq.Kep.7}
\end{equation}

We should point out that the above seven FIs are not all independent because they are related by
\begin{equation} \label{eq.RL0}
\mathbf{R} \cdot \mathbf{L} =0 \enskip \text{and} \enskip \mathbf{R}^{2} = k^{2} + 2E\mathbf{L}^{2}.
\end{equation}
From these relations we deduce that there exist only five independent FIs: the total energy $E$, the angular momentum $\mathbf{L}$ and the direction of the Runge-Lenz vector $\mathbf{R}$.

The Kepler problem is a Liouville integrable system because the FIs $E,L_{a},R_{a}$ are functional independent and in involution, i.e. $\{L_{a},E\}=0$, $\{R_{a},E\}=0$ and $\{L_{a},R_{a}\}=0$. It is also superintegrable, because it has dimension $n=3$ and admits $2n-1=5$ independent FIs.

For the components of the Runge-Lenz vector we find that $\{R_{a}, L_{b}\} = \varepsilon_{abc} R^{c}$ and $\{R_{a},R_{b}\} = -2\varepsilon_{abc}L^{c}E$.

\bigskip

c) Case $\ell=2$.

In that case we have
\begin{eqnarray*}
C_{11} &=&\frac{a_{6}}{2}y^{2}+\frac{a_{1}}{2}z^{2}+a_{4}yz+a_{3}  \\
C_{12} &=&\frac{a_{10}}{2}z^{2}-\frac{a_{6}}{2}xy-\frac{a_{4}}{2}xz-\frac{%
a_{14}}{2}yz \\
C_{13} &=&\frac{a_{14}}{2}y^{2}-\frac{a_{4}}{2}xy-\frac{a_{1}}{2}xz-\frac{%
a_{10}}{2}yz \\
C_{22} &=&\frac{a_{6}}{2}x^{2}+\frac{a_{7}}{2}z^{2}+a_{14}xz+a_{3} \\
C_{23} &=&\frac{a_{4}}{2}x^{2}-\frac{a_{14}}{2}xy-\frac{a_{10}}{2}xz-\frac{%
a_{7}}{2}yz \\
C_{33} &=&\frac{a_{1}}{2}x^{2}+\frac{a_{7}}{2}y^{2}+a_{10}xy+a_{3}
\end{eqnarray*}
\[
L_{a}= b_{3}
\left(
  \begin{array}{c}
    x \\
    y \\
    z \\
  \end{array}
\right), \enskip L_{(a;b)}= b_{3}\delta_{ab}.
\]

Substituting in $G_{,a}=2c_{0}C_{ab}V^{,b}$ and integrating with respect to each coordinate we find
\begin{equation*}
G(x,y,z)=-\frac{2ka_{3}}{r^2} - \frac{b_{3}}{2}r^{2}.
\end{equation*}

The first integral is
\begin{eqnarray*}
I_{1} &=& - b_{3} t^{2} \left[ \frac{1}{2}(\dot{x}^{2} + \dot{y}^{2} + \dot{z}^{2}) - \frac{k}{r^{2}}  \right] + \frac{a_{1}}{2}(z\dot{x}-x\dot{z})^{2}+ \frac{a_{6}}{2}(y\dot{x}-x\dot{y})^{2} +\frac{a_{7}}{2}(z\dot{y} -y\dot{z})^{2} + \\
&& + 2a_{3}\left[ \frac{1}{2}(\dot{x}^{2}+ \dot{y}^{2}+\dot{z}^{2})- \frac{k}{r^{2}}\right] + a_{4} (y\dot{x}-x\dot{y}) (z\dot{x}-x\dot{z}) + a_{10} (z\dot{x}-x\dot{z}) (z\dot{y}-y\dot{z}) - \\
&& - a_{14} (z\dot{y}-y\dot{z}) (y\dot{x}-x\dot{y}) + b_{3} t (x\dot{x} + y\dot{y} + z\dot{z}) - \frac{b_{3}}{2} r^{2}.
\end{eqnarray*}

The above FI contains the three components of the angular momentum, the energy $E =\frac{1}{2}(\dot{x}^{2} + \dot{y}^{2} + \dot{z}^{2}) - \frac{k}{r^{2}}$ of the system and in addition the time-dependent FI
\[
I_{1a}(\ell=2)= - E t^{2} + t(x\dot{x} + y\dot{y} + z\dot{z}) - \frac{r^{2}}{2}.
\]

\bigskip

d) Case $\ell \neq -2,1,2$.

In these cases the KT $C_{ab}$ is that of the case $\ell=2$, the vector $L_{a}=0$ and $G(x,y,z)=-\frac{2ka_{3}}{r^{\ell}}$.

The FI is
\begin{eqnarray*}
I_{1} &=& \frac{a_{1}}{2}(z\dot{x}-x\dot{z})^{2}+ \frac{a_{6}}{2}(y\dot{x}-x\dot{y})^{2} +\frac{a_{7}}{2}(z\dot{y} -y\dot{z})^{2} + 2a_{3}\left[ \frac{1}{2}(\dot{x}^{2}+ \dot{y}^{2}+\dot{z}^{2})- \frac{k}{r^{\ell}}\right]+ \\
&& + a_{4} (y\dot{x}-x\dot{y}) (z\dot{x}-x\dot{z}) + a_{10} (z\dot{x}-x\dot{z}) (z\dot{y}-y\dot{z}) - a_{14} (z\dot{y}-y\dot{z}) (y\dot{x}-x\dot{y})
\end{eqnarray*}
which consists of the FIs generated by the three components of the angular momentum and the total energy $E = \frac{1}{2}(\dot{x}^{2}+ \dot{y}^{2}+\dot{z}^{2})- \frac{k}{r^{\ell}}$ of the system.

\bigskip

\textbf{Integral 2.}
\begin{equation*}
I_{2} = -\frac{t^{3}}{3} L_{(a;b)}\dot{q}^{a}\dot{q}^{b} + t^{2} L_{a} \dot{q%
}^{a} + \frac{t^{3}}{3} L_{a}V^{,a} - t B_{(a;b)} \dot{q}^{a}\dot{q}^{b} +
B_{a}\dot{q}^{a} + tB_{a}V^{,a}
\end{equation*}
where $L_{a}$, $B_{a}$ are such that $L_{(a;b)}$, $B_{(a;b)}$ are KTs, $\left(L_{b}V^{,b}\right)_{,a} = -2L_{(a;b)} V^{,b}$ and $\left(B_{b}V^{,b}\right)_{,a} = -2B_{(a;b)} V^{,b} - 2L_{a}$.
\bigskip

Since $L_{(a;b)}$, $B_{(a;b)}$ are KTs we have
\begin{equation*}
L_{a}=\left(
\begin{array}{c}
-a_{15}y^{2}-a_{11}z^{2}+a_{5}xy+a_{2}xz+2(a_{16}+a_{18})yz +a_{3}x +2a_{4}y+2a_{1}z+a_{6}
\\
-a_{5}x^{2}-a_{8}z^{2}+a_{15}xy-2a_{18}xz+a_{12}yz+ 2(a_{17}-a_{4})x+a_{13}y+2a_{7}z+a_{14}
\\
-a_{2}x^{2}-a_{12}y^{2}-2a_{16}xy+a_{11}xz+a_{8}yz+2(a_{19}- a_{1})x+2(a_{20}-a_{7})y+a_{9}z+a_{10}%
\end{array}%
\right)
\end{equation*}
\begin{equation*}
B_{a}=\left(
\begin{array}{c}
-b_{15}y^{2}-b_{11}z^{2}+b_{5}xy+b_{2}xz +2(b_{16}+b_{18})yz+b_{3}x+2b_{4}y+2b_{1}z+b_{6}
\\
-b_{5}x^{2}-b_{8}z^{2}+b_{15}xy-2b_{18}xz+b_{12}yz+ 2(b_{17}-b_{4})x+b_{13}y+2b_{7}z+b_{14}
\\
-b_{2}x^{2}-b_{12}y^{2}-2b_{16}xy+b_{11}xz+b_{8}yz+2(b_{19}- b_{1})x+2(b_{20}-b_{7})y+b_{9}z+b_{10}%
\end{array}%
\right)
\end{equation*}
\begin{equation*}
L_{(a;b)}=\left(
\begin{array}{ccc}
a_{5}y+a_{2}z+a_{3} & -\frac{a_{5}}{2}x-\frac{a_{15}}{2}y+a_{16}z+a_{17} & -%
\frac{a_{2}}{2}x+a_{18}y-\frac{a_{11}}{2}z+a_{19} \\
-\frac{a_{5}}{2}x-\frac{a_{15}}{2}y+a_{16}z+a_{17} & a_{15}x+a_{12}z+a_{13}
& -(a_{16}+a_{18})x-\frac{a_{12}}{2}y-\frac{a_{8}}{2}z+a_{20} \\
-\frac{a_{2}}{2}x+a_{18}y-\frac{a_{11}}{2}z+a_{19} & -(a_{16}+a_{18})x-\frac{%
a_{12}}{2}y-\frac{a_{8}}{2}z+a_{20} & a_{11}x+a_{8}y+a_{9}%
\end{array}%
\right)
\end{equation*}%
and
\begin{equation*}
B_{(a;b)}=\left(
\begin{array}{ccc}
b_{5}y+b_{2}z+b_{3} & -\frac{b_{5}}{2}x-\frac{b_{15}}{2}y+b_{16}z+b_{17} & -%
\frac{b_{2}}{2}x+b_{18}y-\frac{b_{11}}{2}z+b_{19} \\
-\frac{b_{5}}{2}x-\frac{b_{15}}{2}y+b_{16}z+b_{17} & b_{15}x+b_{12}z+b_{13}
& -(b_{16}+b_{18})x-\frac{b_{12}}{2}y-\frac{b_{8}}{2}z+b_{20} \\
-\frac{b_{2}}{2}x+b_{18}y-\frac{b_{11}}{2}z+b_{19} & -(b_{16}+b_{18})x-\frac{%
b_{12}}{2}y-\frac{b_{8}}{2}z+b_{20} & b_{11}x+b_{8}y+b_{9}%
\end{array}
\right).
\end{equation*}

From the constraint $\left(L_{b}V^{,b}\right)_{,a} = -2L_{(a;b)} V^{,b}$ we find that $(\ell-2)a_{3}=0$, $L_{a}= a_{3}
\left(
  \begin{array}{c}
    x \\
    y \\
    z \\
  \end{array}
\right)$ and $L_{(a;b)}= a_{3}\delta_{ab}$.

Substituting in the remaining constraint $\left(B_{b}V^{,b} \right)_{,a} = -2B_{(a;b)} V^{,b} - 2L_{a}$ we obtain $a_{3}=0 \implies$ $L_{a}=0$,  $(\ell-2)b_{3}=0$, $B_{a}= b_{3}
\left(
  \begin{array}{c}
    x \\
    y \\
    z \\
  \end{array}
\right)$ and $B_{(a;b)}= b_{3}\delta_{ab}$.

For $\ell\neq 2$ we have $B_{a}=0$ and the FI $I_{2}=0$. On the other hand, for $\ell=2$ we get the non-trivial time-dependent FI
\[
I_{2}(\ell=2) = -Et + \frac{1}{2} (x\dot{x} + y\dot{y} + z\dot{z}).
\]

\bigskip

\textbf{Integral 3.}
\begin{equation*}
I_{3} = -e^{\lambda t} L_{(a;b)}\dot{q}^{a}\dot{q}^{b} + \lambda e^{\lambda
t} L_{a} \dot{q}^{a} + e^{\lambda t} L_{a} V^{,a}
\end{equation*}
where $\lambda \neq 0$, $L_{a}$ is such that $L_{(a;b)}$ is a KT and $\left(L_{b}V^{,b}\right)_{,a} = -2L_{(a;b)} V^{,b} - \lambda^{2} L_{a}$.

\bigskip

Since $L_{(a;b)}$ is a KT we have
\begin{equation*}
L_{a}=\left(
\begin{array}{c}
-a_{15}y^{2}-a_{11}z^{2}+a_{5}xy+a_{2}xz+2(a_{16}+a_{18})yz +a_{3}x +2a_{4}y+2a_{1}z+a_{6}
\\
-a_{5}x^{2}-a_{8}z^{2}+a_{15}xy-2a_{18}xz+a_{12}yz+ 2(a_{17}-a_{4})x+a_{13}y+2a_{7}z+a_{14}
\\
-a_{2}x^{2}-a_{12}y^{2}-2a_{16}xy+a_{11}xz+a_{8}yz+2(a_{19}- a_{1})x+2(a_{20}-a_{7})y+a_{9}z+a_{10}%
\end{array}%
\right)
\end{equation*}
and
\begin{equation*}
L_{(a;b)}=\left(
\begin{array}{ccc}
a_{5}y+a_{2}z+a_{3} & -\frac{a_{5}}{2}x-\frac{a_{15}}{2}y+a_{16}z+a_{17} & -%
\frac{a_{2}}{2}x+a_{18}y-\frac{a_{11}}{2}z+a_{19} \\
-\frac{a_{5}}{2}x-\frac{a_{15}}{2}y+a_{16}z+a_{17} & a_{15}x+a_{12}z+a_{13}
& -(a_{16}+a_{18})x-\frac{a_{12}}{2}y-\frac{a_{8}}{2}z+a_{20} \\
-\frac{a_{2}}{2}x+a_{18}y-\frac{a_{11}}{2}z+a_{19} & -(a_{16}+a_{18})x-\frac{%
a_{12}}{2}y-\frac{a_{8}}{2}z+a_{20} & a_{11}x+a_{8}y+a_{9}%
\end{array}%
\right).
\end{equation*}

Substituting in the constraint $\left(L_{b}V^{,b}\right)_{,a} = -2L_{(a;b)} V^{,b} - \lambda^{2} L_{a}$ we find that the vector $L_{a}$ does not vanish only for $\ell=-2$. Therefore in what it follows we consider only that case.

We compute
\begin{equation*}
\left( L_{b}V^{,b}\right) _{,a}=-2k\left(
\begin{array}{c}
2a_{3}x+2a_{17}y+2a_{19}z+a_{6} \\
2a_{13}y+2a_{17}x+2a_{20}z+a_{14} \\
2a_{9}z+2a_{19}x+2a_{20}y+a_{10}%
\end{array}%
\right)
\end{equation*}

\begin{equation*}
2L_{(a;b)}V^{,b}=-2k\left(
\begin{array}{c}
-a_{15}y^{2}-a_{11}z^{2}+a_{5}xy+a_{2}xz+2(a_{16}+a_{18})yz+2a_{3}x+2a_{17}y+2a_{19}z
\\
-a_{5}x^{2}-a_{8}z^{2}+a_{15}xy-2a_{18}xz+a_{12}yz+2a_{17}x+2a_{13}y+2a_{20}z
\\
-a_{2}x^{2}-a_{12}y^{2}-2a_{16}xy+a_{11}xz+a_{8}yz+2a_{19}x+2a_{20}y+2a_{9}z%
\end{array}%
\right)
\end{equation*}%
and
\begin{equation*}
\lambda ^{2}L_{a}=\lambda ^{2}\left(
\begin{array}{c}
-a_{15}y^{2}-a_{11}z^{2}+a_{5}xy+a_{2}xz+2(a_{16}+a_{18})yz+a_{3}x+2a_{4}y+2a_{1}z+a_{6}
\\
-a_{5}x^{2}-a_{8}z^{2}+a_{15}xy-2a_{18}xz+a_{12}yz+2(a_{17}-a_{4})x+a_{13}y+2a_{7}z+a_{14}
\\
-a_{2}x^{2}-a_{12}y^{2}-2a_{16}xy+a_{11}xz+a_{8}yz+2(a_{19}-a_{1})x+2(a_{20}-a_{7})y+a_{9}z+a_{10}%
\end{array}%
\right).
\end{equation*}
Then the condition $\lambda ^{2}L_{a}+\left( L_{b}V^{,b}\right)
_{,a}+2L_{(a;b)}V^{,b}=0$ gives the following set of equations
\begin{eqnarray*}
0 &=&-a_{15}(\lambda ^{2}-2k)y^{2}-a_{11}(\lambda
^{2}-2k)z^{2}+a_{5}(\lambda ^{2}-2k)xy+a_{2}(\lambda
^{2}-2k)xz+2(a_{16}+a_{18})(\lambda ^{2}-2k)yz+ \\
&&+a_{3}(\lambda ^{2}-8k)x+2(\lambda ^{2}a_{4}-4ka_{17})y+2(\lambda
^{2}a_{1}-4ka_{19})z+(\lambda ^{2}-2k)a_{6}
\end{eqnarray*}

\begin{eqnarray*}
0 &=& - a_5 (\lambda^2 - 2k)x^2 - a_8 (\lambda^2 - 2k)z^2 + a_{15}
(\lambda^2 - 2k) xy - 2a_{18} (\lambda^2 - 2k) xz + 2a_{12} (\lambda^2 - 2k)
yz + \\
&& + 2 \left[ a_{17} (\lambda^2 -4k) - \lambda^2a_4\right] x +
a_{13}(\lambda^2 - 8k)y + 2(\lambda^2 a_7 - 4ka_{20}) z +
(\lambda^2-2k)a_{14}
\end{eqnarray*}

\begin{eqnarray*}
0 &=& - a_2 (\lambda^2 - 2k)x^2 - a_{12} (\lambda^2 - 2k)y^2 - 2a_{16}
(\lambda^2 - 2k) xy + a_{11} (\lambda^2 - 2k) xz + a_8 (\lambda^2 - 2k) yz +
\\
&& + 2 \left[ a_{19} (\lambda^2 -4k) - \lambda^2a_1\right] x + 2 \left[
a_{20} (\lambda^2 -4k) - \lambda^2a_7\right] y + a_9(\lambda^2 - 8k)z +
(\lambda^2-2k)a_{10}.
\end{eqnarray*}

We consider the cases:
\bigskip

a) For $\lambda ^{2}=2k$ we have
\begin{equation*}
a_{1}=a_{3}=a_{4}=a_{7}=a_{9}=a_{13}=a_{17}=a_{19}=a_{20}=0.
\end{equation*}%
Then
\begin{equation*}
L_{a}=\left(
\begin{array}{c}
-a_{15}y^{2}-a_{11}z^{2}+a_{5}xy+a_{2}xz+2(a_{16}+a_{18})yz+a_{6} \\
-a_{5}x^{2}-a_{8}z^{2}+a_{15}xy-2a_{18}xz+a_{12}yz+a_{14} \\
-a_{2}x^{2}-a_{12}y^{2}-2a_{16}xy+a_{11}xz+a_{8}yz+a_{10}%
\end{array}%
\right)
\end{equation*}%
and
\begin{equation*}
L_{(a;b)}=\left(
\begin{array}{ccc}
a_{5}y+a_{2}z & -\frac{a_{5}}{2}x-\frac{a_{15}}{2}y+a_{16}z & -\frac{a_{2}}{2%
}x+a_{18}y-\frac{a_{11}}{2}z \\
-\frac{a_{5}}{2}x-\frac{a_{15}}{2}y+a_{16}z & a_{15}x+a_{12}z &
-(a_{16}+a_{18})x-\frac{a_{12}}{2}y-\frac{a_{8}}{2}z \\
-\frac{a_{2}}{2}x+a_{18}y-\frac{a_{11}}{2}z & -(a_{16}+a_{18})x-\frac{a_{12}%
}{2}y-\frac{a_{8}}{2}z & a_{11}x+a_{8}y%
\end{array}%
\right) .
\end{equation*}

The first integral is
\begin{eqnarray*}
I_{3a}(\ell=-2) &=&-\frac{a_{2}}{\lambda }e^{\lambda t}(z\dot{x}-x%
\dot{z})(\dot{x}-\lambda x)-\frac{a_{5}}{\lambda }e^{\lambda t}(y\dot{x}-x%
\dot{y})(\dot{x}-\lambda x)-\frac{a_{8}}{\lambda }e^{\lambda t}(y\dot{z}-z%
\dot{y})(\dot{z}-\lambda z)- \\
&&-\frac{a_{11}}{\lambda }e^{\lambda t}(x\dot{z}-z\dot{x})(\dot{z}-\lambda
z)-\frac{a_{12}}{\lambda }e^{\lambda t}(z\dot{y}-y\dot{z})(\dot{y}-\lambda
y)-\frac{a_{15}}{\lambda }e^{\lambda t}(x\dot{y}-y\dot{x})(\dot{y}-\lambda
y)- \\
&&-\frac{2a_{16}}{\lambda }e^{\lambda t}(z\dot{x}-x\dot{z})(\dot{y}-\lambda
y)-\frac{2a_{18}}{\lambda }e^{\lambda t}(y\dot{x}-x\dot{y})(\dot{z}-\lambda
z)+a_{6}e^{\lambda t}(\dot{x}-\lambda x)+a_{10}e^{\lambda t}(\dot{z}-\lambda z)+ \\
&& +a_{14}e^{\lambda t}(\dot{y}-\lambda y).
\end{eqnarray*}
From this expression we have the following irreducible time-dependent FIs
\[
I_{3a1}= e^{\lambda t}(\dot{x}-\lambda x), \enskip I_{3a2}=e^{\lambda t}(\dot{y}-\lambda y), \enskip I_{3a3}= e^{\lambda t}(\dot{z}-\lambda z).
\]

If $k>0$, then $\lambda= \pm \sqrt{2k}$; and if $k<0$, then $\lambda= \pm i \sqrt{-2k}$. Therefore for all values of the non-zero parameter $k$ of the system there exist two constants $\lambda_{\pm}$ each generating three independent FIs of the system. We have\footnote{The calculations are the same for either $k>0$ or $k<0$. We continue for $k<0$ which is the case of the 3d harmonic oscillator.}
\[
I_{3a1\pm}= e^{\pm i \sqrt{-2k} t}(\dot{x} \mp i \sqrt{-2k} x), \enskip I_{3a2\pm}=e^{\pm i \sqrt{-2k} t}(\dot{y} \mp i \sqrt{-2k} y), \enskip I_{3a3\pm}= e^{\pm i \sqrt{-2k} t}(\dot{z} \mp i \sqrt{-2k} z).
\]
Using the above six FIs we can derive all the FIs found in the case \textbf{Integral 1} for $\ell=-2$. We compute
\[
I_{3a1+}I_{3a1-}=B_{11}, \enskip I_{3a2+}I_{3a2-}=B_{22}, \enskip I_{3a3+}I_{3a3-}=B_{33},
\]
\[
I_{3a1\pm}I_{3a2\mp}=B_{12} \mp i\sqrt{-2k}L_{3}, \enskip I_{3a1\pm}I_{3a3\mp}=B_{13} \pm i\sqrt{-2k}L_{2}, \enskip I_{3a2\pm}I_{3a3\mp}=B_{23} \mp i\sqrt{-2k}L_{1}.
\]
Therefore, all the components of the Jauch-Hill-Fradkin tensor $B_{ij}$ can be constructed by the three components of the angular momentum and the six time-dependent FIs $I_{3a1\pm}$, $I_{3a2\pm}$, $I_{3a3\pm}$.

\bigskip

b) For $\lambda^2=4k$ we get $L_a=0$ and hence the FI vanishes. \bigskip

c) Finally, for $\lambda^2 = 8k$ we have
\begin{equation*}
a_2=a_5=a_6=a_8=a_{10}=a_{11}=a_{12}=a_{14}=a_{15}=a_{16}= a_{18} = 0, %
\enskip a_{17}=2a_4, \enskip a_{19}=2a_1, \enskip a_{20}= 2a_7.
\end{equation*}
Then
\begin{equation*}
L_a = \left(
\begin{array}{c}
a_3x + a_{17}y + a_{19}z \\
a_{17}x + a_{13}y + a_{20}z \\
a_{19}x + a_{20}y + a_9z%
\end{array}
\right), \enskip L_{(a;b)} = \left(
\begin{array}{ccc}
a_3 & a_{17} & a_{19} \\
a_{17} & a_{13} & a_{20} \\
a_{19} & a_{20} & a_9%
\end{array}
\right).
\end{equation*}

The first integral is
\begin{eqnarray*}
I_{3c}(\ell=-2) &=&-\frac{a_{3}}{\lambda }e^{\lambda t}\left( \dot{x}-\frac{\lambda }{2}x\right)^{2}-\frac{a_{9}}{\lambda }e^{\lambda t}\left(
\dot{z}-\frac{\lambda }{2}z\right) ^{2}-\frac{a_{13}}{\lambda }e^{\lambda
t}\left( \dot{y}-\frac{\lambda }{2}y\right) ^{2}- \\
&&-\frac{a_{17}}{\lambda }e^{\lambda t}\left[ 2\dot{x}\dot{y}+\frac{\lambda
^{2}}{2}xy-\lambda (y\dot{x}+x\dot{y})\right] -\frac{a_{19}}{\lambda }%
e^{\lambda t}\left[ 2\dot{x}\dot{z}+\frac{\lambda ^{2}}{2}xz-\lambda (z\dot{x%
}+x\dot{z})\right] - \\
&&-\frac{a_{20}}{\lambda }e^{\lambda t}\left[ 2\dot{y}\dot{z}+\frac{\lambda
^{2}}{2}yz-\lambda (y\dot{z}+z\dot{y})\right].
\end{eqnarray*}
This expression consists of the time-dependent FIs
\[
I_{3b1}= e^{\lambda t}\left( \dot{x}-\frac{\lambda }{2}x\right)^{2}, \enskip I_{3b2}= e^{\lambda t}\left( \dot{y}-\frac{\lambda }{2}y\right)^{2}, \enskip I_{3b3}= e^{\lambda t}\left( \dot{z}-\frac{\lambda }{2}z\right)^{2}, \enskip I_{3b4}= e^{\lambda t}\left[ \dot{x}\dot{y}+ \frac{\lambda^{2}}{4} xy- \frac{\lambda}{2} (y\dot{x}+x\dot{y})\right],
\]
\[
I_{3b5} = e^{\lambda t}\left[ \dot{x}\dot{z}+ \frac{\lambda^{2}}{4} xz - \frac{\lambda}{2} (z\dot{x}+x\dot{z})\right], \enskip I_{3b6}= e^{\lambda t}\left[ \dot{y}\dot{z}+ \frac{\lambda^{2}}{4} yz - \frac{\lambda}{2} (y\dot{z}+z\dot{y})\right].
\]
If $k>0$, $\lambda= \pm 2\sqrt{2k}$; and if $k<0$, $\lambda =\pm 2i\sqrt{-2k}$. Similarly with the calculations of the case a) we find that (we continue for $k<0$ and adopt the notation of the case a) for the FIs)
\[
I_{3b1\pm}=(I_{3a1\pm})^{2}, \enskip I_{3b2\pm}=(I_{3a2\pm})^{2}, \enskip I_{3b3\pm}=(I_{3a3\pm})^{2}, \enskip I_{3b4\pm}= I_{3a1\pm} I_{3a2\pm},
\]
\[
I_{3b5\pm} = I_{3a1\pm} I_{3a3\pm}, \enskip I_{3b6\pm}= I_{3a2\pm}I_{3a3\pm}.
\]
Therefore this case gives again the six time dependent FIs $I_{3a1\pm}$, $I_{3a2\pm}$, $I_{3a3\pm}$ of the case a).

We collect the results of this section in the following Table (we write $q^{i}=(x,y,z)$ ).
\bigskip

\begin{tabular}{|l|l|}
\multicolumn{2}{l}{Table 5: The QFIs of the general Kepler problem.} \\
\hline
$V=-\frac{k}{r^{\ell}}$ & LFIs and QFIs \\ \hline
$\forall$ $\ell$ & $L_{1} = y\dot{z} - z\dot{y}$, $L_{2}= z\dot{x} - x\dot{z}$, $L_{3}= x\dot{y} - y\dot{x}$, $H= \frac{1}{2}(\dot{x}^{2} + \dot{y}^{2} + \dot{z}^{2}) - \frac{k}{r^{\ell}}$ \\
$\ell=-2$ & $B_{ij} = \dot{q}_{i} \dot{q}_{j} - 2k q_{i}q_{j}$ \\
$\ell=-2$, $k>0$ & $I_{3a\pm}= e^{\pm \sqrt{2k} t}(\dot{q}_{a} \mp \sqrt{2k} q_{a})$ \\
$\ell=-2$, $k<0$ & $I_{3a\pm}= e^{\pm i \sqrt{-2k} t}(\dot{q}_{a} \mp i \sqrt{-2k} q_{a})$ \\
$\ell=1$ & $R_{i}= (\dot{q}^{j} \dot{q}_{j}) q_{i} - (\dot{q}^{j}q_{j})\dot{q}_{i}- \frac{k}{r}q_{i}$ \\
$\ell=2$ & $I_{1}= - Ht^{2} + t(\dot{q}^{i}q_{i}) - \frac{r^{2}}{2}$, $I_{2}= - Ht + \frac{1}{2} (\dot{q}^{i}q_{i})$
 \\ \hline
\end{tabular}

\bigskip

\section{The time dependent FIs}
\label{subsec.Kep.3}

As it has been shown Theorem \ref{The first integrals of an autonomous holonomic dynamical system}  produces all FIs of the autonomous conservative dynamical equations i.e. the autonomous  and the time dependent FIs, the latter being equally important as the former. Furthermore   this is achieved in a way which is independent of the dimension, the signature and the curvature of the kinetic metric, defined by the kinetic energy/Lagrangian  of the specific dynamical system. On the contrary the standard methods determine mainly the autonomous FIs, usually for low degrees of freedom and consider principally the `usual' dynamical systems.

The time-dependent FIs can be used to test the integrability of a dynamical system and of course they can be used to obtain the solution of the dynamical equations in terms of quadratures. The Liouville integrability theorem\footnote{See p. 271, section 49 in V.I. Arnold, \emph{`Mathematical Methods of Classical Mechanics'}, Springer, (1989), proof in pp. 272-284. \label{Arnold 1989}} requires $n$ functionally independent FIs in involution of the form $I(q,p)$. However it has been pointed out that we can also use time-dependent FIs of the form\footnote{See Theorem 1, p.17, chapter II, paragraph 2 in V.V. Kozlov, Russ. Math. Surv., Turpion, \textbf{38}(1), pp. 1-76 (1983). \label{Kozlov 1983}}$^{,}$\footnote{See Theorem 3.4 in T.G. Vozmishcheva, J. Math. Sc. \textbf{125}(4), 419 (2005). \label{Vozmishcheva 2005}} $I(q,p,t)$ for the same purpose. It is to be noticed that both Theorems in Refs. \ref{Kozlov 1983}, \ref{Vozmishcheva 2005} refer to non-autonomous Hamiltonians $H(q,p,t)$. Moreover, the usefulness of the time-dependent\footnote{G.H. Katzin and J. Levine, J. Math. Phys. \textbf{26}(12), 3080 (1985). \label{KatzinLev1985}} FIs can be seen from the examples I, II of section VII in Ref. \ref{KatzinLev1985}.

In order to show the use of the time dependent FIs in the solution of the dynamical equations we consider two cases of the general Kepler equations (\ref{eq.GKep.1a}) considered in the section \ref{sec.GKepler}.\\
\\
Example 1. In the case of potential $V=-kr^{2}$ ($\ell=-2$, $k>0$) we found the six time-dependent FIs $I_{3a\pm}= e^{\pm \sqrt{2k} t}(\dot{q}_{a} \mp \sqrt{2k} q_{a})$. We use these FIs to obtain the solution of the corresponding equations. We have
\[
\begin{cases}
e^{\sqrt{2k} t}(\dot{x} - \sqrt{2k}x) = A_{+} \\
e^{-\sqrt{2k} t}(\dot{x} + \sqrt{2k}x) = A_{-}
\end{cases}
\implies
\begin{cases}
\dot{x} - \sqrt{2k}x = A_{+}e^{-\sqrt{2k} t} \\
\dot{x} + \sqrt{2k}x = A_{-}e^{\sqrt{2k} t}
\end{cases} \implies
\]
\[
\dot{x} = \frac{1}{2} \left( A_{+}e^{-\sqrt{2k} t} + A_{-} e^{\sqrt{2k} t}\right)
\implies
x(t)= \frac{1}{2} \left( -\frac{A_{+}}{\sqrt{2k}} e^{-\sqrt{2k} t} + \frac{A_{-}}{\sqrt{2k}} e^{\sqrt{2k} t}\right)
\]
where $A_{\pm}$ are arbitrary constants. Similarly from the other FIs we find
\[
y(t)= \frac{1}{2} \left( -\frac{B_{+}}{\sqrt{2k}} e^{-\sqrt{2k} t} + \frac{B_{-}}{\sqrt{2k}} e^{\sqrt{2k} t}\right), \enskip z(t)= \frac{1}{2} \left( -\frac{C_{+}}{\sqrt{2k}} e^{-\sqrt{2k} t} + \frac{C_{-}}{\sqrt{2k}} e^{\sqrt{2k} t}\right)
\]
where $B_{\pm}$, $C_{\pm}$ are arbitrary constants.\\

\bigskip
Example 2.  For the case of the 3d harmonic oscillator (i.e. $\ell=-2$, $k<0$) using the time-dependent FIs $I_{3a\pm}= e^{\pm i \sqrt{-2k} t}(\dot{q}_{a} \mp i \sqrt{-2k} q_{a})$ we find working in the same way
\[
x(t)= \frac{1}{2} \left( \frac{iD_{+}}{\sqrt{-2k}} e^{-i \sqrt{-2k}t} - \frac{iD_{-}}{\sqrt{-2k}} e^{i \sqrt{-2k}t} \right), \enskip y(t)= \frac{1}{2} \left( \frac{iE_{+}}{\sqrt{-2k}} e^{-i \sqrt{-2k}t} - \frac{iE_{-}}{\sqrt{-2k}} e^{i \sqrt{-2k}t} \right),
\]
\[ z(t)= \frac{1}{2} \left( \frac{iF_{+}}{\sqrt{-2k}} e^{-i \sqrt{-2k}t} - \frac{iF_{-}}{\sqrt{-2k}} e^{i \sqrt{-2k}t} \right)
\]
where $D_{\pm}$, $E_{\pm}$, $F_{\pm}$ are arbitrary constants.

\section{Conclusions}

\label{Conclusions}

The usefulness of FIs in the solution of the dynamical equations  is well-known. Therefore it is important that one has a
systematic method to compute them for a given dynamical system. In Theorem %
\ref{The first integrals of an autonomous holonomic dynamical system} we
have developed such a method for  the case of autonomous conservative dynamical systems. It has
been shown that these integrals are closely related to the KTs and the
symmetries of the kinetic metric, which is defined by the kinetic energy or
the Lagrangian for the particular dynamical system.

From Theorem \ref{The first integrals of an autonomous holonomic dynamical
system} follows that the determination of a QFI/LFI of an
autonomous conservative dynamical system consists of two parts. One part
which is entirely characteristic of the kinetic metric and it is common to
all dynamical systems which share this metric; and a second part which consists
of constraints which involve in addition the potential which defines the specific dynamical
system. The constraints of the first part concern the determination of the
first integrals in terms of the symmetries of the kinetic metric. For example
we have `solved' the constraints of first part for the cases of $E^{2}$ and $E^{3}$ which
concern the majority of the Newtonian dynamical systems as a whole.

With each FI we have associated in a natural manner a gauged Noether symmetry whose
Noether integral is this particular first integral. This implies that all
the quadratic first integrals of a conservative autonomous dynamical system
are Noetherian. This result agrees with the conclusion of the Ref. \ref{kalotas}
concerning the Runge-Lenz vector of the Kepler potential. In this sense we
have managed to geometrize the generalized Noether symmetries which was the secondary
purpose to the present work.

We have applied the Theorem \ref{The first integrals of an autonomous
holonomic dynamical system} in two well-known cases; the case of the
geodesics and the case of the generalized Kepler potential. The latter is the
reference example because it contains the harmonic oscillator as well as the
Kepler potential both being studied in detail in the past. As it was
expected we recovered all existing previous results plus the fact that we
obtained all possible time dependent QFIs.

Finally we have discussed briefly the importance of the time dependent FIs
and we have demonstrated their use in the integration of  of the dynamical equations for two special cases
of the general Kepler potential.

\appendix

\section{Proof of the theorem}

We look for solutions in which $g(t),f(t)$ are analytic functions so that
they can be represented by polynomial functions of $t$:
\begin{equation*}
g(t)=\sum_{k=0}^{n}c_{k}t^{k}=c_{0}+c_{1}t+...+c_{n}t^{n}
\end{equation*}
\begin{equation*}
f(t)=\sum_{k=0}^{m}d_{k}t^{k}=d_{0}+d_{1}t+...+d_{m}t^{m}
\end{equation*}
where $n,m\in \mathbb{N}$ (or infinite) and $c_{k},d_{k}\in \mathbb{R}$.

We consider various cases\footnote{
Equation (\ref{FL.1.e}) is not necessary, because the integrability condition $K_{,[ab]}=0$ does not intervene in the calculations.
}.
\vspace{12pt}

\textbf{I. For both $\mathbf{m, n}$ finite.} \vspace{12pt}

\underline{\textbf{I.1. Case $\mathbf{n=m}$:}} \vspace{12pt}

\underline{\textbf{Subcase $\mathbf{(n=0, m=0)}$.}} $g=c_{0}$, $f=d_{0}$.

\begin{equation*}
\begin{cases}
(\ref{FL.1.a1}) \implies c_0 C_{(ab;c)} = 0 \\
(\ref{FL.1.a}) \implies d_0 L_{(a;b)} + B_{(a;b)} = 0 \\
(\ref{FL.1.b}) \implies - 2 c_0 C_{ab} V^{,b} + K_{,a} = 0 \\
(\ref{FL.1.c}) \implies K_{,t} - d_0 L_b V^{,b} - B_b V^{,b} = 0 \\
(\ref{FL.1.d}) \implies d_0 \left( L_b V^{,b} \right)_{;a} + \left( B_b
V^{,b} \right)_{;a} = 0%
\end{cases}%
\end{equation*}

We define the vector field $\tilde{L}_a \equiv d_0 L_a + B_a$.

From \eqref{FL.1.a} we find that $\tilde{L}_{(a;b)} = 0$ which means that $\tilde{L}_a$ is a KV.

From \eqref{FL.1.d} we have that $\tilde{L}_a V^{,a} = s_0 = const$.

Solving equation \eqref{FL.1.c} we get $K = s_0 t + G(q)$ which into %
\eqref{FL.1.b} gives $G_{,a} = 2 c_0 C_{ab} V^{,b}$.

The first integral is
\begin{equation*}
I_{00} = c_{0}C_{ab}\dot{q}^{a}\dot{q}^{b} + \tilde{L}_{a}\dot{q}^{a} + s_{0}t+G(q)
\end{equation*}%
where $c_0C_{ab}$ is a KT, $\tilde{L}_{a}$ is a KV such that $\tilde{L}_a V^{,a} = s_0$ and $G(q)=2c_{0} \int C_{ab} V^{,b} dq^{a}$.

Therefore the FI $I_{00}$ consists of the independent FIs
\begin{equation*}
Q_{1} = C_{ab}\dot{q}^{a}\dot{q}^{b} + G(q), \qquad Q_{2} = L_{a}\dot{q}^{a} + s_{1}t
\end{equation*}
where $C_{ab}$ is a KT, $L_{a}$ is a KV such that $L_{a}V^{,a} =s_{1}$ and $G_{,a}=2C_{ab}V^{,b}$.

\underline{\textbf{Subcase $\mathbf{(n=1, m=1)}$.}} $g = c_0 + c_1 t$, $f =
d_0 + d_1 t$ with $c_1 \neq 0$ and $d_1 \neq 0$.

\begin{equation*}
\begin{cases}
\eqref{FL.1.a1} \implies C_{(ab;c)} = 0 \\
\eqref{FL.1.a} \implies c_1 C_{ab} + (d_0 + d_1t) L_{(a;b)} + B_{(a;b)} = 0
\\
\eqref{FL.1.b} \implies - 2 c_1 C_{ab} V^{,b} t - 2 c_0 C_{ab} V^{,b} + d_1
L_a + K_{,a} = 0 \\
\eqref{FL.1.c} \implies K_{,t} = \left( d_0 + d_1 t \right) L_a V^{,a} + B_a
V^{,a} \\
\eqref{FL.1.d} \implies \left( d_0 + d_1 t \right) \left( L_b V^{,b}
\right)_{;a} + \left( B_b V^{,b} \right)_{;a} - 2 c_1 C_{ab} V^{,b} = 0.%
\end{cases}%
\end{equation*}

The first equation implies that $C_{ab}$ is a KT.

From \eqref{FL.1.a} $L_a$ is a KV and $c_1 C_{ab} + B_{(a;b)} = 0$.

From \eqref{FL.1.d} we find that $L_aV^{,a}=s_{1}$ and $\left(
B_{b}V^{,b}\right) _{;a}=2c_{1}C_{ab}V^{,b}$. Then \eqref{FL.1.c} gives
\begin{equation*}
K=s_{1}\left( d_{0}t+\frac{d_{1}}{2}t^{2}\right) +B_{a}V^{,a}t+G(q)
\end{equation*}%
which into \eqref{FL.1.b} yields $G_{,a}=2c_{0}C_{ab}V^{,b}-d_{1}L_{a}$.
Using the relation $\left( B_{b}V^{,b}\right)_{;a}=2c_{1}C_{ab}V^{,b}$ we
find that
\begin{equation*}
G_{,a}=\frac{c_{0}}{c_{1}}\left( B_{b}V^{,b}\right) _{;a}-d_{1}L_{a}
\implies L_{a} = \frac{c_0}{c_1d_1} \left(B_b V^{,b}\right)_{,a} - \frac{1}{%
d_1} G_{,a}.
\end{equation*}

The first integral is
\begin{equation*}
I_{11}= - \frac{1}{c_{1}} \left( c_{0}+c_{1}t\right) B_{(a;b)}\dot{q}^{a}\dot{q}^{b}+\left(
d_{0}+d_{1}t\right) L_{a}\dot{q}^{a}+B_{a}\dot{q}^{a}+s_{1}\left( d_{0}t+%
\frac{d_{1}}{2}t^{2}\right) +B_{a}V^{,a}t+G(q)
\end{equation*}%
where $B_{(a;b)}$ is a KT, $L_{a}=\frac{c_{0}}{%
c_{1}d_{1}}\left( B_{b}V^{,b}\right) _{,a}-\frac{1}{d_{1}}G_{,a} \equiv \Phi_{,a}$ is a
gradient KV such that\footnote{%
A brief comment on the commutation (Lie bracket) between a vector field $%
B^{a}$ and a gradient vector $V^{,a}$.
\begin{equation*}
\lbrack \mathbf{B},\mathbf{\nabla }%
V]^{a}=B^{b}V^{,a}{}_{,b}-V^{,b}B^{a}{}_{,b}= B^{b}V^{,a}{}_{;b}-V^{,b}B^{a}{}_{;b}\implies
\end{equation*}%
\begin{equation*}
\lbrack \mathbf{B},\mathbf{\nabla }%
V]_{a}=B^{b}V_{;ab}-V^{,b}B_{a;b}=B^{b}V_{;ba}-B_{a;b}V^{,b}=\left(
B_{b}V^{,b}\right) _{;a}-B_{b;a}V^{,b}-B_{a;b}V^{,b}=\left(
B_{b}V^{,b}\right) _{;a}-2B_{(a;b)}V^{,b}.
\end{equation*}%

Therefore
\begin{equation*}
\lbrack \mathbf{B},\mathbf{\nabla }V]_{a}=0\iff \left( B_{b}V^{,b}\right)
_{;a}=2B_{(a;b)}V^{,b}.
\end{equation*}%

If $B^{a}$ is a KV such that $B_{a}V^{,a}=const$, then $B^{a}$ and $V^{,a}$
commute, i.e. $[\mathbf{B},\mathbf{\nabla }V]^{a}=0$.} $L_{a}V^{,a}=s_{1}$ and
the vector $B_{a}$ is such that $\left( B_{b}V^{,b}\right)
_{;a}=-2B_{(a;b)}V^{,b}$.

First of all
\begin{equation*}
L_{a} = \Phi_{,a} = \frac{c_{0}}{c_{1}d_{1}} \left(B_{b} V^{,b}\right)_{,a}
- \frac{1}{d_{1}} G_{,a} \implies G(q) = \frac{c_{0}}{c_{1}} B_{a}V^{,a} - d_{1}\Phi(q).
\end{equation*}
Therefore
\begin{eqnarray*}
I_{11} &=& \frac{c_{0}}{c_{1}} Q_{3} + Q_{4} + d_{0}Q_{2} + d_{1} Q_{5}
\end{eqnarray*}
where
\begin{equation*}
Q_{3} = - B_{(a;b)} \dot{q}^{a} \dot{q}^{b} + B_{a}V^{,a}, \enskip Q_{4} = - t B_{(a;b)} \dot{q}^{a} \dot{q}^{b} + B_{a}\dot{q}^{a} + t
B_{a}V^{,a}, \enskip
Q_{5} = tL_{a}\dot{q}^{a} + \frac{s_{1}}{2}t^{2} - \Phi(q)
\end{equation*}
are independent FIs.

\underline{\textbf{Subcase $\mathbf{(n=2, m=2)}$.}} $g = c_0 + c_1 t + c_2
t^2$, $f = d_0 + d_1 t + d_2 t^2$ with $c_2 \neq 0$ and $d_2 \neq 0$.

\begin{equation*}
\begin{cases}
\eqref{FL.1.a1} \implies C_{(ab;c)} = 0 \\
\eqref{FL.1.a} \implies (c_1 + 2c_2 t) C_{ab} + (d_0 + d_1t + d_2 t^2)
L_{(a;b)} + B_{(a;b)} = 0 \\
\eqref{FL.1.b} \implies - 2 (c_0 + c_1 t + c_2 t^2) C_{ab} V^{,b} + (d_1 +
2d_2 t) L_a + K_{,a} = 0 \\
\eqref{FL.1.c} \implies K_{,t} = \left( d_0 + d_1 t + d_2 t^2 \right) L_a
V^{,a} + B_a V^{,a} \\
\eqref{FL.1.d} \implies 2d_2 L_a + \left( d_0 + d_1 t + d_2 t^2 \right)
\left( L_b V^{,b} \right)_{;a} + \left( B_b V^{,b} \right)_{;a} - 2 (c_1 + 2
c_2 t) C_{ab} V^{,b} = 0.%
\end{cases}%
\end{equation*}

From \eqref{FL.1.a} $C_{ab} = 0$ and $L_a$, $B_a$ are KVs.

From \eqref{FL.1.d} we find that $L_aV^{,a}=s_{1}$ and $L_a = - \frac{1}{2d_2%
} \left(B_{b}V^{,b}\right)_{;a}$, that is $L_a$ is a gradient KV.

The solution of \eqref{FL.1.c} is
\begin{equation*}
K = s_{1} \left( d_{0}t + \frac{d_{1}}{2}t^{2} + \frac{d_2}{3} t^3 \right) +
B_{a}V^{,a}t + G(q)
\end{equation*}
which into \eqref{FL.1.b} gives
\begin{equation*}
G_{,a} + d_1 L_a \underbrace{+ 2d_2 L_a t + \left( B_{b}V^{,b} \right)_{,a} t%
}_{=0} = 0 \implies G_{,a} = -d_1 L_a = \frac{d_1}{2d_2} \left(B_b
V^{,b}\right)_{,a} \implies G(q) = \frac{d_1}{2d_2} B_a V^{,a}.
\end{equation*}

The first integral is
\begin{equation*}
I_{22} = \left(d_{0}+d_{1}t+d_2t^2\right) L_{a}\dot{q}^{a} + B_{a}\dot{q}%
^{a} + s_{1} \left( d_{0}t + \frac{d_{1}}{2}t^{2} + \frac{d_2}{3} t^3
\right) + B_{a}V^{,a}t + \frac{d_1}{2d_2} B_a V^{,a}
\end{equation*}
where $L_{a} = - \frac{1}{2d_2} \left(B_{b}V^{,b}\right)_{;a}$ is a gradient
KV such that $L_{a}V^{,a}=s_{1}$ and $B_a$ is a KV.

The FI $I_{22}= d_{0}Q_{2} + d_{1}Q_{5} + F_{1}$ where
\[
F_{1} = t^{2} X_{a}\dot{q}^{a} + \frac{s}{3} t^{3} + B_{a}\dot{q}^{a} + B_{a}V^{,a}t, \quad X_{a}\equiv d_{2}L_{a}, \enskip s\equiv d_{2}s_{1}
\]
is a new independent FI.

\underline{\textbf{Subcase $\mathbf{(n=m>2)}$.}} $c_n \neq 0$ and $d_n \neq
0 $.

\begin{equation*}
\begin{cases}
\eqref{FL.1.a1} \implies C_{(ab;c)} = 0 \\
\eqref{FL.1.a} \implies (c_1 + 2c_2 t + ... + n c_n t^{n-1}) C_{ab} + (d_0 +
d_1t + .. + d_n t^n) L_{(a;b)} + B_{(a;b)} = 0 \\
\eqref{FL.1.b} \implies - 2 (c_0 + c_1 t + ... + c_n t^n) C_{ab} V^{,b} +
(d_1 + 2d_2 t + ... + n d_n t^{n-1}) L_a + K_{,a} = 0 \\
\eqref{FL.1.c} \implies K_{,t} = \left( d_0 + d_1 t + ... + d_n t^n \right)
L_a V^{,a} + B_a V^{,a} \\
\eqref{FL.1.d} \implies \left[2d_2 + 3 \cdot 2 d_3 t + ... + n (n-1) d_n
t^{n-2}\right] L_a + \left( d_0 + d_1 t + ... + d_n t^n \right) \left( L_b
V^{,b} \right)_{;a} + \left( B_b V^{,b} \right)_{;a} - \\
\qquad \qquad - 2 (c_1 + 2c_2 t + ... + n c_n t^{n-1}) C_{ab} V^{,b} = 0.%
\end{cases}%
\end{equation*}

From \eqref{FL.1.a} $C_{ab} = 0$ and $L_a$, $B_a$ are KVs.

From \eqref{FL.1.d} we find that $L_a = 0$ and $B_a V^{,a} = s_2$.

The solution of \eqref{FL.1.c} is $K = s_2 t + G(q)$ which into %
\eqref{FL.1.b} gives $G = const$. Such a constant is ignored because any
constant can be added to $I$ without changing the condition $\frac{dI}{dt}=0$%
.

The first integral is (of the form $Q_{2}$)
\begin{equation*}
I_{nn}(n>2)=B_{a}\dot{q}^{a}+s_{2}t
\end{equation*}%
where $B_{a}$ is a KV such that $B_{a}V^{,a}=s_{2}$.\vspace{16pt}

We continue with the case $n>m$. This case is broken down equivalently into
the cases $n=m+1$ and $n>m+1$. Both cases are analyzed below\footnote{%
It is much more convenient to follow this line of analysis because for $n >
m+1$ equation \eqref{FL.1.a} implies directly that $C_{ab} = 0$

and the
derived first integrals are non-quadratic. To be more specific these are the
only non-quadratic first integrals for

the case $n > m$, as for $n=m+1$ all
the derived integrals are quadratic.}. \vspace{16pt}

\underline{\textbf{I.2. Case $\mathbf{n=m+1}$.}} \vspace{12pt}

\underline{\textbf{Subcase $\mathbf{(n=1, m=0)}$.}} $g = c_0 + c_1 t$, $f =
d_0$ with $c_1 \neq 0$.

\begin{equation*}
\begin{cases}
\eqref{FL.1.a1} \implies C_{(ab;c)} = 0 \\
\eqref{FL.1.a} \implies c_1 C_{ab} + \tilde{L}_{(a;b)} = 0 \\
\eqref{FL.1.b} \implies - 2 c_1 C_{ab} V^{,b} t - 2 c_0 C_{ab} V^{,b} +
K_{,a} = 0 \\
\eqref{FL.1.c} \implies K_{,t} - \tilde{L}_a V^{,a} = 0 \\
\eqref{FL.1.d} \implies \left( \tilde{L}_b V^{,b} \right)_{;a} - 2 c_1 C_{ab}
V^{,b} = 0.%
\end{cases}%
\end{equation*}

From \eqref{FL.1.a1} since $c_{1}\neq 0$ follows that $C_{ab}$ is a KT.

Solving \eqref{FL.1.c} we get $K = \bar{L}_a V^{,a} t + G(q)$ which into %
\eqref{FL.1.b} gives $G_{,a} = 2 c_0 C_{ab} V^{,b}$. But
\begin{equation*}
2 C_{ab} V^{,b} = \frac{1}{c_1} \left( \tilde{L}_b V^{,b} \right)_{;a}.
\end{equation*}

Therefore
\begin{equation*}
G_{,a} = \frac{c_0}{c_1} \left( \tilde{L}_b V^{,b} \right)_{,a} \implies G(q)= \frac{c_0}{c_1} \tilde{L}_a V^{,a}.
\end{equation*}

The first integral is
\begin{equation*}
I_{10} = - \frac{1}{c_1}\left( c_{0} + c_{1}t \right) \tilde{L}_{(a;b)}\dot{q}^{a}\dot{q}^{b} + \tilde{L}_{a}\dot{q}^{a}+\left( t + \frac{c_{0}}{c_{1}}\right) \tilde{L}_{a}V^{,a}
\end{equation*}%
where $\tilde{L}_{a}$ is a vector such that $\tilde{L}_{(a;b)}$ is a KT and $\left( \tilde{L}_b V^{,b}\right)_{;a} = - 2 \tilde{L}_{(a;b)} V^{,b}$.

We note that $I_{10} = \frac{c_{0}}{c_{1}} Q_{3}(\tilde{L}_{a}) + Q_{4}(\tilde{L}_{a})$.

\underline{\textbf{Subcase $\mathbf{(n=2, m=1)}$.}} $g = c_0 + c_1 t + c_2
t^2$, $f = d_0 + d_1 t$ with $c_2 \neq 0$ and $d_1 \neq 0$.

\begin{equation*}
\begin{cases}
\eqref{FL.1.a1} \implies C_{(ab;c)} = 0 \\
\eqref{FL.1.a} \implies \left( c_1 + 2 c_2 t \right) C_{ab} + \left( d_0 +
d_1 t \right) L_{(a;b)} + B_{(a;b)} = 0 \\
\eqref{FL.1.b} \implies - 2 \left( c_0 + c_1 t + c_2 t^2 \right) C_{ab}
V^{,b} + d_1 L_a + K_{,a} = 0 \\
\eqref{FL.1.c} \implies K_{,t} = \left( d_0 + d_1 t \right) L_a V^{,a} + B_a
V^{,a} \\
\eqref{FL.1.d} \implies \left( d_0 + d_1 t \right) \left( L_b V^{,b}
\right)_{;a} + \left( B_b V^{,b} \right)_{;a} - 2 \left( c_1 + 2 c_2 t
\right) C_{ab} V^{,b} = 0.%
\end{cases}%
\end{equation*}

From the first equation $C_{ab}$ is a KT.

Equation \eqref{FL.1.a} gives $2c_2 C_{ab} + d_1 L_{(a;b)} = 0$ and $c_1
C_{ab} + d_0 L_{(a;b)} + B_{(a;b)} = 0$.

From \eqref{FL.1.d} we have that $d_0 \left( L_b V^{,b} \right)_{;a} +
\left( B_b V^{,b} \right)_{;a} - 2 c_1 C_{ab} V^{,b} = 0$ and $d_1 \left(
L_b V^{,b} \right)_{;a} = 4 c_2 C_{ab} V^{,b}$.

Solving \eqref{FL.1.c} we find that
\begin{equation*}
K=\left( d_{0}t+\frac{d_{1}}{2}t^{2}\right) L_{a}V^{,a}+B_{a}V^{,a}t+G(q)
\end{equation*}%
which into \eqref{FL.1.b} and using the last relations gives
\begin{equation*}
G_{,a}=2c_{0}C_{ab}V^{,b}-d_{1}L_{a} \implies G_{,a}= \frac{c_0 d_1}{2c_2}
\left( L_b V^{,b} \right)_{,a} - d_1 L_a \implies G(q) = \frac{c_0 d_1}{2c_2}
L_a V^{,a} - d_1 \int L_a dq^a.
\end{equation*}

Note also that
\begin{equation*}
\begin{cases}
2c_2 C_{ab} + d_1 L_{(a;b)} = 0 \\
c_1 C_{ab} + d_0 L_{(a;b)} + B_{(a;b)} = 0%
\end{cases}
\implies B_{(a;b)} = \left( \frac{2 c_2 d_0}{d_1} - c_1 \right) C_{ab}
\end{equation*}

\begin{equation*}
\begin{cases}
d_0 \left( L_b V^{,b} \right)_{;a} + \left( B_b V^{,b} \right)_{;a} - 2 c_1
C_{ab} V^{,b} = 0 \\
d_1 \left(L_b V^{,b} \right)_{;a} = 4 c_2 C_{ab} V^{,b}%
\end{cases}
\implies \left( B_b V^{,b} \right)_{;a} = 2 c_1 C_{ab} V^{,b} - \frac{4c_2d_0%
}{d_1} C_{ab} V^{,b}
\end{equation*}
and $\frac{c_1 d_1}{2c_2} \left(L_b V^{,b} \right)_{;a} = d_0 \left(L_b
V^{,b} \right)_{;a} + \left(B_b V^{,b} \right)_{;a}$. Therefore
\begin{equation*}
\begin{cases}
\left( B_b V^{,b} \right)_{;a} = - 2 \left( \frac{2c_2d_0}{d_1} - c_1
\right) C_{ab} V^{,b} \\
B_{(a;b)} = \left( \frac{2 c_2 d_0}{d_1} - c_1 \right) C_{ab}%
\end{cases}
\implies \left( B_b V^{,b} \right)_{;a} = - 2 B_{(a;b)} V^{,b} \implies [
B^a, V^{,a} ] \equiv [ \mathbf{B}, \mathbf{\nabla}V ]^a \neq 0.
\end{equation*}

The first integral is
\begin{eqnarray*}
I_{21} &=&-\frac{d_{1}}{2c_{2}} \left( c_{0}+c_{1}t+c_{2}t^{2}\right) L_{(a;b)} \dot{q}^{a}\dot{q}%
^{b}+\left( d_{0}+d_{1}t\right) L_{a}\dot{q}^{a}+B_{a}\dot{q}^{a}+\left(
d_{0}t+\frac{d_{1}}{2}t^{2}\right) L_{a}V^{,a}+B_{a}V^{,a}t+G(q)
\end{eqnarray*}
where $L_a$ is a vector
such that $L_{(a;b)}$ is a KT and $\left( L_{b}V^{,b}\right)_{;a} = - 2 L_{(a;b)} V^{,b}$; $B_a$ is
a vector satisfying the relations $\left( B_b V^{,b} \right)_{;a} = - 2
B_{(a;b)} V^{,b}$ and $B_{(a;b)}= \frac{2c_2d_{0} - c_{1}d_{1}}{d_{1}}
C_{ab} $; and $G(q)= \frac{c_{0}d_{1}}{2c_{2}}L_{a} V^{,a} - d_{1} \int
L_{a}dq^{a}$.

First of all
\begin{equation*}
G_{,a} = \frac{c_{0}d_{1}}{2c_{2}} \left(L_{b} V^{,b}\right)_{,a} - d_{1}
L_{a}.
\end{equation*}
From this $L_{a}=\Phi_{,a}$, i.e. $L_{a}$ is a gradient and hence
\begin{equation*}
G(q)= \frac{c_{0}d_{1}}{2c_{2}}L_{a} V^{,a} - d_{1}\Phi(q).
\end{equation*}

Moreover $B_{(a;b)}= \left(\frac{ c_{1}d_{1}}{2c_{2}}- d_{0}\right)
L_{(a;b)} $ implies
\begin{equation*}
- \frac{ c_{1}d_{1}}{2c_{2}} L_{(a;b)} = - B_{(a;b)} - d_{0}L_{(a;b)}
\end{equation*}
and $B_{(a;b)}$ is a KT.

Substituting the above results into $I_{21}$ we find
\begin{equation*}
I_{21} = \frac{c_{0}d_{1}}{2c_{2}} Q_{3}(L_{a}) + Q_{4} + d_{0}Q_{4}(L_{a})
+ d_{1}Q_{6}
\end{equation*}
where
\begin{equation*}
Q_{6} = - \frac{t^{2}}{2} L_{(a;b)} \dot{q}^{a} \dot{q}^{b} + t L_{a} \dot{q}^{a} + \frac{t^{2}}{2} L_{a}V^{,a} - \Phi(q)
\end{equation*}
is a new independent FI.

We note that the expression
\[
Q_{1} + Q_{6} = - \frac{t^{2}}{2} L_{(a;b)} \dot{q}^{a} \dot{q}^{b} + C_{ab} \dot{q}^{a} \dot{q}^{b} + t L_{a} \dot{q}^{a} + \frac{t^{2}}{2} L_{a}V^{,a} - \Phi(q) + G(q)
\]
where $\Phi_{,a}=L_{a}$ and $G_{,a}=2C_{ab}V^{,b}$ leads to the new independent FI
\[
Q_{16} = - \frac{t^{2}}{2} L_{(a;b)} \dot{q}^{a} \dot{q}^{b} + C_{ab} \dot{q}^{a} \dot{q}^{b} + t L_{a} \dot{q}^{a} + \frac{t^{2}}{2} L_{a}V^{,a} + H(q)
\]
where now $H_{,a} = 2C_{ab}V^{,b} - L_{a}$. From $Q_{16}$ the FIs $Q_{1}$, $Q_{6}$ are derived as subcases, that is, $Q_{16}(C_{ab}=0)=Q_{6}$ and $Q_{16}(L_{a}=0)=Q_{1}$.

\underline{\textbf{Subcase $\mathbf{(n=3, m=2)}$.}} $g = c_0 + c_1 t + c_2
t^2 + c_3 t^3$, $f = d_0 + d_1 t + d_2 t^2$ with $c_3 \neq 0$ and $d_2 \neq
0 $.

\begin{equation*}
\begin{cases}
\eqref{FL.1.a1} \implies C_{(ab;c)} = 0 \\
\eqref{FL.1.a} \implies \left( c_1 + 2 c_2 t + 3c_3t^2 \right) C_{ab} +
\left( d_0 + d_1 t + d_2 t^2\right) L_{(a;b)} + B_{(a;b)} = 0 \\
\eqref{FL.1.b} \implies - 2 \left( c_0 + c_1 t + c_2 t^2 + c_3 t^3 \right)
C_{ab} V^{,b} + (d_1 + 2d_2 t) L_a + K_{,a} = 0 \\
\eqref{FL.1.c} \implies K_{,t} = \left( d_0 + d_1 t + d_2 t^2 \right) L_a
V^{,a} + B_a V^{,a} \\
\eqref{FL.1.d} \implies 2d_2 L_a + \left( d_0 + d_1 t + d_2 t^2 \right)
\left( L_b V^{,b} \right)_{;a} + \left( B_b V^{,b} \right)_{;a} - 2 \left(
c_1 + 2 c_2 t + 3 c_3 t^2 \right) C_{ab} V^{,b} = 0.%
\end{cases}%
\end{equation*}

From the first equation $C_{ab}$ is a KT.

From \eqref{FL.1.a} we have that $3c_3 C_{ab} + d_2 L_{(a;b)} = 0$, $2c_2
C_{ab} + d_1 L_{(a;b)} = 0$ and $c_1 C_{ab} + d_0 L_{(a;b)} + B_{(a;b)} = 0$.

From \eqref{FL.1.d} we find that $d_2 \left( L_b V^{,b} \right)_{;a} = 6 c_3
C_{ab} V^{,b}$, $d_1 \left( L_b V^{,b} \right)_{;a} = 4 c_2 C_{ab} V^{,b}$
and $2d_2 L_a + d_0 \left( L_b V^{,b} \right)_{;a} + \left( B_b V^{,b}
\right)_{;a} - 2 c_1 C_{ab} V^{,b} = 0$.

The solution of \eqref{FL.1.c} is
\begin{equation*}
K = \left( d_{0}t + \frac{d_{1}}{2}t^{2} + \frac{d_2}{3}t^3 \right)
L_{a}V^{,a} + B_{a}V^{,a}t + G(q)
\end{equation*}
which into \eqref{FL.1.b} and using the above derived conditions gives
\begin{equation*}
G_{,a}= 2c_{0}C_{ab}V^{,b}-d_{1}L_{a} \implies G_{,a}= \frac{c_0 d_2}{3c_3}
\left( L_b V^{,b} \right)_{,a} - d_1 L_a \implies G(q) = \frac{c_0 d_2}{3c_3}
L_a V^{,a} - d_1 \int L_a dq^a.
\end{equation*}

From the first set of conditions we get
\begin{equation*}
\begin{cases}
3c_3 C_{ab} + d_2 L_{(a;b)} = 0 \\
2c_2 C_{ab} + d_1 L_{(a;b)} = 0 \\
c_1 C_{ab} + d_0 L_{(a;b)} + B_{(a;b)} = 0%
\end{cases}
\implies
\begin{cases}
C_{ab} = - \frac{d_2}{3c_3} L_{(a;b)} \\
\left( d_1 - \frac{2c_2 d_2}{3c_3} \right) L_{(a;b)} = 0 \\
B_{(a;b)} = \left( \frac{3 c_3 d_0}{d_2} - c_1 \right) C_{ab}%
\end{cases}%
\end{equation*}
and from the second
\begin{equation*}
\begin{cases}
d_2 \left( L_b V^{,b} \right)_{;a} = 6 c_3 C_{ab} V^{,b} \\
d_1 \left(L_b V^{,b} \right)_{;a} = 4 c_2 C_{ab} V^{,b} \\
2d_2 L_a + d_0 \left( L_b V^{,b} \right)_{;a} + \left( B_b V^{,b}
\right)_{;a} - 2 c_1 C_{ab} V^{,b} = 0%
\end{cases}
\implies
\begin{cases}
\left( L_b V^{,b} \right)_{;a} = \frac{6 c_3}{d_2} C_{ab} V^{,b} \\
\left( \frac{6 c_3 d_1}{d_2} - 4c_2 \right) C_{ab} V^{,b} = 0 \\
L_a = \left( \frac{c_1}{6c_3} - \frac{d_0}{2d_2} \right) \left( L_b V^{,b}
\right)_{;a} - \frac{1}{2d_2} \left( B_b V^{,b} \right)_{;a}.%
\end{cases}%
\end{equation*}
Therefore $L_a$ is a gradient vector and the function $G(q)$ becomes
\begin{equation*}
G(q) = \left(\frac{c_0 d_2}{3c_3} - \frac{c_1 d_1}{6c_3} + \frac{d_0d_1}{2d_2%
} \right) L_a V^{,a} + \frac{d_1}{2d_2} B_a V^{,a}.
\end{equation*}

Finally, the first integral is
\begin{eqnarray*}
I_{32} &=& - \frac{d_2}{3c_3} \left( c_{0}+c_{1}t+c_{2}t^{2} + c_3t^3 \right) L_{(a;b)} \dot{q}^{a}%
\dot{q}^{b}+ \left( d_{0}+d_{1}t + d_2t^2\right) L_{a}\dot{q}^{a} + B_{a}%
\dot{q}^{a} + \\
&& + \left( d_{0}t + \frac{d_{1}}{2}t^{2} + \frac{d_2}{3}t^3
\right) L_{a}V^{,a} + B_a V^{,a}t + \left(\frac{2c_0 d_2 - c_1 d_1}{6c_3} + \frac{d_0d_1}{2d_2%
} \right) L_a V^{,a} + \frac{d_1}{2d_2} B_a V^{,a}.
\end{eqnarray*}
where the vector $L_{a}=\left( \frac{c_{1}}{6c_{3}}-\frac{d_{0}}{2d_{2}}\right)
\left(L_{b}V^{,b}\right) _{;a}-\frac{1}{2d_{2}}\left( B_{b}V^{,b}\right)_{;a}$ is a gradient
such that $L_{(a;b)}$ is a KT, $\left( \frac{2c_{2}d_{2}}{3c_{3}}- d_{1}\right) L_{(a;b)} =0$ and
$\left( L_{b}V^{,b}\right) _{;a}=-2L_{(a;b)}V^{,b}$; and $B_{a}$ is a vector satisfying the relation $B_{(a;b)}=\left( \frac{c_{1}d_{2}}{3c_{3}}-d_{0}\right) L_{(a;b)}$.

The vector $L_{a}=\Psi_{,a}$ where
\begin{equation*}
\Psi(q) = \left( \frac{c_{1}}{6c_{3}}-\frac{d_{0}}{2d_{2}}\right)
L_{a}V^{,a} - \frac{1}{2d_{2}} B_{a}V^{,a}.
\end{equation*}
Observe also that
\begin{equation*}
- \frac{c_{2}d_{2}}{3c_{3}} L_{(a;b)} = - \frac{d_{1}}{2} L_{(a;b)}, \enskip
- \frac{c_{1}d_{2}}{3c_{3}} L_{(a;b)} = - B_{(a;b)} - d_{0} L_{(a;b)}.
\end{equation*}
From the last $B_{(a;b)}$ is a KT.

Another useful relation is the following (condition for $Q_{7}$ being a FI)
\begin{equation*}
L_{a} = \Psi_{,a} \implies 2d_{2}L_{a} = \left( \frac{c_{1}d_{2}}{3c_{3}}%
-d_{0}\right) \left(L_{b}V^{,b}\right) _{,a}-\left( B_{b}V^{,b}\right)_{,a}
\implies
\end{equation*}
\begin{equation*}
\left( B_{b}V^{,b}\right)_{,a} = -2B_{(a;b)}V^{,b} -2d_{2}L_{a}.
\end{equation*}

Substituting the above relations in the FI $I_{32}$ we find
\begin{equation*}
I_{32} = \frac{d_{2}c_{0}}{3c_{3}}Q_{3}(L_{a}) + Q_{7} + d_{0}Q_{4}(L_{a}) +
d_{1}Q_{6}(\Psi)
\end{equation*}
where
\begin{equation*}
Q_{7} = - \frac{t^{3}}{3} d_{2}L_{(a;b)}\dot{q}^{a}\dot{q}^{b} + t^{2}
d_{2}L_{a}\dot{q}^{a} + \frac{t^{3}}{3} d_{2}L_{a}V^{,a} - t B_{(a;b)} \dot{q%
}^{a}\dot{q}^{b} + B_{a}\dot{q}^{a} + t B_{a}V^{,a}
\end{equation*}
is a new independent FI.

\underline{\textbf{Subcase $\mathbf{(n=m+1,m>2)}$.}} $c_{n}\neq
0$ and $d_{m}\neq 0$.

\begin{equation*}
\begin{cases}
\eqref{FL.1.a1} \implies C_{(ab;c)} = 0 \\
\eqref{FL.1.a} \implies \left[ c_1 + 2 c_2 t + ... + (m+1) c_n t^{m} \right]
C_{ab} + \left( d_0 + d_1 t + ... + d_m t^m \right) L_{(a;b)} + B_{(a;b)} = 0
\\
\eqref{FL.1.b} \implies - 2 \left( c_0 + c_1 t + ... + c_n t^{m+1} \right)
C_{ab} V^{,b} + (d_1 + 2d_2 t + ... + m d_m t^{m-1}) L_a + K_{,a} = 0 \\
\eqref{FL.1.c} \implies K_{,t} = \left( d_0 + d_1 t + ... + d_m t^m \right)
L_a V^{,a} + B_a V^{,a} \\
\eqref{FL.1.d} \implies \left[ 2d_2 + 3 \cdot 2 d_3 t + ... + m (m-1) d_m
t^{m-2} \right] L_a + \left( d_0 + d_1 t + ... + d_m t^m \right) \left( L_b
V^{,b} \right)_{;a} + \left( B_b V^{,b} \right)_{;a} - \\
\qquad \qquad - 2 \left[ c_1 + 2 c_2 t + ... + (m+1) c_n t^{m} \right]
C_{ab} V^{,b} = 0.%
\end{cases}%
\end{equation*}

From the first equation $C_{ab}$ is a KT.

From \eqref{FL.1.a} we find the conditions $(k+1) c_{k+1} C_{ab} + d_k
L_{(a;b)} = 0$ where $k = 1,2,...,m$ and $c_1 C_{ab} + d_0 L_{(a;b)} +
B_{(a;b)} = 0$. For $k=m$ we get
\begin{equation*}
C_{ab} = - \frac{d_m}{(m+1)c_{m+1}} L_{(a;b)}
\end{equation*}
and the remaining equations become
\begin{equation*}
\left[ d_k - \frac{(k+1)c_{k+1}d_m}{(m+1)c_{m+1}} \right] L_{(a;b)} = 0, %
\enskip k=1,2,...,m-1
\end{equation*}
and
\begin{equation*}
B_{(a;b)} = \left[ \frac{c_1d_m}{(m+1)c_{m+1}} - d_0 \right] L_{(a;b)}.
\end{equation*}

From \eqref{FL.1.d} we find that $2(k+1)c_{k+1} C_{ab} V^{,b} = d_k \left(
L_b V^{,b} \right)_{;a}$ where $k=m-1,m$, $(k+2)(k+1)d_{k+2} L_a + d_k
\left( L_b V^{,b} \right)_{;a} - 2 (k+1) c_{k+1} C_{ab} V^{,b} = 0$ where $%
k=1,...,m-2$ and $2d_2 L_a + d_0 \left(L_b V^{,b} \right)_{;a} + \left( B_b
V^{,b} \right)_{;a} - 2 c_1 C_{ab} V^{,b} = 0$, i.e. $L_a$ is a gradient
vector. The first set of equations gives for $k=m$
\begin{equation*}
\left( L_b V^{,b} \right)_{,a} = \frac{2(m+1)c_{m+1}}{d_m} C_{ab} V^{,b}
\implies \left( L_b V^{,b} \right)_{,a} = - 2 L_{(a;b)} V^{,b}
\end{equation*}
and for $k=m-1$
\begin{equation*}
\left[ d_{m-1} - \frac{m c_m d_m}{(m+1)c_{m+1}}\right] \left( L_b V^{,b}
\right)_{,a} = 0.
\end{equation*}
The second set of equations for $k=m-2$ gives
\begin{equation*}
L_a = \left[ \frac{c_{m-1}}{m(m+1)c_{m+1}} - \frac{d_{m-2}}{m (m-1) d_m} %
\right] \left( L_b V^{,b} \right)_{,a}
\end{equation*}
and the remaining equations (exist only for $m>3$) are
\begin{equation*}
\left\{ d_k + (k+2)(k+1)d_{k+2} \left[ \frac{c_{m-1}}{m(m+1)c_{m+1}} - \frac{%
d_{m-2}}{m (m-1) d_m} \right] - \frac{(k+1) c_{k+1} d_m}{(m+1)c_{m+1}}
\right\} \left( L_b V^{,b} \right)_{,a} = 0, \enskip k=1,2,...,m-3.
\end{equation*}
From the last condition we get
\begin{equation*}
\left( B_b V^{,b} \right)_{,a} = \left[ \frac{c_1 d_m}{(m+1) c_{m+1}} -
\frac{2 d_2 c_{m-1}}{m(m+1)c_{m+1}} + \frac{2d_2 d_{m-2}}{m (m-1) d_m} - d_0 %
\right] \left( L_b V^{,b} \right)_{,a}.
\end{equation*}

The solution of \eqref{FL.1.c} is
\begin{equation*}
K = \left( d_{0}t + \frac{d_{1}}{2}t^{2} + ... + \frac{d_m}{m+1}t^{m+1}
\right) L_{a}V^{,a} + B_{a}V^{,a}t + G(q)
\end{equation*}
which into \eqref{FL.1.b} and using the conditions $2d_2 L_a + d_0 \left(L_b
V^{,b} \right)_{;a} + \left( B_b V^{,b} \right)_{;a} - 2 c_1 C_{ab} V^{,b} =
0$, $2(k+1)c_{k+1} C_{ab} V^{,b} = d_k \left( L_b V^{,b} \right)_{;a}$ where
$k=m-1,m$ and $(k+2)(k+1)d_{k+2} L_a + d_k \left( L_b V^{,b} \right)_{;a} -
2 (k+1) c_{k+1} C_{ab} V^{,b} = 0$ where $k=1,...,m-2$ gives
\begin{eqnarray*}
G_{,a} &=& - \left( d_{0}t + \frac{d_{1}}{2}t^{2}+ ... + \frac{d_{m-1}}{m}%
t^{m} + \frac{d_m}{m+1}t^{m+1} \right) \left( L_{b}V^{,b} \right)_{,a} + 2
\left( c_0 + c_1 t + ...+ c_m t^m + c_{m+1} t^{m+1} \right) C_{ab} V^{,b} -
\\
&& - \left( B_{b} V^{,b} \right)_{,a} t - (d_1 + 2d_2 t + ... + m d_m
t^{m-1}) L_a \\
&=& 2 c_0 C_{ab} V^{,b} - d_1 L_a \\
&=& \left[ \frac{c_0d_m}{(m+1)c_{m+1}} - \frac{c_{m-1} d_1}{m(m+1)c_{m+1}} +
\frac{d_1 d_{m-2}}{m (m-1) d_m} \right] \left( L_{b}V^{,b} \right)_{,a}
\implies
\end{eqnarray*}
\begin{eqnarray*}
G(q) &=& \left[ \frac{c_0d_m}{(m+1)c_{m+1}} - \frac{c_{m-1} d_1}{%
m(m+1)c_{m+1}} + \frac{d_1 d_{m-2}}{m (m-1) d_m} \right] L_a V^{,a}.
\end{eqnarray*}
Therefore
\begin{equation*}
K = \left( d_{0}t + \frac{d_{1}}{2}t^{2} + ... + \frac{d_m}{m+1}t^{m+1}
\right) L_{a}V^{,a} + B_{a}V^{,a}t + \left[ \frac{c_0d_m}{(m+1)c_{m+1}} -
\frac{c_{m-1} d_1}{m(m+1)c_{m+1}} + \frac{d_1 d_{m-2}}{m (m-1) d_m} \right]
L_a V^{,a}.
\end{equation*}

The first integral is
\begin{eqnarray*}
I_{(m+1)m}(m>2) &=&-\frac{d_{m}}{(m+1)c_{m+1}}\left(
c_{0}+c_{1}t+...+c_{m+1}t^{m+1}\right)L_{(a;b)} \dot{q}^{a} \dot{q}^{b}
+\left( d_{0}+d_{1}t+...+d_{m}t^{m}\right) L_{a} \dot{q}^{a}+ \\
&&+B_{a}\dot{q}^{a} +\left( d_{0}t+\frac{d_{1}}{2}t^{2}+...+\frac{d_{m}}{m+1}%
t^{m+1}\right) L_{a}V^{,a}+B_{a}V^{,a}t+G(q).
\end{eqnarray*}%
where $c_{m+1}d_{m}\neq 0$ for a finite $m>2$; the vector
\begin{equation*}
L_{a}=\left[ \frac{c_{m-1}}{m(m+1)c_{m+1}}-\frac{d_{m-2}}{m(m-1)d_{m}}\right]
\left( L_{b}V^{,b}\right) _{,a}
\end{equation*}
is a gradient such that $L_{(a;b)}$ is a KT, $\left( L_{b}V^{,b}\right) _{,a}=-2L_{(a;b)}V^{,b}$, $\left[ d_{k}-%
\frac{(k+1)c_{k+1}d_{m}}{(m+1)c_{m+1}}\right] L_{(a;b)}$ $=0$, where $%
k=1,2,...,m-1$, \newline
$\left[ d_{m-1}-\frac{mc_{m}d_{m}}{(m+1)c_{m+1}}\right] \left(
L_{b}V^{,b}\right) _{,a}$ $=0$ and
\begin{equation*}
\left\{ d_{k}+(k+2)(k+1)d_{k+2}\left[ \frac{c_{m-1}}{m(m+1)c_{m+1}}-\frac{%
d_{m-2}}{m(m-1)d_{m}}\right] -\frac{(k+1)c_{k+1}d_{m}}{(m+1)c_{m+1}}\right\}
\left( L_{b}V^{,b}\right) _{,a}=0,
\end{equation*}%
where $k=1,2,...,m-3$; $B_{a}$ is a vector which must satisfy the conditions
$B_{(a;b)}=\left[ \frac{c_{1}d_{m}}{(m+1)c_{m+1}}-d_{0}\right] L_{(a;b)}$
and \newline
\begin{equation*}
\left( B_{b}V^{,b}\right) _{,a}=\left[ \frac{c_{1}d_{m}}{(m+1)c_{m+1}}%
-\right. \left. \frac{2d_{2}c_{m-1}}{m(m+1)c_{m+1}}+\frac{2d_{2}d_{m-2}}{%
m(m-1)d_{m}}-d_{0}\right] \left( L_{b}V^{,b}\right) _{,a};
\end{equation*}%
and the function
\begin{equation*}
G(q)=\left[ \frac{c_{0}d_{m}}{(m+1)c_{m+1}}-\frac{c_{m-1}d_{1}}{m(m+1)c_{m+1}%
}+\frac{d_{1}d_{m-2}}{m(m-1)d_{m}}\right] L_{a}V^{,a}.
\end{equation*}

In that case $m>2$ which implies that $m-1>1$ and $m-2>0$. Therefore, there
always exist $k$-values for $m-1$, $m-2$.

For $k=m-2$ the condition $\left[ d_{k}-\frac{(k+1) c_{k+1}d_{m}}{%
(m+1)c_{m+1}}\right] L_{(a;b)}=0$ gives
\begin{equation*}
\left[ d_{m-2}-\frac{(m-1) c_{m-1}d_{m}}{(m+1)c_{m+1}}\right] L_{(a;b)}=0
\implies
\left[ \frac{c_{m-1}}{m(m+1)c_{m+1}}-\frac{d_{m-2}}{m(m-1)d_{m}}\right]
\left( L_{b}V^{,b}\right) _{,a} = 0 \implies L_{a}=0
\end{equation*}
because $\left( L_{b}V^{,b}\right)_{,a}=-2L_{(a;b)}V^{,b}$.

Since $L_{a}=0$ we have $G=0$, $B_{(a;b)}=0$, i.e. $B_{a}$ is a KV, and $B_{a}V^{,a}=s_{1}=const$. Therefore
\begin{equation*}
I_{(m+1)m}(m>2) = B_{a}\dot{q}^{a} + st = Q_{2}(B_{a}).
\end{equation*}
\vspace{12pt}

\underline{\textbf{I.3. Case $\mathbf{n>m+1}$}.}
\vspace{12pt}

\underline{\textbf{Subcase $\mathbf{(n>1, m=0)}$.}} $c_n \neq 0$.
\begin{equation*}
\begin{cases}
\eqref{FL.1.a1} \implies C_{(ab;c)} = 0 \\
\eqref{FL.1.a} \implies (c_1 + 2 c_2 t + ... + n c_n t^{n-1}) C_{ab} + \tilde{L}_{(a;b)} = 0 \\
\eqref{FL.1.b} \implies -2 (c_0 + c_1 t + ... + c_n t^n) C_{ab} V^{,b} +
K_{,a} = 0 \\
\eqref{FL.1.c} \implies K_{,t} = \tilde{L}_b V^{,b} \\
\eqref{FL.1.d} \implies \left( \tilde{L}_b V^{,b} \right)_{;a} - 2 (c_1 + 2c_2 t + ... + n c_n t^{n-1}) C_{ab} V^{,b} = 0.
\end{cases}
\end{equation*}

From \eqref{FL.1.a} we find that $C_{ab} = 0$; and $\tilde{L}_{(a;b)} = 0$
that is $\tilde{L}_a$ is a KV. Then equation \eqref{FL.1.d} gives $\tilde{L}_aV^{,a} = s_0$ and \eqref{FL.1.c} yields $K = s_0 t + G(q)$.

The last result into \eqref{FL.1.b} gives $G_{,a} = 0$, i.e. $G$ is a
constant which is ignored because any arbitrary constant can be added to a
first integral $I$ without changing the condition $\frac{dI}{dt} = 0$.

The first integral is (of the form $Q_{2}$)
\begin{equation*}
I_{n0}(n>1)=\tilde{L}_{a}\dot{q}^{a}+s_{0}t.
\end{equation*}
where $\tilde{L}_{a}\equiv d_{0}L_{a}+B_{a}$ is a KV such that $\tilde{L}_aV^{,a} = s_0$.

\underline{\textbf{Subcase $\mathbf{(n>2, m=1)}$.}} $c_n \neq 0$ and $d_1
\neq 0$.

\begin{equation*}
\begin{cases}
\eqref{FL.1.a1} \implies C_{(ab;c)} = 0 \\
\eqref{FL.1.a} \implies \left( c_1 + 2 c_2 t + ... + n c_n t^{n-1} \right)
C_{ab} + \left( d_0 + d_1 t \right) L_{(a;b)} + B_{(a;b)} =0 \\
\eqref{FL.1.b} \implies - 2 \left( c_0 + c_1 t + ... + c_n t^n \right)
C_{ab} V^{,b} + d_1 L_a + K_{,a} = 0 \\
\eqref{FL.1.c} \implies K_{,t} = \left( d_0 + d_1 t \right) L_a V^{,a} + B_a
V^{,a} \\
\eqref{FL.1.d} \implies \left( d_0 + d_1 t \right) \left( L_b V^{,b}
\right)_{;a} + \left( B_b V^{,b} \right)_{;a} - 2 \left( c_1 + 2 c_2 t + ...
+ n c_n t^{n-1} \right) C_{ab} V^{,b} = 0.%
\end{cases}%
\end{equation*}

From \eqref{FL.1.a} we find that $C_{ab} = 0$ and $L_a$, $B_a$ are KVs.

From \eqref{FL.1.d} we have that $L_a V^{,a} = s_1$ and $B_a V^{,a} = s_2$.

Then equation \eqref{FL.1.c} gives
\begin{equation*}
K = s_1 \left( d_0t + \frac{d_1}{2} t^2 \right) + s_2 t + G(q)
\end{equation*}
which into \eqref{FL.1.b} yields $G_{,a} = - d_1 L_a$ that is $L_a$ is a
gradient KV.

The first integral is (consists of FIs of the form $Q_{2}$, $Q_{5}$)
\begin{equation*}
I_{n1}(n>2)=(d_{0}+d_{1}t)L_{a}\dot{q}^{a}+B_{a}\dot{q}^{a}+ \frac{s_{1}d_{1}%
}{2}t^{2}+(s_{1}d_{0}+s_{2})t+G(q)
\end{equation*}%
where $L_{a}$, $B_{a}$ are KVs such that $L_{a}V^{,a}=s_{1}$ and $%
B_{a}V^{,a}=s_{2}$; and $G(q)=-d_{1}\int L_{a}dq^{a}$.

\underline{\textbf{Subcase $\mathbf{(n>3, m=2)}$.}} $c_n \neq 0$ and $d_2
\neq 0$.

\begin{equation*}
\begin{cases}
\eqref{FL.1.a1} \implies C_{(ab;c)} = 0 \\
\eqref{FL.1.a} \implies \left( c_1 + 2 c_2 t + ... + n c_n t^{n-1} \right)
C_{ab} + \left( d_0 + d_1 t + d_2 t^2 \right) L_{(a;b)} + B_{(a;b)} =0 \\
\eqref{FL.1.b} \implies - 2 \left( c_0 + c_1 t + ... + c_n t^n \right)
C_{ab} V^{,b} + (d_1 + 2d_2t) L_a + K_{,a} = 0 \\
\eqref{FL.1.c} \implies K_{,t} = \left( d_0 + d_1 t + d_2 t^2 \right) L_a
V^{,a} + B_a V^{,a} \\
\eqref{FL.1.d} \implies 2d_2 L_a + \left( d_0 + d_1 t + d_2t^2 \right)
\left( L_b V^{,b} \right)_{;a} + \left( B_b V^{,b} \right)_{;a} - \\
\qquad \qquad - 2 \left( c_1 + 2 c_2 t + ... + n c_n t^{n-1} \right) C_{ab}
V^{,b} = 0.%
\end{cases}%
\end{equation*}

From \eqref{FL.1.a} we find that $C_{ab} = 0$ and $L_a$, $B_a$ are KVs.

From \eqref{FL.1.d} we have that $L_a V^{,a} = s_1$ and $L_a = - \frac{1}{%
2d_2} \left( B_b V^{,b} \right)_{;a}$, that is $L_a$ is a gradient KV.

Then equation \eqref{FL.1.c} gives
\begin{equation*}
K = s_1 \left( d_0t + \frac{d_1}{2} t^2 + \frac{d_2}{3} t^3 \right) + B_a
V^{,a} t + G(q)
\end{equation*}
which into \eqref{FL.1.b} yields
\begin{equation*}
G_{,a} = - d_1 L_a = \frac{d_1}{2d_2} \left( B_b V^{,b} \right)_{,a}
\implies G(q) = \frac{d_1}{2d_2} B_a V^{,a}.
\end{equation*}

The first integral is (consists of FIs of the form $Q_{2}$, $Q_{5}$, $Q_{7}$)
\begin{equation*}
I_{n2}(n>3)=(d_{0}+d_{1}t+d_2t^2) L_{a}\dot{q}^{a} + B_{a} \dot{q}^{a} + s_1
\left( d_0t + \frac{d_1}{2} t^2 + \frac{d_2}{3} t^3 \right) + B_a V^{,a} t +
\frac{d_1}{2d_2} B_a V^{,a}
\end{equation*}
where $L_a = - \frac{1}{2d_2} \left( B_b V^{,b} \right)_{;a}$ is a gradient
KV such that $L_a V^{,a} = s_1$ and $B_{a}$ is a KV.

\underline{\textbf{Subcase $\mathbf{(n>m+1, m>2)}$.}} $c_n \neq 0$ and $d_m
\neq 0$. Note that $n > n-1 > m > 2$.

\begin{equation*}
\begin{cases}
\eqref{FL.1.a1} \implies C_{(ab;c)} = 0 \\
\eqref{FL.1.a} \implies \left( c_1 + 2 c_2 t + ... + n c_n t^{n-1} \right)
C_{ab} + \left( d_0 + d_1 t + ... + d_m t^m \right) L_{(a;b)} + B_{(a;b)} = 0
\\
\eqref{FL.1.b} \implies - 2 \left( c_0 + c_1 t + ... + c_n t^{n} \right)
C_{ab} V^{,b} + (d_1 + 2d_2 t + ... + m d_m t^{m-1}) L_a + K_{,a} = 0 \\
\eqref{FL.1.c} \implies K_{,t} = \left( d_0 + d_1 t + ... + d_m t^m \right)
L_a V^{,a} + B_a V^{,a} \\
\eqref{FL.1.d} \implies \left[ 2d_2 + 3 \cdot 2 d_3 t + ... + m (m-1) d_m
t^{m-2} \right] L_a + \left( d_0 + d_1 t + ... + d_m t^m \right) \left( L_b
V^{,b} \right)_{;a} + \left( B_b V^{,b} \right)_{;a} - \\
\qquad \qquad - 2 \left( c_1 + 2 c_2 t + ... + n c_n t^{n-1} \right) C_{ab}
V^{,b} = 0.%
\end{cases}%
\end{equation*}

From \eqref{FL.1.a} we find that $C_{ab} = 0$ and $L_a$, $B_a$ are KVs.

From \eqref{FL.1.d} we have that $L_a = 0$ and $B_a V^{,a} = s_2$.

Then the solution of \eqref{FL.1.c} is $K = s_2 t + G(q)$ which into %
\eqref{FL.1.b} gives $G=const$.

The first integral is (again of the form $Q_{2}$)
\begin{equation*}
I_{nm}(n>m+1, m>2)= B_a \dot{q}^a + s_2 t
\end{equation*}
where $B_{a}$ is a KV such that $B_a V^{,a} = s_2$. \vspace{12pt}

\underline{\textbf{I.4. Case $\mathbf{n<m}$.}} \vspace{12pt}

\underline{\textbf{Subcase $\mathbf{(n=0, m=1)}$.}} $g = c_0$, $f = d_0 +
d_1 t$ with $d_1 \neq 0$.

\begin{equation*}
\begin{cases}
\eqref{FL.1.a1} \implies c_0 C_{(ab;c)} = 0 \\
\eqref{FL.1.a} \implies (d_0 + d_1 t) L_{(a;b)} + B_{(a;b)} = 0 \\
\eqref{FL.1.b} \implies -2 c_0 C_{ab} V^{,b} + d_1 L_a + K_{,a} = 0 \\
\eqref{FL.1.c} \implies K_{,t} = (d_0 + d_1t) L_b V^{,b} + B_b V^{,b} \\
\eqref{FL.1.d} \implies (d_0 + d_1t) \left( L_b V^{,b} \right)_{;a} + \left(
B_b V^{;b} \right)_{;a} = 0.%
\end{cases}%
\end{equation*}

Equation \eqref{FL.1.a} implies that $L_{(a;b)} = 0$ and $B_{(a;b)} = 0$,
that is $L_a$, $B_a$ are KVs. Similarly, equation \eqref{FL.1.d} gives that $%
L_b V^{,b} = s_1 = const$ and $B_b V^{,b} = s_2 =const$.

Then \eqref{FL.1.c} gives
\begin{equation*}
K = s_1 \left(d_0 t + \frac{d_1}{2}t^2\right) + s_2 t + G(q)
\end{equation*}
which into \eqref{FL.1.b} yields $G_{,a} = 2c_0C_{ab}V^{,b} - d_1 L_a$.

The first integral is ($c_{0}$ is absorbed by $C_{ab}$)
\begin{equation*}
I_{01} = C_{ab}\dot{q}^{a}\dot{q}^{b}+ \left( d_{0}+d_{1}t\right)
L_{a}\dot{q}^{a} + B_{a}\dot{q}^{a} + s_{1}\left( d_{0}t + \frac{d_{1}}{2}%
t^{2} \right)+ s_{2}t + G\left( q\right)
\end{equation*}
where $d_1 \neq 0$, $L_{a},B_{a}$ are KVs such that $L_{a}V^{,a}=s_{1}$, $B_{a}V^{,a}=s_{2} $; and $C_{ab}$ is a KT such that $G_{,a}-2 C_{ab} V^{,b}+ d_1L_{a} =0$.

We have
\begin{equation*}
I_{01} = Q_{2}(B_{a}) + d_{0}Q_{2} + Q_{8}
\end{equation*}
where
\begin{equation*}
Q_{8} = C_{ab}\dot{q}^{a}\dot{q}^{b} + d_{1}tL_{a}\dot{q}^{a} + d_{1} \frac{%
s_{1}}{2} t^{2} + G(q)
\end{equation*}
is just a subcase of $Q_{16}$ since $d_{1}L_{a}$ is a KV.

\underline{\textbf{Subcase $\mathbf{(n=0, m=2)}$.}} $g = c_0$, $f = d_0 +d_1 t + d_2 t^2 $ with $d_2 \neq 0$.

\begin{equation*}
\begin{cases}
\eqref{FL.1.a1} \implies c_0 C_{(ab;c)} = 0 \\
\eqref{FL.1.a} \implies \left( d_0 + d_1 t + d_2 t^2 \right) L_{(a;b)} +
B_{(a;b)} = 0 \\
\eqref{FL.1.b} \implies -2 c_0 C_{ab} V^{,b} + \left( d_1 + 2d_2 t \right)
L_a + K_{,a} = 0 \\
\eqref{FL.1.c} \implies K_{,t} = \left( d_0 + d_1t + d_2 t^2 \right) L_b
V^{,b} + B_b V^{,b} \\
\eqref{FL.1.d} \implies 2d_2 L_a + \left( d_0 + d_1t + d_2 t^2 \right)
\left( L_b V^{,b} \right)_{;a} + \left( B_b V^{,b} \right)_{;a}= 0.%
\end{cases}%
\end{equation*}

From \eqref{FL.1.a} we have that $L_a$, $B_a$ are KVs.

From \eqref{FL.1.d} we get $L_{b}V^{,b}=s_{1}$; and $L_{a}= -\frac{1}{2d_{2}}
\left( B_{b}V^{,b}\right)_{,a}$, that is $L_{a}$ is a gradient KV.

Equation \eqref{FL.1.c} yields
\begin{equation*}
K = s_1 \left( d_0t + \frac{d_1}{2}t^2 + \frac{d_2}{3} t^3 \right) + B_b
V^{,b}t + G(q)
\end{equation*}
and \eqref{FL.1.b} gives $G_{,a} = 2 c_0 C_{ab} V^{,b} - d_1L_a$. Using the
relation $L_{a}= -\frac{1}{2d_{2}} \left( B_{b}V^{,b} \right)_{,a}$ we find
that
\begin{equation*}
G(q) = \frac{d_1}{2d_{2}} B_{a} V^{,a} + 2 c_0 \int C_{ab} V^{,b} dq^a.
\end{equation*}

The first integral is ($c_{0}$ is absorbed by $C_{ab}$)
\begin{equation*}
I_{02} = C_{ab}\dot{q}^{a}\dot{q}^{b} + \left( d_{0} +
d_{1}t+d_{2}t^{2} \right) L_a \dot{q}^{a} + B_{a}\dot{q}^{a} + s_{1} \left(
d_0t + \frac{d_1}{2}t^{2} + \frac{d_2}{3} t^3 \right) + B_{a}V^{,a}t +
G\left( q\right)
\end{equation*}
where $d_{2}\neq 0$, $B_{a}$ is a KV, $L_{a} = - \frac{1}{2d_{2}} \left(
B_{b}V^{,b} \right)_{,a}$ is a gradient KV such that $L_{a}V^{,a} = s_{1}$;
and the $C_{ab}$ is a KT satisfying the relation $G_{,a} - 2C_{ab}V^{,b} +
d_{1}L_{a} = 0$.

We find that $I_{02} = Q_{8} + d_{0}Q_{2} + Q_{7}(L_{a}=KV,B_{a}=KV)$.

\underline{\textbf{Subcase $\mathbf{(n=0, m>2)}$.}} $d_m \neq 0$.

\begin{equation*}
\begin{cases}
\eqref{FL.1.a1} \implies c_0 C_{(ab;c)} = 0 \\
\eqref{FL.1.a} \implies \left( d_0 + d_1 t + ... + d_m t^m \right) L_{(a;b)}
+ B_{(a;b)} = 0 \\
\eqref{FL.1.b} \implies -2 c_0 C_{ab} V^{,b} + \left( d_1 + 2d_2 t + ... + m
d_m t^{m-1} \right) L_a + K_{,a} = 0 \\
\eqref{FL.1.c} \implies K_{,t} = \left( d_0 + d_1t + ... + d_m t^m \right)
L_b V^{;b} + B_b V^{;b} = 0 \\
\eqref{FL.1.d} \implies \left[ 2d_2 + 3 \cdot 2 d_3 t + ... + m (m-1) d_m
t^{m-2} \right] L_a + \left( d_0 + d_1t + ... + d_m t^m \right) \left( L_b
V^{;b} \right)_{;a} + \left( B_b V^{;b} \right)_{;a}= 0.%
\end{cases}%
\end{equation*}

From \eqref{FL.1.a} $L_a$, $B_a$ are KVs.

From \eqref{FL.1.d} $L_a = 0$ and $B_a V^{,b} = s_2$.

Solving \eqref{FL.1.c} we find $K = s_2t + G(q)$ which into \eqref{FL.1.b}
gives $G_{,a} = 2 c_0 C_{ab} V^{,b}$.

The first integral is (of the form $Q_{8}$)
\begin{equation*}
I_{0m}(m>2)=c_{0}C_{ab}\dot{q}^{a}\dot{q}^{b}+B_{a}\dot{q}^{a}+ s_{2}t+G(q)
\end{equation*}%
where if $c_{0}\neq 0$ the $C_{ab}$ is a KT, $B_{a}$ is a KV such that $%
B_{a}V^{,a}=s_{2}$ and $G(q)=2c_{0}\int C_{ab}V^{,b}dq^{a}$.

\underline{\textbf{Subcase $\mathbf{(n=1, m=2)}$.}} $g = c_0 + c_1 t$, $f =d_0 + d_1 t + d_2 t^2$ with $c_1 \neq 0$ and $d_2 \neq 0$.

\begin{equation*}
\begin{cases}
\eqref{FL.1.a1} \implies C_{(ab;c)} = 0 \\
\eqref{FL.1.a} \implies c_1 C_{ab} + (d_0 + d_1t + d_2t^2) L_{(a;b)} +
B_{(a;b)} = 0 \\
\eqref{FL.1.b} \implies - 2 \left( c_0 + c_1 t \right) C_{ab} V^{,b} +
\left( d_1 + 2 d_2 t \right) L_a + K_{,a} = 0 \\
\eqref{FL.1.c} \implies K_{,t} = \left( d_0 + d_1 t + d_2 t^2 \right) L_a
V^{,a} + B_a V^{,a} \\
\eqref{FL.1.d} \implies 2 d_2 L_a + \left( d_0 + d_1 t + d_2 t^2 \right)
\left( L_b V^{,b} \right)_{;a} + \left( B_b V^{,b} \right)_{;a} - 2 c_1
C_{ab} V^{,b} = 0.%
\end{cases}%
\end{equation*}

From the first equation $C_{ab}$ is a KT.

Equation \eqref{FL.1.a} implies that $L_a$ is a KV and $c_1 C_{ab} = -
B_{(a;b)}$.

From \eqref{FL.1.d} we have $L_a V^{,a} = s_1$ and $2 d_2 L_a + \left( B_b V^{,b} \right)_{;a} - 2 c_1 C_{ab} V^{,b} = 0$.

The solution of \eqref{FL.1.c} is written
\begin{equation*}
K = s_1 \left( d_0t + \frac{d_1}{2} t^2 + \frac{d_2}{3} t^3 \right) + B_a
V^{,a} t + G(q)
\end{equation*}
where from \eqref{FL.1.b} we find
\begin{equation*}
- 2 \left( c_0 + c_1 t \right) C_{ab} V^{,b} + \left( d_1 + 2 d_2 t \right)
L_a + G_{,a} + \left( B_b V^{,b} \right)_{,a} t = 0 \implies
\end{equation*}
\begin{equation*}
G_{,a} - 2 c_0 C_{ab} V^{,b} + d_1 L_a + \underbrace{\left[ -2 c_1 C_{ab}
V^{,b} + 2d_2 L_a + \left( B_b V^{,b} \right)_{,a} \right]}_{=0} t = 0
\implies
\end{equation*}
\begin{equation*}
G_{,a} = \underbrace{2 c_0 C_{ab} V^{,b}} - d_1 L_a = c_0 \underbrace{\frac{%
2d_2c_0}{c_1}L_a + \frac{c_0}{c_1} \left( B_b V^{,b} \right)_{;a}} - d_1 L_a
.
\end{equation*}

The first integral is
\begin{eqnarray*}
I_{12} &=& -\frac{1}{c_{1}} \left( c_{0}+c_{1}t \right) B_{(a;b)}%
\dot{q}^{a} \dot{q}^{b} + \left( d_{0}+d_{1}t+d_{2}t^{2} \right) L_{a}\dot{q}%
^{a} + B_{a}\dot{q}^{a} + s_1 \left( d_{0}t + \frac{d_{1}}{2}t^{2} + \frac{%
d_{2}}{3}t^{3} \right) + \\
&& + B_{a} V^{,a}t + G\left(q\right)
\end{eqnarray*}
where $c_{1}d_{2}\neq 0$, $B_{a}$ is the vector (\ref{FL.20}), $L_{a} = -%
\frac{1}{2d_{2}} \left[ \left( B_{b}V^{,b} \right)_{,a} + 2B_{(a;b)} V^{,b} %
\right]$ is a KV such that $L_{a}V^{,a}=s_{1}$ and $G\left(q\right)$ is
defined by the condition $G_{,a} - \frac{c_0}{c_1} \left( B_{b}V^{,b}
\right)_{,a} + \left( d_{1} - \frac{2d_{2}c_{0}}{c_{1}} \right) L_{a}$ $=0$.

\bigskip

The condition $G_{,a} - \frac{c_0}{c_1} \left( B_{b}V^{,b} \right)_{,a} +
\left( d_{1} - \frac{2d_{2}c_{0}}{c_{1}} \right) L_{a}$ $=0$ generates the
following two cases: \bigskip

1) For $d_{1} \neq \frac{2d_{2}c_{0}}{c_{1}}$. Then $L_{a}$ is a gradient
KV, that is
\begin{equation*}
L_{a} = \Phi_{,a} \implies G(q) = \frac{c_0}{c_1} B_{a}V^{,a} - d_{1}
\Phi(q) + \frac{2d_{2}c_{0}}{c_{1}} \Phi(q).
\end{equation*}
Then
\begin{equation*}
I_{12(1)} = \frac{c_{0}}{c_{1}} Q_{9} + d_{0}Q_{2} + d_{1} Q_{6}(L_{a}=KV) +
Q_{7}(L_{a}=KV)
\end{equation*}
where
\begin{equation*}
Q_{9} = -B_{(a;b)}\dot{q}^{a}\dot{q}^{b} + B_{a}V^{,a} + 2d_{2} \Phi(q).
\end{equation*}
Observe that $Q_{3}=Q_{9}(\Phi=0)$ and $\left( B_{b}V^{,b} \right)_{,a} = -
2B_{(a;b)} V^{,b} -2d_{2}\Phi_{,a}$.

2) For $d_{1} = \frac{2d_{2}c_{0}}{c_{1}}$ we have
\begin{equation*}
G(q) = \frac{c_0}{c_1} B_{a}V^{,a}.
\end{equation*}
Then
\begin{equation*}
I_{12(2)} = \frac{c_{0}}{c_{1}} Q_{3} + d_{0}Q_{2} + Q_{8}(C_{ab}= -
B_{(a;b)}) + Q_{7}(L_{a}=KV).
\end{equation*}

\underline{\textbf{Subcase $\mathbf{(n=1,m>2)}$.}} $g=c_{0}+c_{1}t$, $c_{1}\neq 0$ and $d_{m}\neq 0$.

\begin{equation*}
\begin{cases}
\eqref{FL.1.a1} \implies C_{(ab;c)} = 0 \\
\eqref{FL.1.a} \implies c_1 C_{ab} + (d_0 + d_1t + ... + d_mt^m) L_{(a;b)} +
B_{(a;b)} = 0 \\
\eqref{FL.1.b} \implies - 2 \left( c_0 + c_1 t \right) C_{ab} V^{,b} +
\left( d_1 + 2 d_2 t + ... + m d_m t^{m-1} \right) L_a + K_{,a} = 0 \\
\eqref{FL.1.c} \implies K_{,t} = \left( d_0 + d_1 t + ... + d_m t^m \right)
L_a V^{,a} + B_a V^{,a} \\
\eqref{FL.1.d} \implies \left[ 2d_2 + 2 \cdot 3 d_3 t + ... + (m-1) m d_m
t^{m-2} \right] L_a + \\
\qquad \qquad + \left( d_0 + d_1 t + ... + d_m t^m \right) \left( L_b V^{,b}
\right)_{;a} + \left( B_b V^{,b} \right)_{;a} - 2 c_1 C_{ab} V^{,b} = 0.%
\end{cases}%
\end{equation*}

From the first equation $C_{ab}$ is a KT.

Equation \eqref{FL.1.a} implies that $L_a$ is a KV and $c_1 C_{ab} =
-B_{(a;b)}$.

From \eqref{FL.1.d} we have $L_a = 0$ and $\left( B_b V^{,b} \right)_{;a} =2 c_1 C_{ab} V^{,b}$.

The solution of \eqref{FL.1.c} is $K = B_a V^{,a} t + G(q)$ which into %
\eqref{FL.1.b} gives
\begin{equation*}
- 2 c_0 C_{ab} V^{,b} \underbrace{- 2 c_1 C_{ab} V^{,b} t + \left( B_b
V^{,b} \right)_{;a} t}_{=0} + G_{,a} = 0 \implies G_{,a} = 2 c_0 C_{ab}
V^{,b} .
\end{equation*}
But $\left( B_b V^{,b} \right)_{;a} = 2 c_1 C_{ab} V^{,b}$. Therefore
\begin{equation*}
G_{,a} = \frac{c_0}{c_1} \left( B_b V^{,b} \right)_{;a} \implies G(q) =\frac{c_0}{c_1} B_a V^{,a}.
\end{equation*}

The first integral is (consists of FIs of the form $Q_{1}$, $Q_{4}$)
\begin{equation*}
I_{1m}(m>2)=\left( c_{0}+c_{1}t\right) C_{ab}\dot{q}^{a}\dot{q}^{b}+B_{a}%
\dot{q}^{a}+B_{a}V^{,a}t+\frac{c_{0}}{c_{1}}B_{a}V^{,a}
\end{equation*}%
where $C_{ab}=-\frac{1}{c_{1}}B_{(a;b)}$ is a KT and $B_{a}$ is a vector
such that $\left( B_{b}V^{,b}\right) _{;a}+2B_{(a;b)}V^{,b}=0$.

\underline{\textbf{Subcase $\mathbf{(n>1, m>n)}$.}} $c_n \neq 0$ and $d_m
\neq 0$.

\begin{equation*}
\begin{cases}
\eqref{FL.1.a1} \implies C_{(ab;c)} = 0 \\
\eqref{FL.1.a} \implies \left( c_1 + 2 c_2 t + ... + n c_n t^{n-1} \right)
C_{ab} + \left( d_0 + d_1 t + ... + d_m t^m \right) L_{(a;b)} + B_{(a;b)} = 0
\\
\eqref{FL.1.b} \implies - 2 \left( c_0 + c_1 t + ... + c_n t^{n} \right)
C_{ab} V^{,b} + (d_1 + 2d_2 t + ... + m d_m t^{m-1}) L_a + K_{,a} = 0 \\
\eqref{FL.1.c} \implies K_{,t} = \left( d_0 + d_1 t + ... + d_m t^m \right)
L_a V^{,a} + B_a V^{,a} \\
\eqref{FL.1.d} \implies \left[ 2d_2 + 3 \cdot 2 d_3 t + ... + m (m-1) d_m
t^{m-2} \right] L_a + \left( d_0 + d_1 t + ... + d_m t^m \right) \left( L_b
V^{,b} \right)_{;a} + \left( B_b V^{,b} \right)_{;a} - \\
\qquad \qquad - 2 \left( c_1 + 2 c_2 t + ... + n c_n t^{n-1} \right) C_{ab}
V^{,b} = 0.%
\end{cases}%
\end{equation*}

From \eqref{FL.1.a} we find that $C_{ab} = 0$ and $L_a$, $B_a$ are KVs.

From \eqref{FL.1.d} we have that $L_a = 0$ and $B_a V^{,a} = s_2$.

Then the solution of \eqref{FL.1.c} is $K = s_2 t + G(q)$ which into %
\eqref{FL.1.b} gives $G=const$.

The first integral is (of the form $Q_{2}$)
\begin{equation*}
I_{nm}(n>1,m>n)= B_a \dot{q}^a + s_2 t
\end{equation*}
where $B_{a}$ is a KV such that $B_a V^{,a} = s_2$.

\vspace{12pt}

\textbf{II. For $\mathbf{n=\infty}$ and $\mathbf{m}$ finite.} \vspace{12pt}

We find the equivalences
\begin{equation*}
(n=\infty, m=0) \equiv (n>1, m=0) \equiv (g=e^{\lambda t}, m=0), \enskip %
(n=\infty, m=1) \equiv (n>2, m=1) \equiv (g=e^{\lambda t}, m=1),
\end{equation*}
\begin{equation*}
\enskip (n=\infty, m=2) \equiv (n>3, m=2) \equiv (g=e^{\lambda t}, m=2), %
\enskip (n=\infty, m>2) \equiv (n>m+1, m>2) \equiv (g=e^{\lambda t}, m>2).
\end{equation*}

Then for each case we have. \vspace{12pt}

\underline{\textbf{II.1. Case $\mathbf{(g = e^{\lambda t}}$, $\mathbf{f = d_0)}$.}} $\lambda \neq 0$.

\begin{equation*}
\begin{cases}
\eqref{FL.1.a1} \implies C_{(ab;c)} = 0 \\
\eqref{FL.1.a} \implies \lambda e^{\lambda t} C_{ab} + d_0 L_{(a;b)} +
B_{(a;b)} = 0 \\
\eqref{FL.1.b} \implies - 2 e^{\lambda t} C_{ab} V^{,b} + K_{,a} = 0 \\
\eqref{FL.1.c} \implies K_{,t} = d_0 L_a V^{,a} + B_a V^{,a} \\
\eqref{FL.1.d} \implies d_0 \left( L_b V^{,b} \right)_{;a} + \left( B_b
V^{,b} \right)_{;a} - 2 \lambda e^{\lambda t} C_{ab} V^{,b} = 0.%
\end{cases}%
\end{equation*}

From \eqref{FL.1.a} we get $C_{ab}=0$ and $\tilde{L}_{a}\equiv
d_{0}L_{a}+B_{a} $ is a KV.

From \eqref{FL.1.d} we have that $\tilde{L}_a V^{,a} = s_0$.

Equation \eqref{FL.1.c} gives $K = s_0t + G(q)$ which into \eqref{FL.1.b}
yields $G = const$.

The first integral is (of the form $Q_{2}$)
\begin{equation*}
I_{e0}=\tilde{L}_{a}\dot{q}^{a}+s_{0}t
\end{equation*}
where $\tilde{L}_{a}$ is a KV such that $\tilde{L}_{a}V^{,a}=s_{0}$.

\underline{\textbf{II.2. Case $\mathbf{(g=e^{\lambda t}}$, $\mathbf{f=d_{0}+d_{1}t)}$.}} $\lambda \neq 0$ and $d_{1}\neq 0$.

\begin{equation*}
\begin{cases}
\eqref{FL.1.a1} \implies C_{(ab;c)} = 0 \\
\eqref{FL.1.a} \implies \lambda e^{\lambda t} C_{ab} + \left( d_0 + d_1 t
\right) L_{(a;b)} + B_{(a;b)} = 0 \\
\eqref{FL.1.b} \implies - 2 e^{\lambda t} C_{ab} V^{,b} + d_1 L_a + K_{,a} =
0 \\
\eqref{FL.1.c} \implies K_{,t} = \left( d_0 + d_1 t \right) L_a V^{,a} + B_a
V^{,a} \\
\eqref{FL.1.d} \implies \left( d_0 + d_1 t \right) \left( L_b V^{,b}
\right)_{;a} + \left( B_b V^{,b} \right)_{;a} - 2 \lambda e^{\lambda t}
C_{ab} V^{,b} = 0.%
\end{cases}%
\end{equation*}

From \eqref{FL.1.a} we have that $C_{ab} = 0$ and $L_a$, $B_a$ are KVs.

From \eqref{FL.1.d} we get that $L_a V^{,a} = s_1$ and $B_a V^{,a} = s_2$.

Then equation \eqref{FL.1.c} gives
\begin{equation*}
K = s_1 \left( d_0t + \frac{d_1}{2} t^2 \right) + s_2 t + G(q)
\end{equation*}
which into \eqref{FL.1.b} gives $G_{,a} = - d_1 L_a$.

The first integral is (consists of $Q_{2}$, $Q_{5}$)
\begin{equation*}
I_{e1}=\left( d_{0}+d_{1}t\right) L_{a}\dot{q}^{a}+B_{a}\dot{q}%
^{a}+(s_{1}d_{0}+s_{2})t+\frac{s_{1}d_{1}}{2}t^{2}+G(q)
\end{equation*}%
where $L_{a}=-\frac{1}{d_{1}}G_{,a}$ is a gradient KV such that $%
L_{a}V^{,a}=s_{1}$, $B_{a}$ is a KV such that $B_{a}V^{,a}=s_{2}$ and $%
G(q)=-d_{1}\int L_{a}dq^{a}$.

\underline{\textbf{II.3. Case $\mathbf{(g = e^{\lambda t}}$, $\mathbf{f = d_0 + d_1 t + d_2t^2)}$.}} $\lambda \neq 0$ and $d_2 \neq 0$.

\begin{equation*}
\begin{cases}
\eqref{FL.1.a1} \implies C_{(ab;c)} = 0 \\
\eqref{FL.1.a} \implies \lambda e^{\lambda t} C_{ab} + \left( d_0 + d_1 t +
d_2t^2 \right) L_{(a;b)} + B_{(a;b)} = 0 \\
\eqref{FL.1.b} \implies - 2 e^{\lambda t} C_{ab} V^{,b} + (d_1 + 2d_2t) L_a
+ K_{,a} = 0 \\
\eqref{FL.1.c} \implies K_{,t} = \left( d_0 + d_1 t + d_2t^2 \right) L_a
V^{,a} + B_a V^{,a} \\
\eqref{FL.1.d} \implies 2d_2 L_a + \left( d_0 + d_1 t + d_2 t^2 \right)
\left( L_b V^{,b} \right)_{;a} + \left( B_b V^{,b} \right)_{;a} - 2 \lambda
e^{\lambda t} C_{ab} V^{,b} = 0.%
\end{cases}%
\end{equation*}

From \eqref{FL.1.a} we have that $C_{ab} = 0$ and $L_a$, $B_a$ are KVs.

From \eqref{FL.1.d} we get that $L_a V^{,a} = s_1$ and $\left( B_b V^{,b}
\right)_{;a} = -2d_2 L_a$, that is $L_a$ is a gradient KV.

Then equation \eqref{FL.1.c} gives
\begin{equation*}
K = s_1 \left( d_0t + \frac{d_1}{2} t^2 + \frac{d_2}{3} t^3 \right) + B_a
V^{,a} t + G(q)
\end{equation*}
which into \eqref{FL.1.b} gives $G_{,a} = - d_1 L_a$.

Observe that
\begin{equation*}
\begin{cases}
G_{,a} = - d_1 L_a \\
L_a = - \frac{1}{2 d_2} \left( B_b V^{,b} \right)_{;a}%
\end{cases}
\implies G_{,a} = \frac{d_1}{2 d_2} \left( B_b V^{,b} \right)_{,a} \implies
G = \frac{d_1}{2 d_2} B_b V^{,b} + const.
\end{equation*}

The first integral is (consists of $Q_{2}$, $Q_{5}$, $Q_{7}$)
\begin{equation*}
I_{e2}=\left( d_{0}+d_{1}t+d_{2}t^{2}\right) L_{a}\dot{q}^{a}+B_{a}\dot{q}%
^{a}+s_{1}\left( d_{0}t+\frac{d_{1}}{2}t^{2}+\frac{d_{2}}{3}t^{3}\right)
+B_{a}V^{,a}t+\frac{d_{1}}{2d_{2}}B_{b}V^{,b}
\end{equation*}%
where $L_{a}=-\frac{1}{2d_{2}}\left( B_{b}V^{,b}\right) _{;a}$ is a gradient
KV such that $L_{a}V^{,a}=s_{1}$ and $B_{a}$ is a KV.

\underline{\textbf{II.4. Case $\mathbf{(g = e^{\lambda t}}$, $\mathbf{m>2)}$.}} $\lambda \neq 0$ and $d_m \neq 0$.

\begin{equation*}
\begin{cases}
\eqref{FL.1.a1} \implies C_{(ab;c)} = 0 \\
\eqref{FL.1.a} \implies \lambda e^{\lambda t} C_{ab} + \left( d_0 + d_1 t +
... + d_m t^m \right) L_{(a;b)} + B_{(a;b)} = 0 \\
\eqref{FL.1.b} \implies - 2 e^{\lambda t} C_{ab} V^{,b} + \left( d_1 + 2 d_2
t + ... + m d_m t^{m-1} \right) L_a + K_{,a} = 0 \\
\eqref{FL.1.c} \implies K_{,t} = \left( d_0 + d_1 t + ... + d_m t^m \right)
L_a V^{,a} + B_a V^{,a} \\
\eqref{FL.1.d} \implies \left[ 2 d_2 + 3 \cdot 2 t + ... + m (m-1) d_m
t^{m-2} \right] L_a + \left( d_0 + d_1 t + ... + d_m t^m \right) \left( L_b
V^{,b} \right)_{;a} + \\
\qquad \qquad + \left( B_b V^{,b} \right)_{;a} - 2 \lambda e^{\lambda t}
C_{ab} V^{,b} = 0.%
\end{cases}%
\end{equation*}

From \eqref{FL.1.a} we have that $C_{ab} = 0$ and $L_a$, $B_a$ are KVs.

From \eqref{FL.1.d} we get that $L_a = 0$ and $B_a V^{,a} = s_2$.

Then equation \eqref{FL.1.c} gives $K = s_2 t + G(q)$ which into %
\eqref{FL.1.b} gives $G = const$.

Therefore the first integral is (of the form $Q_{2}$)
\begin{equation*}
I_{em}(m>2)=B_{a}\dot{q}^{a}+s_{2}t
\end{equation*}%
where $B_{a}$ is a KV such that $B_{a}V^{,a}=s_{2}$. This is not a quadratic
integral.

\vspace{12pt}

\textbf{III. For $\mathbf{n}$ finite and $\mathbf{m=\infty}$.} \vspace{12pt}

We distinguish between two cases because in the condition \eqref{FL.1.d} we
have to compare polynomial coefficients of the infinite sums $f_{,tt}$ and $%
f $. \vspace{12pt}

\textbf{III.1. Case with $\mathbf{f_{,tt} \neq \lambda^2 f}$.}

\begin{equation*}
(n=0, m=\infty) \equiv (n=0, m>2), \enskip (n=1, m=\infty) \equiv (n=1,
m>2), \enskip (n>1, m=\infty) \equiv (n>1, m>n).
\end{equation*}

\textbf{III.2. Case with $\mathbf{f_{,tt} = \lambda^2 f}$.}

\begin{equation*}
(n=0, m=\infty) \equiv (n=0, f=e^{\lambda t}), \enskip (n=1, m=\infty)
\equiv (n=1, f=e^{\lambda t}), \enskip (n>1, m=\infty) \equiv (n>1,
f=e^{\lambda t}).
\end{equation*}

For each subcase we have. \vspace{12pt}

\underline{\textbf{Subcase $\mathbf{(g=c_{0}}$, $\mathbf{f=e^{\lambda t})}$.}}  $\lambda \neq 0$.

\begin{equation*}
\begin{cases}
\eqref{FL.1.a1} \implies c_0 C_{(ab;c)} = 0 \\
\eqref{FL.1.a} \implies e^{\lambda t} L_{(a;b)} + B_{(a;b)} = 0 \\
\eqref{FL.1.b} \implies -2 c_0 C_{ab} V^{,b} + \lambda e^{\lambda t} L_a +
K_{,a} = 0 \\
\eqref{FL.1.c} \implies K_{,t} = e^{\lambda t} L_b V^{,b} + B_b V^{,b} \\
\eqref{FL.1.d} \implies \lambda^2 e^{\lambda t} L_a + e^{\lambda t} \left(
L_b V^{,b} \right)_{;a} + \left( B_b V^{,b} \right)_{;a} = 0.%
\end{cases}%
\end{equation*}

Equation \eqref{FL.1.a} implies that $L_a$, $B_a$ are KVs.

From \eqref{FL.1.d} we find that $B_a V^{,b} = s_2$; and $L_a = - \frac{1}{%
\lambda^2} \left( L_b V^{,b} \right)_{,a}$ that is $L_a$ is a gradient KV.

From \eqref{FL.1.c} we get that
\begin{equation*}
K = \frac{1}{\lambda} e^{\lambda t} L_b V^{,b} + s_2 t + G(q)
\end{equation*}
which into \eqref{FL.1.b} gives $G_{,a} = 2 c_0 C_{ab} V^{,b}$.

The first integral is ($c_{0}$ is absorbed by $C_{ab}$)
\begin{equation*}
I_{0e} = C_{ab}\dot{q}^{a}\dot{q}^{b} + e^{\lambda t} L_{a}\dot{q}%
^{a} +B_{a}\dot{q}^{a} + \frac{1}{\lambda } e^{\lambda t} L_{a}V^{,a} +
s_{2}t +G\left(q\right)
\end{equation*}%
where $\lambda \neq 0$, $L_{a} = - \frac{1}{\lambda^2} \left( L_bV^{,b}
\right)_{,a}$ is a gradient KV, $B_{a}$ is a KV such that $B_{a}V^{;a}=s_{2}$%
; and $C_{ab}$ is a KT such that $G_{,a}-2C_{ab}V^{,b}=0$.

The above FI is written
\begin{equation*}
I_{0e} = Q_{1} + Q_{2}(B_{a}) + Q_{10}
\end{equation*}
where
\begin{equation*}
Q_{10} = e^{\lambda t} \left( L_{a}\dot{q}^{a} + \frac{1}{\lambda}
L_{a}V^{,a} \right)
\end{equation*}
is a new independent FI.

\underline{\textbf{Subcase $\mathbf{(g = c_0 + c_1 t}$, $\mathbf{f = e^{\lambda t})}$.}} $\lambda \neq 0$ and $c_1 \neq 0$.

\begin{equation*}
\begin{cases}
\eqref{FL.1.a1} \implies C_{(ab;c)} = 0 \\
\eqref{FL.1.a} \implies c_1 C_{ab} + e^{\lambda t} L_{(a;b)} + B_{(a;b)} = 0
\\
\eqref{FL.1.b} \implies - 2 \left( c_0 + c_1 t \right) C_{ab} V^{,b} +
\lambda e^{\lambda t} L_a + K_{,a} = 0 \\
\eqref{FL.1.c} \implies K_{,t} = e^{\lambda t} L_b V^{,b} + B_b V^{,b} \\
\eqref{FL.1.d} \implies \lambda^2 e^{\lambda t} L_a + e^{\lambda t} \left(
L_b V^{,b} \right)_{;a} + \left( B_b V^{,b} \right)_{;a} - 2 c_1 C_{ab}
V^{,b} = 0.%
\end{cases}%
\end{equation*}

The first equation implies that $C_{ab}$ is a KT.

From \eqref{FL.1.a} we find that $L_a$ is a KV and $c_1 C_{ab} = - B_{(a;b)}$%
.

From \eqref{FL.1.d} we get the conditions $L_{a}=-\frac{1}{\lambda ^{2}}%
\left( L_{b}V^{,b}\right) _{;a}$ and $\left( B_{b}V^{,b}\right)
_{;a}=2c_{1}C_{ab}V^{,b}$.

Equation \eqref{FL.1.c} gives
\begin{equation*}
K=\frac{1}{\lambda }e^{\lambda t}L_{b}V^{,b}+B_{b}V^{,b}t+G(q)
\end{equation*}%
and by substituting into \eqref{FL.1.b} we end up with the relation (use
above conditions)
\begin{equation*}
G_{,a} = 2c_0C_{ab}V^{,b} = \frac{c_0}{c_1} \left( B_b V^{,b} \right)_{;a}
\implies G(q) = \frac{c_{0}}{c_{1}} B_{b} V^{,b}.
\end{equation*}

The first integral is
\begin{equation*}
I_{1e} = -\frac{1}{c_{1}}\left( c_{0}+c_{1}t \right) B_{(a;b)}\dot{%
q}^{a} \dot{q}^{b} + e^{\lambda t} L_{a}\dot{q}^{a} + B_{a}\dot{q}^{a} +
\frac{1}{\lambda} e^{\lambda t} L_{a}V^{,a} + B_{a}V^{,a} t + \frac{c_{0}}{%
c_{1}} B_{a}V^{,a}
\end{equation*}
where $\lambda c_{1} \neq 0$, $L_{a}=-\frac{1}{\lambda^{2}} \left(
L_{b}V^{,b}\right)_{,a}$ is a gradient KV and $B_a$ is such that $B_{(a;b)}$ is a KT and $\left(B_{b}V^{,b}\right)_{,a} = -2B_{(a;b)}V^{,b}$.

We compute $I_{1e} = \frac{c_{0}}{c_{1}}Q_{3} + Q_{4} + Q_{10}$.

\underline{\textbf{Subcase $\mathbf{(n>1}$, $\mathbf{f = e^{\lambda t})}$.}} $\lambda \neq 0$ and $c_n\neq 0$.

\begin{equation*}
\begin{cases}
\eqref{FL.1.a1} \implies C_{(ab;c)} = 0 \\
\eqref{FL.1.a} \implies (c_1 + 2c_2t + ... + nc_n t^{n-1}) C_{ab} +
e^{\lambda t} L_{(a;b)} + B_{(a;b)} = 0 \\
\eqref{FL.1.b} \implies - 2 \left( c_0 + c_1 t + ... + c_n t^n \right)
C_{ab} V^{,b} + \lambda e^{\lambda t} L_a + K_{,a} = 0 \\
\eqref{FL.1.c} \implies K_{,t} = e^{\lambda t} L_b V^{,b} + B_b V^{,b} \\
\eqref{FL.1.d} \implies \lambda^2 e^{\lambda t} L_a + e^{\lambda t} \left(
L_b V^{,b} \right)_{;a} + \left( B_b V^{,b} \right)_{;a} - 2 (c_1 + 2c_2t +
... + nc_n t^{n-1}) C_{ab} V^{,b} = 0.%
\end{cases}%
\end{equation*}

From \eqref{FL.1.a} we find that $C_{ab} = 0$ and $L_a$, $B_a$ are KVs. Then
equation \eqref{FL.1.d} implies that $B_a V^{,a} = s_2$ and $L_a = - \frac{1%
}{\lambda^2} \left( L_b V^{,b} \right)_{;a}$, i.e. $L_a$ is a gradient KV.

The solution of \eqref{FL.1.c} gives
\begin{equation*}
K = \frac{1}{\lambda} e^{\lambda t} L_b V^{,b} + s_2t + G(q)
\end{equation*}
which into \eqref{FL.1.b} gives
\begin{equation*}
\underbrace{\lambda e^{\lambda t} L_a + \frac{1}{\lambda} e^{\lambda t}
\left( L_b V^{,b} \right)_{,a}}_{=0} + G_{,a} = 0 \implies G = const.
\end{equation*}

The first integral is (consists of $Q_{2}$, $Q_{10}$)
\begin{equation*}
I_{ne}(n>1)=e^{\lambda t}L_{a}\dot{q}^{a}+B_{a}\dot{q}^{a}+\frac{1}{\lambda }%
e^{\lambda t}L_{a}V^{,a}+s_{2}t
\end{equation*}%
where $L_{a}=-\frac{1}{\lambda ^{2}}\left( L_{b}V^{,b}\right) _{;a}$ is a
gradient KV and $B_{a}$ is a KV such that $B_{a}V^{,a}=s_{2}$.

\vspace{12pt}

\textbf{IV. Both $\mathbf{m}$, $\mathbf{n}$ are infinite.} \vspace{12pt}

We consider three cases. \vspace{12pt}

\textbf{IV.1. Case where $\mathbf{f_{,tt} = \lambda^2 f}$ and $\mathbf{g_{,t} \neq \lambda f}$.}

\begin{equation*}
(n=\infty, m=\infty) \equiv (g=e^{\mu t}, f=e^{\lambda t}, \lambda \neq \mu)
\equiv (n>1, f=e^{\lambda t}).
\end{equation*}

We consider the general subcase. \vspace{12pt}

\underline{\textbf{Subcase $\mathbf{(g = e^{\lambda t}}$, $\mathbf{f = e^{\mu t})}$.}} $\lambda \neq 0$ and $\mu \neq 0$.

\begin{equation*}
\begin{cases}
\eqref{FL.1.a1} \implies C_{(ab;c)} = 0 \\
\eqref{FL.1.a} \implies \lambda e^{\lambda t} C_{ab} + e^{\mu t} L_{(a;b)} +
B_{(a;b)} = 0 \\
\eqref{FL.1.b} \implies - 2 e^{\lambda t} C_{ab} V^{,b} + \mu e^{\mu t} L_a
+ K_{,a} = 0 \\
\eqref{FL.1.c} \implies K_{,t} = e^{\mu t} L_a V^{,a} + B_a V^{,a} \\
\eqref{FL.1.d} \implies \mu^2 e^{\mu t} L_a + e^{\mu t} \left( L_b V^{,b}
\right)_{;a} + \left( B_b V^{,b} \right)_{;a} - 2 \lambda e^{\lambda t}
C_{ab} V^{,b} = 0.%
\end{cases}%
\end{equation*}
\vspace{12pt}

a) \underline{For $\lambda \neq \mu$:}

From \eqref{FL.1.a} we have that $C_{ab} = 0$ and $L_a$, $B_a$ are KVs.

From \eqref{FL.1.d} we find that $\mu^2 L_a + \left( L_b V^{,b} \right)_{;a}
= 0$ and $B_b V^{,b} = s_2$.

The solution of \eqref{FL.1.c} is
\begin{equation*}
K = \frac{1}{\mu} e^{\mu t} L_a V^{,a} + s_2 t + G(q)
\end{equation*}
which into \eqref{FL.1.b} using the relation $\mu^2 L_a + \left( L_b V^{,b}
\right)_{;a} = 0$ gives $G = const$.

The first integral is (consists of $Q_{2}$, $Q_{10}$)
\begin{equation*}
I_{ee}(\lambda \neq \mu )=e^{\mu t}L_{a}\dot{q}^{a}+B_{a}\dot{q}^{a}+\frac{1%
}{\mu }e^{\mu t}L_{a}V^{a,}+s_{2}t
\end{equation*}%
where $L_{a}=-\frac{1}{\mu ^{2}}\left( L_{b}V^{,b}\right) _{;a}$ is a
gradient KV and $B_{a}$ is a KV such that $B_{b}V^{,b}=s_{2}$. This is not a
quadratic integral.\vspace{12pt}

b) \underline{For $\lambda = \mu$:}

From \eqref{FL.1.a} we have that $\lambda C_{ab} + L_{(a;b)} = 0$ and $B_a$
is a KV.

From \eqref{FL.1.d} we find that $\lambda^2 L_a + \left( L_b V^{,b}
\right)_{;a} - 2 \lambda C_{ab} V^{,b} = 0$ and $B_b V^{,b} = s_2$.

The solution of \eqref{FL.1.c} is
\begin{equation*}
K = \frac{1}{\lambda} e^{\lambda t} L_a V^{,a} + s_2 t + G(q)
\end{equation*}
which into \eqref{FL.1.b} using the relation $\lambda^2 L_a + \left( L_b
V^{,b} \right)_{;a} - 2 \lambda C_{ab} V^{,b} = 0$ gives $G = const$.

The first integral is
\begin{equation*}
I_{ee}(\lambda = \mu) = -\frac{1}{\lambda } e^{\lambda t}
L_{(a;b)} \dot{q}^a \dot{q}^b + e^{\lambda t} L_a \dot{q}^a + B_a \dot{q}^a
+ \frac{1}{\lambda} e^{\lambda t} L_a V^{,a} + s_2 t
\end{equation*}
where $\lambda \neq 0$, $L_a$ is such that $L_{(a;b)}$ is a KT, $%
\lambda^{2}L_{a}+\left( L_{b}V^{,b}\right)_{,a} + 2L_{(a;b)}V^{,b}=0$ and $%
B_{a}$ is a KV with $B_{a}V^{,a}=s_{2}$.

We compute $I_{ee}(\lambda=\mu) = Q_{2}(B_{a}) + Q_{11}$ where
\begin{equation*}
Q_{11} = e^{\lambda t} \left( -\frac{1}{\lambda} L_{(a;b)} \dot{q}^a \dot{q}%
^b + L_{a}\dot{q}^{a} + \frac{1}{\lambda} L_{a}V^{,a} \right)
\end{equation*}
is a new independent FI. Observe that $Q_{10}=Q_{11}(L_{a}=KV)$.

\vspace{12pt}

\textbf{IV.2. Case where $\mathbf{f_{,tt} = \lambda^2 f}$ and $\mathbf{g_{,t} = \lambda f}$.}
\begin{equation*}
(n=\infty, m=\infty) \equiv (g=e^{\lambda t}, f=e^{\lambda t}).
\end{equation*}
\vspace{12pt}

\textbf{IV.3. Case where $\mathbf{f_{,tt} \neq \lambda^2 f}$ and $\mathbf{g_{,t} \neq \lambda f}$ or $\mathbf{g_{,t} = \lambda f}$.}
\begin{equation*}
(n=\infty, m=\infty) \equiv (n>1, m>n) \equiv (n>m+1, m>2) \equiv (g=
e^{\lambda t}, m>2) \equiv (n=m, m>2).
\end{equation*}

\bigskip

By collecting all the above FIs $Q_{A}$ the derivation of the Theorem \ref%
{The first integrals of an autonomous holonomic dynamical system} is
straightforward. Specifically, we cover all the FIs mentioned in the Theorem %
\ref{The first integrals of an autonomous holonomic dynamical system} as
follows:
\[
I_{1}=Q_{16}, \enskip I_{2}=Q_{7}, \enskip I_{3}= Q_{11}.
\]
The FIs $Q_{1}$, $Q_{3}$, $Q_{5}$, $Q_{6}$, $Q_{8}$, $Q_{9}$ are subcases of $I_{1}$; the $Q_{2}$, $Q_{4}$ are subcases of $I_{2}$; and finally $Q_{10}$ is a subcase of $I_{3}$.

Not all the above FIs are independent. After a careful and exhausted study we have shown that all these FIs can be produced by the three parameterized FIs listed in Theorem \ref{The first integrals of an autonomous holonomic dynamical system} section \ref{sec.tables.theorem}.

\theendnotes

\end{document}